\documentclass[epsfig,usegraphicx,useAMS,usenatbib]{mn2e}

\def\sqdeg{\,deg$^2$}
\def\etal{{et~al.~}}
\def\msolh{\,{\it h}^{-1}\, {\rm M_{\odot}}}
\def\msol{\,{\rm M_{\odot}}}
\def\mpch{\,{\it h}^{-1}\, {\rm Mpc}}
\def\mpc{\,{\rm Mpc}}

\def\Mpch{$\mpch$}
\def\kms{{\rm \,km \, s^{-1}}}

\title[GAMA: The GAMA Galaxy Group Catalogue (G$^3$Cv1)]{Galaxy and Mass Assembly (GAMA): The GAMA Galaxy Group Catalogue (G$^3$Cv1)}
\author[A.S.G.~Robotham~\etal]{A.S.G.~Robotham$^{1}$\thanks{E-mail:asgr@st-and.ac.uk},
P.~Norberg,$^2$
S.P.~Driver,$^{1,3}$
I.K.~Baldry,$^4$
S.P.~Bamford,$^5$
\newauthor A.M.~Hopkins,$^6$
J.~Liske,$^7$
J.~Loveday,$^8$
A.~Merson,$^9$
J.A.~Peacock,$^2$
S.~Brough,$^6$
\newauthor E.~Cameron,$^{10}$
C.J.~Conselice,$^5$
S.M.~Croom,$^{11}$
C.S.~Frenk,$^{9}$
M.~Gunawardhana,$^{11}$
\newauthor D.T.~Hill,$^1$
D.H.~Jones,$^{12}$
L.S.~Kelvin,$^1$
K.~Kuijken,$^{13}$
R.C.~Nichol,$^{14}$
\newauthor H.R.~Parkinson,$^2$
K.A.~Pimbblet,$^{12}$
S.~Phillipps,$^{15}$
C.C.~Popescu,$^{16}$
M.~Prescott,$^4$
\newauthor R.G.~Sharp,$^{17}$
W.J.~Sutherland,$^{18}$
E.N.~Taylor,$^{11}$
D.~Thomas,$^{14}$
R.J.~Tuffs,$^{19}$
\newauthor E.~van~Kampen,$^7$
D.~Wijesinghe,$^{11}$\\\\
$^1$SUPA\thanks{Scottish Universities Physics Alliance}, School of Physics \& Astronomy, University of St Andrews, North Haugh, St Andrews, KY16 9SS, UK\\
$^2$SUPA, Institute for Astronomy, University of Edinburgh, Royal Observatory, Blackford Hill, Edinburgh EH9 3HJ, UK\\
$^3$ICRAR\thanks{International Centre for Radio Astronomy Research}, The University of Western Australia, 35 Stirling Highway, Crawley, WA 6009, Australia\\
$^4$Astrophysics Research Institute, Liverpool John Moores University, Egerton Wharf, Birkenhead, CH41 1LD, UK\\
$^5$Centre for Astronomy and Particle Theory, University of Nottingham, University Park, Nottingham NG7 2RD, UK\\
$^6$Australian Astronomical Observatory, PO Box 296, Epping, NSW 1710, Australia\\
$^7$European Southern Observatory, Karl-Schwarzschild-Str.~2, 85748 Garching, Germany\\
$^8$Astronomy Centre, University of Sussex, Falmer, Brighton BN1 9QH, UK\\
$^9$Institute for Computational Cosmology, Department of Physics, Durham University, South Road, Durham DH1 3LE, UK\\
$^{10}$Department of Physics, Swiss Federal Institute of Technology (ETH-Z{\" u}rich), 8093 Z{\" u}rich, Switzerland\\
$^{11}$Sydney Institute for Astronomy, School of Physics, University of Sydney, NSW 2006, Australia\\
$^{12}$School of Physics, Monash University, Clayton, Victoria 3800, Australia\\
$^{13}$Leiden University, P.O.~Box 9500, 2300 RA Leiden, The Netherlands\\
$^{14}$Institute of Cosmology and Gravitation (ICG), University of Portsmouth, Dennis Sciama Building, Portsmouth PO1 3FX, UK\\
$^{15}$HH Wills Physics Laboratory, University of Bristol, Tyndall Avenue, Bristol, BS8 1TL, UK\\
$^{16}$Jeremiah Horrocks Institute, University of Central Lancashire, Preston PR1 2HE, UK\\
$^{17}$Research School of Astronomy \& Astrophysics, Mount Stromlo Observatory, Cotter Road, Western Creek, ACT 2611, Australia\\
$^{18}$Astronomy Unit, Queen Mary University London, Mile End Rd, London E1 4NS, UK\\
$^{19}$Max Planck Institute for Nuclear Physics (MPIK), Saupfercheckweg 1, 69117 Heidelberg, Germany\\
\vspace*{-3em}
}

\begin{document}

\date{\vspace*{-5em}\noindent12/04/2011}

\pagerange{\pageref{firstpage}--\pageref{lastpage}} \pubyear{2011}

\maketitle

\label{firstpage}

\begin{abstract}
Using the complete GAMA-I survey covering $\sim142$~\sqdeg\ to $r_{\rm AB}=19.4$,
of which $\sim47$~\sqdeg\ is to $r_{\rm AB}=19.8$, we create 
the GAMA-I galaxy group catalogue (G$^3$Cv1), generated using a
friends-of-friends (FoF) based grouping algorithm. Our algorithm has
been tested extensively on one family of mock GAMA lightcones, 
constructed from $\Lambda$CDM N-body simulations populated with 
semi-analytic galaxies. Recovered group properties are robust to 
the effects of interlopers and are median unbiased in the most important 
respects.
G$^3$Cv1 contains 14,388 galaxy groups (with multiplicity $\ge 2$), 
including 44,186 galaxies out of a possible 110,192 galaxies,
implying $\sim$40\% of all galaxies are assigned to a group. 
The similarities of the mock group catalogues and G$^3$Cv1 are 
multiple: global characteristics are in general well recovered.
However, we do find a noticeable deficit in the number of high
multiplicity groups in GAMA compared to the mocks. 
Additionally, despite exceptionally good local spatial completeness, 
G$^3$Cv1 contains significantly fewer compact groups with 5 or more members, this effect becoming most evident for high multiplicity systems. 
These two differences are most likely due to limitations  
in the physics included of the current GAMA lightcone mock. 
Further studies using a variety of galaxy formation models are required 
to confirm their exact origin.
The G$^3$Cv1 catalogue will be made publicly available as and when the
relevant GAMA redshifts are made available at
{\tt http://www.gama-survey.org}.
\end{abstract}


\section{Introduction}
 
Galaxy group and cluster catalogues have a long history in
astronomy. Early attempts at creating associations of galaxies were
quite qualitative in nature \citep[e.g.~][]{abel58,zwic61}, but 
more recently significant effort has been devoted to 
robustly detecting grouped structures 
\citep[e.g.~][]{huch82,moor93,eke04,gerk05,yang05,berl06,brou06,knob09}. 
The pioneering application of this process was by \citet{huch82},
where the catalogue of \citet{de-v75}, the earliest reasonably
complete attempt at a group catalogue, was reconstructed using fully
quantitative means--- i.e.\ by a method that was reproducible and not
subjective.

The power of group catalogues resides in their relation to the theoretically motivated dark matter haloes.
$\Lambda$CDM, the literatures current favoured structure formation paradigm, makes very strong predictions about the self similar
hierarchical merging process that occurs between haloes of dark matter
\citep{spri05}. Galaxy groups are the observable equivalent 
of dark matter haloes, and thus offer a direct insight into the physics that has
occurred in the dark matter haloes in the Universe up to the present day.
Further to offering a route to studying dark matter dynamics \citep[e.g.][]{plio06,robo08}, analysis of galaxy groups
opens the way to understanding how galaxies
populate haloes \citep[e.g.][]{coor02,yang03,coor06,robo06,robo10b}.

The strongest differentials between competing physical models of dark
matter are found at the extremes of the halo mass function (HMF),
i.e.\ on cluster scales \citep[e.g.][]{eke96} and on low mass scales. 
The HMF describes the comoving number density distribution of dark matter haloes as a function of halo mass. 
Low mass groups are 
highly sensitive to the temperature of the CDM. We either expect to see
a continuation of the near power-law prediction for the HMF down to Local Group 
mass haloes \citep[see][and references therein]{jenk01} for a cold dark matter
Universe, or, as the dark matter becomes warmer, the slope should become
suppressed significantly. 

The Galaxy and Mass Assembly project (GAMA) is a major new
multi-wavelength spectroscopic galaxy survey \citep{driv11}. The final
redshift survey will contain $\sim$400,000 redshifts to $r_{\rm AB}=19.8$
over $\sim360$~\sqdeg, with a survey design aimed at providing an
exceptionally uniform spatial completeness
\citep{robo10a,bald10,driv11}. One of the principal science goals of
GAMA is to make a statistically significant analysis of low mass
groups ($M \leq 10^{13}\msolh$), helping to constrain the low
mass regime of the dark matter HMF
and galaxy formation efficiency in Local Group like haloes. 

As well as allowing us to determine galaxy group dynamics and composition at the highest fidelity possible due to the increased redshift density, GAMA will also provide mult$i$-band photometry spanning the UV (GALEX), visible (SDSS; VST-KIDS), near-IR (UKIDSS-LAS, VIKING), mid-IR (WISE), far-IR (ATLAS) and radio (GMRT, ASKAP). By combining a GAMA Galaxy Group Catalogue (G$^3$C) constructed with spatially near-complete redshifts and 21 band photometry, the GAMA project is in a unique position to answer many of the most pressing questions that exist in extra-galactic astronomy today. Crucially, the interplay between Star Formation Rate (SFR), stellar mass, morphology, QSO activity and Star Formation Efficiency (SFE) with environment can be probed in unprecedented detail. The group catalogue presented here will also enable galaxy evolution to be investigated as a function of halo mass, rather than with coarse environmental markers, in statistically significant low mass regimes for the first time. This is a huge advance on the capabilities of current large spectroscopic surveys like SDSS and 2dFGRS that are almost single pass and hence suffer seriously from spectroscopic incompleteness in clustered regions. GAMA, by being at least 6 pass in every unit of sky, is exceptionally complete on all angular scales \citep{robo10a,driv11}.

The catalogue and group analyses presented here is based on the first
three years of spectroscopic observations (February 2008 to May 2010)
made at the Anglo-Australian Telescope (AAT). Within the GAMA project,
this period is referred to as GAMA-I, since the deeper, larger area,
continuation of the GAMA survey is commonly referred to as GAMA-II. 

The paper is organized as follows. \S\ref{sec:algo} describes
the precise FoF grouping algorithm, the GAMA data and the lightcone
mocks used for the present analysis. The testing and grouping
parameter optimisation using the mocks are considered
in~\S\ref{sec:group_para}. Group properties 
(i.e. velocity dispersion, radius, dynamical mass and total
luminosity) and their estimates are presented in~\S\ref{sec:prop_est}.
\S\ref{sec:properties} presents global group properties for G$^ 3$C
and corresponding mock group catalogues. A few GAMA group examples are
discussed in \S\ref{sec:examples}, with conclusions presented in
\S\ref{sec:conclusions}. We assume throughout an $\Omega_{\rm m}=0.25$,
$\Omega_\Lambda=0.75$, H$_0 = h \, 100 \kms \, \mpc^{-1}$
cosmological model, corresponding to the cosmology of the Millennium
N-body simulation used to construct the GAMA lightcone mocks.

\section{Galaxy grouping: algorithm, data and mocks}
\label{sec:algo}

There are many subtle differences in the specific algorithm used to
construct groups from spectroscopic surveys, but the major dichotomy
occurs at the scale of association considered: galaxy-galaxy links or
halo-galaxy links. Here we adopt galaxy-galaxy linking via a
Friends-of-Friends (FoF) algorithm (\S\ref{sec:fof}), having also
explored a halo-galaxy grouping and found it to be less successful at
recovering small mass groups from our mock galaxy catalogues. The halo
method implemented was a variant of the Voronoi tessellation scheme
used in \citet{gerk05}, which worked reasonably well for larger groups and
clusters, but was not competitive compared to our FoF implementation
in the low halo mass regime.

\subsection{Friends of Friends}
\label{sec:fof}

\begin{figure}
\centerline{\mbox{\includegraphics[width=3.5in]{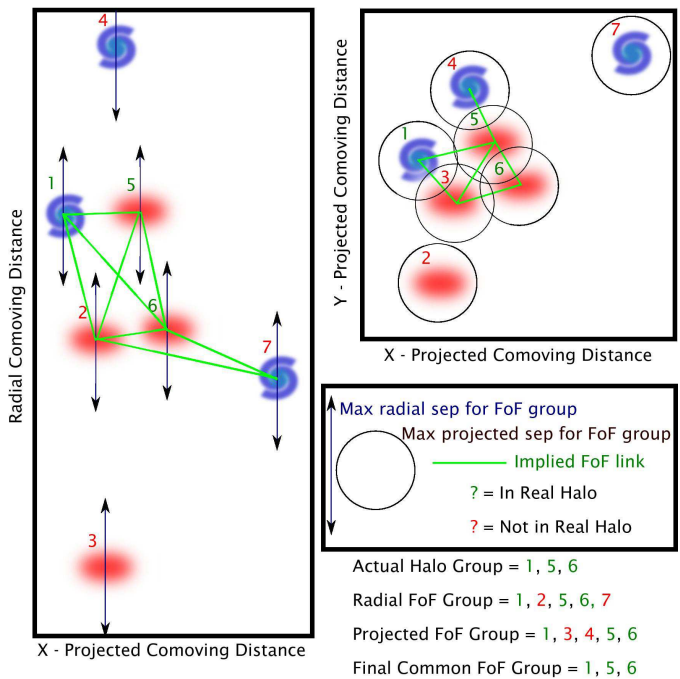}}}
\caption{\small Schematic of the two step process used when associating galaxies via FoF algorithm on 
  redshift survey data. The same set of galaxies are
  shown in two panels: along the line of sight (left) and projected on
  the sky (right). Both the radial and projected separations are used
  to disentangle projection effects and recover the underlying group
  (galaxies 1, 5 and 6 in this example). The radial linking length has
  to be significantly larger than the projected one to properly
  account for peculiar velocities along the line of sight.}
\label{fig:FoFscheme}
\end{figure}

A standard Friends-of-Friends algorithm creates links between galaxies
based on their separation as a measure of the local density. In 
practice the projected and radial separations are treated separately,
due to significant line-of-sight effects from peculiar velocities
within groups and clusters. The comoving radial separations within a group
appear larger than the projected ones, because radial distances
inferred from galaxy redshifts contain peculiar velocity
information along the line of sight on top of their underlying Hubble
distance away from the observer.
Fig.~\ref{fig:FoFscheme} shows schematically how the radial and
projected separations are used to detect a group. This shows that
neither the radial nor the projected separation provide enough
information to unambiguously detect a group, but their combination
generate a secure grouping.

\subsubsection{Projected linking condition}

In its simplest form we can say that two galaxies are associated in
projection when the following condition is met:

\begin{equation}
\tan[\theta_{1,2}] (D_{\rm com,1} + D_{\rm com,2})/2 \leq b_{\rm i,j} (D_{\rm lim,1}+D_{\rm lim,2})/2 \; ,
\label{eq:D_perp_12}
\end{equation}

\noindent where $\theta_{1,2}$ is the angular separation of the two
galaxies, $D_{\rm com,i}$ is the radial distance in comoving
coordinates to galaxy $i$, $b_{\rm i,j}$ the mean required linking overdensity and $D_{\rm lim,i}$
is the mean comoving inter-galaxy separation at the position of galaxy
$i$, here defined as

\begin{equation}
D_{{\rm lim},i}=\left[ \int^{M_{{\rm lim},i}}_{-\infty}\phi(M)dM \right]^{-1/3} \; ,
\label{eq:D_lim}
\end{equation}

\noindent where $M_{{\rm lim},i}$ is the effective absolute magnitude
limit of the survey at the position of galaxy $i$, $\phi(M)$ the
survey galaxy luminosity function (LF).

$b$ is used to specify the overdensity with respect to the mean
required to define a group. 
The approximate overdensity contour that this linking would recover
in a simulation (Universe) with equal mass particles (galaxies) is given by $\rho / \bar{\rho} \sim 3/(2 \pi b^3) $ \citep{cole96}. For a uniform spherical distribution
of mass the virial radius corresponds to a mean overdensity of 178,
hence the popularity of masses defined as being within 178 and 200 times the mean
overdensity. For an NFW type profile \citep{NFW96} the overdensity within the virial
radius is approximately $178/3\simeq59$. This implies an interparticle linking length of $b\simeq0.2$ in real space,
corresponding to a volume overdensity $1/b^{3}=125$ between galaxies. Linking together
1000s of dark matter particles in a simulation with real-space coordinates is a relatively
simple and robust process, extending this methodology to redshift-space using galaxies that trace the dark matter is
non-trivial. Consequently, it is not simply true to state that $b=0.2$ will return the virial mass limits for
each galaxy group in the GAMA catalogue. Instead, $b$ will be recovered from careful application to mock
catalogues (see below for full details). Since there a subtle effects that vary the precise $b$ used
on a galaxy by galaxy basis $b_{\rm i,j}$ used above is the mean $b$ for galaxy i and j respectively. In general,
for near-by galaxies, $b$ does not vary significantly.

To this standard form of the mean comoving inter-galaxy separation at
the position of galaxy $i$, 
we introduce an extra term, with Eq.~\ref{eq:D_lim} thus becoming:

\begin{equation}
D_{{\rm lim},i}= \left(\frac{\phi(M_{{\rm lim},i})}{\phi(M_{{\rm gal},i})} \right)^{\nu/3} \, \left[\int^{M_{{\rm lim},i}}_{-\infty}\phi(M)dM \right]^{-1/3} \; ,
\label{eq:D_lim_nu}
\end{equation}

\noindent where $M_{{\rm gal},i}$ is the absolute magnitude of galaxy $i$. 
This extra term, 
$(\phi(M_{{\rm lim},i})/\phi(M_{{\rm gal},i}))^{\nu/3}$, 
allows for larger linking distances for intrinsically brighter
galaxies, provided $\nu > 0$ and the LF is strictly increasing (which
is true for GAMA). 
Adjusting $\nu$ allows the algorithm to be more or less sensitive to
the intrinsic brightness of a galaxy, and can be thought of as a
softening power. The principle behind introducing this term is that
associations should be more significant between brighter galaxies, and
tests on mocks show that this generates notably better quality group
catalogues as determined from the cost function (see \S\ref{sec:cost_fct}).

\subsubsection{Line-of-sight linking condition}

With Eq.~\ref{eq:D_perp_12} we have established an association in
projection, but we also require that a given pair of galaxies are
associated along the line-of-sight or radially, i.e.: 

\begin{equation}
|D_{\rm com,1}-D_{\rm com,2}| \leq b \, R \, (D_{\rm lim,1}+D_{\rm lim,2})/2 \; ,
\label{eq:D_ls_12}
\end{equation}

\noindent where $b$ is the linking length of Eq.~\ref{eq:D_perp_12}, $D_{\rm lim,i}$ is given by Eq.~\ref{eq:D_lim_nu} and $R$ is the radial expansion factor to account for peculiar motions of galaxies within groups. With a redshift survey, the measured redshift contains both information on the Hubble flow redshift and any galaxy peculiar velocity along the line of sight.

\subsubsection{Global linking conditions}

To construct a group catalogue we link together all associations that
meet our criteria given by Eq.~\ref{eq:D_perp_12} and
Eq.~\ref{eq:D_ls_12}. Galaxies that are not directly linked to each other
can still be grouped together by virtue of common links between
them. All possible groups are constructed in precisely this manner,
leaving either completely ungrouped galaxies or galaxies in groups
with 2 or more members. 

Despite its apparent simplicity, a FoF algorithm is still a very
parametric approach to grouping. On top of the assumed cosmology, 
it requires the survey selection function,
and values for the linking parameters $b$ and $R$. 
The galaxy LF can be directly estimated from the data 
\citep[e.g.~][]{love95, norb02, blan03}, while the linking
parameters cannot be estimated from the data. Instead they are
commonly determined from either analytic calculations or analyses of
N-body simulations populated with galaxies, with the latter approach
taken here (see \S\ref{sec:mocks} for the description of the GAMA
lightcone mocks).  

Merely using a static combination of $b$ and $R$ is less than optimal
for accurately reconstructing groups in the mock data. An obvious
shortcoming is that galaxies in clusters are significantly spread out
along the line-of-sight, due to their large peculiar velocities a
result of being bound to massive structures. To account for this we
introduce a local environment measure that calculates the density
contrast of a cylinder that is centred on the galaxy of
interest. Similar to the approach of \citet{eke04}, we allow the $b$
and $R$ parameters to scale as a function of the observed density
contrast, leading to position (${\bf r}$) and faint magnitude limit
($m_{\rm lim}$) dependent linking parameters:

\begin{eqnarray}
b({\bf r},m_{\rm lim}) & = & b_0 \, \left(\frac{1}{\Delta}\frac{\rho_{\rm
    emp}({\bf r},m_{\rm lim})}{\bar{\rho}({\bf r},m_{\rm lim})}\right)^{E_{\rm b}} \\
R({\bf r},m_{\rm lim}) & = & R_0 \, \left(\frac{1}{\Delta}\frac{\rho_{\rm
    emp}({\bf r},m_{\rm lim})}{\bar{\rho}({\bf r},m_{\rm lim})}\right)^{E_{\rm R}}
\label{eq:link_para}
\end{eqnarray}

\noindent where $\bar{\rho}$ is the average local density implied by the
selection function, $\rho_{\rm emp}$ is the empirically estimated
density, $m_{\rm lim}$ the apparent magnitude limit at position
${\bf r}$ and $\Delta$ is the density contrast, an
additional free parameter together with $E_{\rm b}$ and $E_{\rm R}$.
For this work $\bar{\rho}$ is estimated from the galaxy selection function
at ${\bf r}$ (i.e.\ it varies with the GAMA survey depth). 
$\rho_{\rm emp}$ is calculated directly from the number density within
a comoving cylinder centred on ${\bf r}$ and of projected
radius $r_{\Delta}$ and radial extent $l_{\Delta}$.
$\Delta$ determines the transition between where the power scaling
reduces or increases the linking lengths, so a galaxy within a local
volume precisely $\Delta$ times overdense will not have its links
altered. The exact values for $E_{\rm b}$, $E_{\rm R}$ and $\Delta$
are determined from the joint optimisation of the group cost function
(see \S\ref{sec:cost_fct}) for all the parameters that affect the
quality of the grouping when tested on the mocks. The parameters
required for the FoF algorithm described above are now: $b_0$, $R_0$,
$\Delta$, $r_{\Delta}$, $l_{\Delta}$, $E_{\rm b}$, $E_{\rm R}$ and
$\nu$. Whilst many parameters, $b_0$ and $R_0$ are the dominant one for
the grouping, the latter 6 merely determining how best to modify the
linking locally, and typically introducing minor perturbations to the
grouping. 

\subsubsection{Completeness corrections}

Since the GAMA
survey is highly complete ($\sim$98\% within the $r$-band limits) the
effect of incompleteness is minor, and tests on the mocks indicate the
final catalogues are extremely similar regardless of whether the
linking length is adjusted based on the local completeness. A number
of definitions of local completeness were investigated: completeness
within a pixel on a mask, completeness on a fixed angular top-hat
scale around each galaxy and a completeness window
function that represents the physical scale of a group on the sky. The
difference between each was quite minor, but defining completeness on
a physical scale produced marginally better grouping costs
(\S\ref{sec:optimisation}). Hence the completeness corrected linking
parameter $b$ at position ${\bf r}$ is given by:

\begin{equation}
b_{\rm comp}({\bf r},m_{\rm lim}) = \frac{b({\bf r},m_{\rm lim})}{c({\bf r})^{1/3}} \; ,
\end{equation}

\noindent where $c({\bf r})$ is the redshift completeness within a
projected comoving radius of $1.0\mpch$ centred on ${\bf r}$. The
effect is to slightly increase the linking length to account for the
loss of (possible) nearby galaxies that it could otherwise be linked
with. Since GAMA was designed to be extremely complete even at small
angular scales \citep{robo10a}, the mean modifications are less than
1\%.

\subsection{Data: GAMA survey}
\label{sec:data}

Extensive details of the GAMA survey characteristics are given in
\citet{driv11}, with the survey input catalogue described in
\citet{bald10} and the spectroscopic tiling algorithm in
\citet{robo10a}.

Briefly, the GAMA-I survey covers three regions  
each $12 \times 4$ degrees centred at 09h, 12h and 14h30m
(respectively G09, G12 and G15 from here). The survey
depths and areas relevant for this study are: $\sim96$~\sqdeg\ to
$r_{\rm AB}=19.4$ (G09 and G15) and $\sim47$~\sqdeg\ to $r_{\rm AB}=19.8$
(G12)\footnote{See \citet{bald10} for additional GAMA-I selections.}.
All regions are more than 98\% complete 
\citep[see][for precise completeness details]{driv11}, with special
emphasis on a high close pair completeness, which is greater than 95\%
for all galaxies with up to 5 neighbours within 40$''$ of them 
\citep[see Fig.~19 of][]{driv11}\footnote{99.8\% of all galaxies
  have 5 or fewer neighbours within 40$''$.}. 
Despite this high global redshift completeness, we still apply
completeness corrections to the FoF algorithm (as described in \S\ref{sec:fof}) and use the
masks described in \citet{bald10} and \citet{driv11}, to account for
areas masked out by bright stars, poor imaging, satellite trails, etc.
{ 
The velocity errors on GAMA redshifts are typically $\sim50\kms$
\citep{driv11}, slightly larger than 
  the nominal SDSS velocity uncertainties of $\sim35\kms$ but
  significantly better than the typical $\sim80\kms$ associated with
  2dFGRS redshifts \citep{coll01}.
}

For this study, we use a global GAMA $(k+e)$-correction, of the form:
\begin{equation}
(k+e)(z)=\sum_{i=0}^{N} a_i(z_{\rm ref},z_{p}) (z-z_{p})^i + Q_{z_{\rm ref}} (z-z_{\rm ref})
\label{eq:kpe_global}
\end{equation}
where $z_{\rm ref}$ is the reference redshift to which all galaxies are
$(k+e)$-corrected, $Q_{z_{\rm ref}}$ is a single luminosity evolution
parameter \citep[as in e.g.][]{lin99}, $z_{p}$ is a reference
redshift for the polynomial fit to median KCORRECT-v4.2
k-correction \citep{blan07} of GAMA-I galaxies, and
$a_i(z_{\rm ref},z_{p})$ the coefficients of that polynomial fit. The
present study uses $z_{\rm ref}=0$, $Q_0=1.75$, $z_{p}=0.2$ and $N=4$,
with $a={0.2085, 1.0226, 0.5237, 3.5902, 2.3843}$, for both data and
mocks. The precise value for $Q_0=1.75$ is not essential, as our
estimate of the luminosity function accounts for any residual 
redshift evolution.

Once the global $(k+e)$-correction have been defined, it is
straightforward to estimate the redshift dependent galaxy luminosity
function using a non-parametric estimator like the  Step-Wise Maximum Likelihood (SWML) of
\citet{efst88}. We perform this analysis in five disjoint redshift
bins, which are all correlated through the global normalisation
constraint. This is set by the cumulative number counts at
$r_{\rm AB}=19.8$ ($\sim1050$ galaxies/\sqdeg), as estimated directly from
the full GAMA survey and compared to $\sim 6250$~\sqdeg\ of SDSS DR6
survey (to account for possible sample variance issues). This LF estimate
is used both to described the survey selection function (as
required by Eqs.~\ref{eq:D_perp_12}--\ref{eq:link_para}), to adjust
the galaxy magnitudes in the GAMA mock catalogues (see
\S\ref{sec:mocks}) and is hereafter referred to as $\phi_{\rm GAMA}$.

\subsection{GAMA mock catalogues}
\label{sec:mocks}

To appropriately test the quality and understand the intrinsic
limitations of a given group finder it is essential to test it
thoroughly on a series of realistic mock galaxy catalogues, for which
the true grouping is known. Those tests should include all the
limitations of the real spectroscopic survey, e.g. spectroscopic
incompleteness, redshift uncertainties, varying magnitude limits,
etc. 

In this first paper on GAMA groups, we limit our tests of the group
finding algorithm to one single type of mock galaxy catalogue,
constructed from the Millennium dark matter simulation
\citep{spri05}, populated with galaxies using the GALFORM~\citet{bowe06}
semi-analytic galaxy formation recipe. The galaxy positions are
interpolated between the Millennium snapshots to best mimic the effect
of a proper lightcone output, enabling the mocks to include the
evolution of the underlying dark matter structures along the line of
sight, key for a survey of the depth of GAMA that spans
$\sim$4Gyr. Finally, the semi-analytic galaxies have their SDSS $r$-band
filter magnitudes modified to give a perfect match to the redshift
dependent GAMA luminosity and selection function (see
\S\ref{sec:data}; Loveday et al., in prep). When adjusting the
magnitudes, we use the global GAMA k+e correction of
Eq.~\ref{eq:kpe_global}. The 9 mock galaxy catalogues 
have the exact GAMA survey geometry, with each mock extracted
from the N-body simulation while preserving the true angular
separation between the three GAMA regions.

The main limitations of this first generation of GAMA mock galaxy
catalogue for the present group study are listed below: 

\begin{itemize}

\item[1)] the luminosity dependent galaxy clustering does not
  perfectly match the data \citep{kim09}, in particular in redshift space (Norberg et al. in prep). By their
  nature, semi-analytic mock galaxy catalogue are not constrained
  precisely to match in any great detail the observed clustering
  signal 
  \citep[as opposed to halo occupation distributions (HOD) or conditional luminosity functions (CLF) mocks, e.g.~][]{coor02, yang03, coor06}.

\item[2)] the GAMA survey is so spectroscopically complete to the
  GAMA-I survey limits (above 98\% on scales relevant for this study)
  that no attempt of modelling any residual survey incompleteness into
  the mocks have been made. 

\item[3)] apparent magnitude uncertainties have a negligible effect on
  the GAMA survey selection and hence are not accounted for in these
  mocks.
  
\item[4)]  velocity measurement uncertainties are not incorporated
  into the mocks.

\item[5)] the 9 GAMA mocks are not statistically independent, as they
  are drawn from a single N-body simulation. However, we ensure in the
  construction of the different mocks that no single galaxy at the
  exact same stage of evolution is found in more than one mock,
  i.e. there is no spatial overlap between the 9 GAMA lightcone mocks
  created. 

\item[6)] { despite the high numerical resolution of the Millennium
  dark matter simulation, the 
          lightcones used for this work, once the shift in magnitudes have
  been accounted for, are not complete below 
  M$_{\rm r_{\rm  AB}} - 5~\log_{10} h \simeq -14.05$.  
  This limit is faint enough to not attempt to address this issue in
  this first generation of GAMA mocks.} 

\item[7)] the halo definition used in these mocks correspond to
  standard halo definition of GALFORM~\citep{cole00,bowe06,bens10},
  i.e. DHalo~\citep{helly03}, as listed in the Millennium GAVO
  database\footnote{\tt http://www.g-vo.org/Millennium}. 
  DHalo is a collection of SubFind subhaloes~\citep{spri01} grouped
  together to make a halo. The differences between DHalo and
  FoFHalo\footnote{FoFHalos are identified with a linking length of
    $b=0.2$ in the underlying dark matter simulation} are subtle. 
  A preliminary analysis on a small fraction
  of the mock data show that the log ratio of the DHalo and
  FoFHalo masses are median unbiased, and exhibit a 1-$\sigma$ scatter
  of 0.05 dex. The 10\% population that exhibits the largest mass
  mismatch are still median unbiased (i.e.\ they will not affect the
  median relationship between the FoF masses we measure and
  the intrinsic dark matter mass of the halo), but can scatter
  more than 1 dex away from the median. Because the two halo mass definitions are not biased w.r.t.\ each other, the DHalo mass 
  can be used safely in this paper as a halo mass definition.

\item[8)] the most luminous galaxy of a halo is nearly always at its
  centre and at rest w.r.t. the dark matter halo.  

\end{itemize}

These mocks are a subset of the first generation of wide and deep mock
galaxy catalogues for the Pan-STARRS PS1 survey. Further details on
their construction are given in Merson et al. (in prep.).

\section{Parameter optimisation using mock catalogues}
\label{sec:group_para}

The minimisation or maximisation of non-analytic functions that depend
on multiple parameters is an intense research area in statistics and
computational science. 
When the dimensionality of the dataset is low, typically 2--4 dimensions,
it is straightforward to completely map out the whole parameter space
on a grid. However, when the number of parameters is large (e.g. up to
8 for our FoF algorithm) then such a computationally intensive
approach is not ideal, especially if each set of parameter values
requires a series of complex calculations. For our data size and
problem considered, each complete grouping takes 10s of seconds, with
a full parameter space not necessarily obvious to define.
Hence we use the Nelder-Mead optimisation technique \citep[i.e.\
downhill simplex, see][]{neld65} that allows for maxima (or minima) to be
investigated for non-differentiable functions. The onus is still on
the user to choose an appropriate function to maximise.
For this work we desire a high group detection rate with a low
interloper fraction in each group, and this is the criteria that defines
the cost function to be minimised.

\subsection{Group cost function}
\label{sec:cost_fct}

One of the defining characteristics of how we decide to determine grouping quality is that the statistics measured should be two-way (bijective). By this we mean that the group catalogue made with this algorithm is an accurate representation on the mock group catalogue, and vice-versa. This is an important distinction since it is possible for the group catalogue to perfectly recover every mock group, but for these to be the minority of the final catalogue, i.e.\ most of the groups are spurious. This has a serious effect on almost any science goal involving use of the GAMA groups since any given group would be more likely to be false than real--- follow up proposals making use of the groups would be highly inefficient, and any science involving the stacking of detections of multiple groups (X-ray, HI) would be hard to achieve.

With this two-way nature of defining grouping quality in mind, there are two global measures that can be ascertained: how well are the groups and the galaxies within them recovered. To retrieve a group accurately we require the joint galaxy population of the FoF groups and mock haloes to include more than 50\% of their respective group members. This is called a bijective match, and it ensures that there is no ambiguity when we associate groups together--- it is impossible for a group to bijectively match more than one group. To turn this into a global grouping efficiency statistic we define the following quantities:

\begin{eqnarray}
E_{\rm FoF}&=&\frac{Ng_{\rm bij}}{Ng_{\rm FoF}} \\
E_{\rm mock}&=&\frac{Ng_{\rm bij}}{Ng_{\rm mock}} \\
E_{\rm tot}&=&E_{\rm FoF}E_{\rm mock}
\end{eqnarray}

\noindent where $Ng_{\rm bij}$, $Ng_{\rm FoF}$ and $Ng_{\rm mock}$ are the number of bijective, FoF and mock groups respectively. $E_{\rm tot}$ is the global halo finding efficiency measurement (or purity product) we want to use in our maximisation statistic, and will be 1 if all groups are bijectively found, and 0 if no groups are determined bijectively.

The second measure of group quality determines how significantly matched individual groups are, in effect it determines the `purity' of the matching groups. The best two-way matching group is the one which has the largest product for the relative membership fractions between the FoF and mock group. Take for example a FoF group with 5 members where 3 of these galaxies are shared with a mock group that has 9 members and the other 2 are shared with a mock group that has 3 members. In this case the two possible purity products are $\frac{3}{5}\times\frac{3}{9}=\frac{9}{45}=0.2$ and $\frac{2}{5}\times\frac{2}{3}=\frac{4}{15} \sim 0.27$, so the latter match would be considered the best quality match. We note in this example that the FoF group is not bijectively matched to any mock group. From the definition of a bijective group above, it is clear that the match quality for a bijective group must always be larger than $\frac{1}{2}\times\frac{1}{2}=0.25$. Globally we define the following statistics:

\begin{eqnarray}
Q_{\rm FoF}&=&\frac{\sum_{\rm i=1}^{Ng_{\rm FoF}} P_{\rm FoF}[i]*Nm_{\rm FoF}[i]}{\sum Nm_{\rm FoF}} \\
Q_{\rm mock}&=&\frac{\sum_{i=1}^{Ng_{\rm mock}} P_{\rm mock}[i]*Nm_{\rm mock}[i]}{\sum Nm_{\rm mock}} \\
Q_{\rm tot}&=&Q_{\rm FoF}Q_{\rm mock}
\label{eq:quality}
\end{eqnarray}

\noindent where $Nm_{\rm FoF}[i]$ and $Nm_{\rm mock}[i]$ are the number of group members in the $i^{th}$ FoF and mock group respectively. $P_{\rm FoF}[i]$ and $P_{\rm mock}[i]$ are the purity products of the $i^{th}$ best matching FoF and mock group respectively. In the example above $P_{\rm FoF}\sim 0.27$ and $Nm_{\rm FoF}=5$. If a halo is perfectly recovered between the FoF and mock then $P_{\rm FoF}$ and $P_{\rm mock}$ both equal 1 for that matching halo. $Q_{\rm tot}$ is the global grouping purity we want to use in our maximisation statistic, and will be 1 if all groups are found perfectly in the FoF catalogue. The lower limit must be more than 0 (since it is always possible to break a catalogue with $N_{\rm gal}$ galaxies into a catalogue of $N_{\rm gal}$ groups), and at worst $Q_{\rm tot}=Ng_{\rm mock}^{2} / N_{\rm gal}^{2}$. 

Using $E_{\rm tot}$ and $Q_{\rm tot}$ we can now calculate our final summary statistic:

\begin{equation}
S_{\rm tot} = E_{\rm tot} Q_{\rm tot},
\end{equation}

\noindent where $S_{\rm tot}$ will span the range 0--1.

\subsection{Optimisation}
\label{sec:optimisation}

Whilst it is possible to optimise the set of grouping parameters such that the absolute maximum value for $S_{\rm tot}$ is obtained, in practice some of the parameters barely affect the returned group catalogue as long as sensible values are chosen. For FoF group finding, $\Delta$, $r_{\Delta}$, $l_{\Delta}$ have a weak affect on the final grouping, and fixing them at 9, 1.5$\,h^{-1}$Mpc and 12 proved to be almost as effective as allowing them to be freely optimised. For expediency they were fixed after this initial determination. The other 5 FoF parameters do require optimisation, the descending order of parameter importance is: $b_0$, $R_0$, $E_{\rm b}$, $E_{\rm R}$ and $\nu$.

As well as choosing the set of parameters to adjust, the set of groups chosen as the basis of optimisation must be considered carefully. The optimisation strategy has to be defined depending on the desired goals. Most further studies will make use of the largest and best fidelity groups, and these groups suffer disproportionately if the optimisation is carried out using smaller systems and then applied to all of the mock data. Because of this only groups with 5 or more members were used to determine the appropriate combination of parameters. Part of the justification for this is that 5 or more members are required to make a meaningful estimate of the dynamical velocity dispersion ($\sigma_{\rm FoF}$) and 50$^{th}$ percentile radius (Rad$_{\rm 50-group}$).

To optimise the overall grouping to maximise the output of $S_{\rm tot}$ we used a standard Nelder-Mead \citep{neld65} approach, using the {\sc optim} function available in the {\sc R} programming environment. We simultaneously attempted to find the optimal combination of the 5 specified parameters for all 9 mock GAMA volumes, a process that took $\sim$ 2 days CPU time. The optimisation was done for 3 different magnitude limits: $r_{\rm AB}\le19.0$ mags, $r_{\rm AB}\le19.4$ mags and $r_{\rm AB}\le19.8$ mags. The returned parameters were extremely similar. The set generated for $r_{\rm AB}\le 19.4$ were the best compromise, producing the highest overall cost for all 3 depths combines. Since the solutions were so similar, we took the parameters found for $r_{\rm AB}\le 19.4$ as the single set to be used for all analysis. Table \ref{bestparamstab} contains the optimal numbers for the 5 parameters investigated.

\begin{table*}
\begin{center}
\begin{tabular}{cc|cccccc}
$b_0$ & $R_0$ & $E_{\rm b}$ & $E_{\rm R}$ & $\nu$ &  $S_{\rm tot}(r_{AB} \le 19.0)$ & $S_{\rm tot}(r_{AB} \le 19.4)$ & $S_{\rm tot}(r_{AB} \le 19.8)$ \\
\hline
0.06 & 18 & -0.00 & -0.02 & 0.63 & 0.40 & 0.42 & 0.41 \\
\end{tabular}
\end{center}
\caption{The optimal global parameters for all groups with $N_{\rm FoF} \ge 5$.}
\label{bestparamstab}
\end{table*}

The most significant fact to highlight in Table \ref{bestparamstab} is that $E_{\rm b}$ and $E_{\rm R}$ are so close to zero that their effect is completely negligible. Interestingly, if we instead attempt the same optimisation problem but remove $\nu$ these parameters become more significant, but the final cost for the optimisation is not as good. This means the 3 parameters adapt in a degenerate manner, but the luminosity based adaptation is the most successful, and the parameter most fundamentally related to optimal galaxy groups. The GAMA galaxy group catalogue will still use all 5 parameters as specified, but we note that in future extensions to this work $E_{\rm b}$ and $E_{\rm R}$ may be removed.
 
It is clear that the chosen set of parameters produce very similar final $S_{\rm tot}$ for all depths ($\sim0.4$). This implies that on average $E_{\rm FoF}$, $E_{\rm mock}$, $Q_{\rm FoF}$ and $Q_{\rm mock}$ are all $\sim0.8$. Even though no restriction is made in terms of which grouping direction has most significance, the breakdown of each global grouping component indicates that the cost is most easily increased by improving the overall halo finding efficiency, where for $N_{\rm FoF} \ge 5$ (a useful selection since largely groups are typically harder to group accurately), $E_{\rm tot}=0.69$ and $Q_{\rm tot}=0.53$. The contribution to the overall cost is also slightly asymmetric from the mock and FoF components: $E_{\rm mock}=0.89$, $E_{\rm FoF}=0.77$, $Q_{\rm mock}=0.73$ and $Q_{\rm FoF}=0.80$. Overall, the cost of mock groups to $S_{\rm tot}$ is 0.65, and from the FoF groups it is 0.62. These numbers indicate that the FoF algorithm must recover, on average, more groups than actually exist in the mock data. Also, the FoF algorithm is slightly better at constructing the groups it finds than it is at recovering haloes from the data. These statistics mean that the most successful algorithm is necessarily a conservative one where real haloes are robustly and unambiguously detected, and interloper rates kept low in these systems. This is required since it is very easy to create spurious group detections once the grouping is more generous.

\subsubsection{Parameter sensitivity}

To assess how sensitive the best parameters found are to perturbations in the volume investigated (sample variance) we made optimisations for each of the 9 GAMA mock volumes. The distribution of the parameters gives us an indication of both how well constrained they are, and how degenerate they are with respect to the other parameters.

A PCA analysis of the outcome for 5 parameters optimised to 9 volumes suggests nearly all the parameter variance is explained with just two principle components. The most significant parameters are $b$ and $\nu$, and these are anti-correlated. $R$ is the only other significant parameter that contributes to component 1, and this is anti-correlated closely with $b$. $E_{\rm b}$ and $E_{\rm R}$ dominate the second component, and they are strongly anti-correlated.

\begin{table*}
\begin{center}
\begin{tabular}{cccccccc}
$\sigma_{b_0}$ & $\sigma_{R_0}$ & $\sigma_{E_{\rm b}}$ & $\sigma_{E_{\rm R}}$ & $\sigma_{\nu}$ & $\sigma_{b_0}/b_0$ & $\sigma_{R_0}/R_0$ & $\sigma_{\nu}/\nu$ \\
\hline
0.00 & 1 & 0.02 & 0.10 & 0.06 & 0.03 & 0.04 & 0.09 \\
\end{tabular}
\end{center}
\caption{The 1-$\sigma$ spread of the optimal grouping parameters found for the 9 different mock GAMA lightcones. For the three most important parameters, their relative spread is indicated as well.}
\label{paramerror}
\end{table*}

Table \ref{paramerror} shows the 1-$\sigma$ spread in optimal parameter values obtained, and gives an indication of how stable our parameters are to the sample selection. The only surprising fact is that $E_{\rm R}$ is prone to vary quite a large amount depending on the volume, however this is precisely because it has least influence on the quality of any grouping outcome, and hence a large change can cause minor improvements in the grouping. $b$ is extremely well constrained, which is important to know since it is comfortably the most significant parameter for any FoF grouping algorithm.

\section{Group properties, reliability and quality of grouping algorithm}
\label{sec:prop_est}

Whilst the primary aim of the grouping algorithm is to maximise the
accuracy of the content of the groups, it is essential to derive well
determined global group properties. The group velocity dispersion
($\sigma_{\rm FoF}$) and radius ($r_{\rm FoF}$), are key properties
to recover accurately, as they form the most directly inferred group
characteristics, together with the group centre and total group
luminosity ($L_{\rm FoF}$).
The importance of their precise recovery is further strengthened
by the expectation that a reasonable dynamical mass estimator is
proportional to $\sigma_{\rm FoF}^2$ and $r_{\rm FoF}$ 
(\S\ref{sec:mass}).

There are many ways to estimate $\sigma_{\rm FoF}$ and $r_{\rm FoF}$,
but it is essential for the estimates to be median
un-biased
and robust to slight perturbations in group membership. Both constraints
are important so as to not make our group properties overly sensitive to
some precise aspect of the grouping algorithm (a process that will never
produce a perfect catalogue).

Hereafter we adopt the following notation. 
$X_{\rm FoF}$ and $X_{\rm halo}$ correspond to a quantity $X$ measured
using galaxies of the Friends-of-Friends mock group and
of the underlying/true Dark Matter haloes respectively. The
estimate of $X$ is done with the same method both times, i.e. only the
galaxy membership changes between the two measurements for
matched FoF and halo groups. Matching in the mocks corresponds to the 
best group matching between FoF groups and intrinsic haloes, defined as the 
two way match that produces the highest $Q_{\rm tot}$ (see \S\ref{sec:cost_fct} for
further details). We refer to group multiplicity, 
$N_{\rm FoF}$, as the number of group members a given FoF group has,
which has to be distinguished from $N_{\rm halo}$ the true number of
group members fo a given halo. $X_{\rm mock}$ is a value based on an output
of the semi-analytic mock groups directly, it is not measured using a
similar method as for the FoF groups. In practice, only the total luminosity
of the galaxies in the mock group ($L_{\rm mock}$) require this notation since they
are found from summing up the flux of all group members beyond the
magnitude limit of the simulated lightcone. Finally, $X_{\rm DM}$ refers to a property
that is measured from the Millennium Simulation dark matter haloes themselves (so not
dependent on the semi-analytics in any manner). In practice, only the total mass of all dark
matter particles within the halo ($M_{\rm DM}$) require this notation.
 
\subsection{Velocity dispersion estimator}
\label{sec:veldisp}

The group velocity dispersion, $\sigma_{\rm FoF}$, is measured with
the {\bf gapper} estimator introduced by \citet{beer90}, and used
for velocity dispersion estimates in e.g.~2PIGG \citep{eke04}. 
This estimator is unbiased, even for low multiplicity systems, and is
robust to weak perturbations in group membership. 

In summary, for a group of multiplicity $N=N_{\rm FoF}$, all recession
velocities are ordered within the group and gaps between each velocity
pair is calculated using $g_{i} = v_{i+1} - v_{i}$ for $i=1,2...,N-1$, 
as well as weights defined by $w_i=i(N-i)$. The velocity dispersion
is then estimated via:

\begin{equation}
\sigma_{\rm gap}=\frac{\sqrt{\pi}}{N(N-1)}\sum_{i=1}^{N-1}w_i g_i \; .
\label{eq:gapper}
\end{equation}

\noindent Based on the fact that in the majority of mock haloes the
brightest galaxy is moving with the halo centre of mass, the velocity
dispersion is increased by an extra factor of 
$\sqrt{N/(N-1)}$ \citep[as implemented in][]{eke04}. 
Eq.~\ref{eq:gapper} assumes no uncertainty on the recession
velocities, while in reality the accuracy of the redshifts (and
therefore recession velocities) depend among other things on the
galaxy survey considered. To account for this the velocity dispersion
is further modified by the total measurement error $\sigma_{\rm err}$
being removed in quadrature, giving:

\begin{equation}
\sigma=\sqrt{\frac{N}{N-1}\sigma_{\rm gap}^2-\sigma_{\rm err}^2} \; .
\label{eq:gapper_err}
\end{equation}

\noindent The total measurement error $\sigma_{\rm err}$ is the result
of adding together the expected velocity error for each individual
galaxy in quadrature, where we account for the survey origin of the 
redshift, the leading source of uncertainty in estimating 
$\sigma_{\rm err}$. The GAMA redshift catalogue is mainly composed of
redshifts from GAMA ($\sim84$\%), SDSS ($\sim12$\%) and 2dFGRS
($\sim3$\%) where the typical errors are $\sim50 \kms$,
$\sim30 \kms$ and $\sim80 \kms$ \citep[see][for further details on the
  redshift uncertainties in the GAMA catalogue]{driv11}.

\begin{figure*}
\centerline{\mbox{\includegraphics[width=7in]{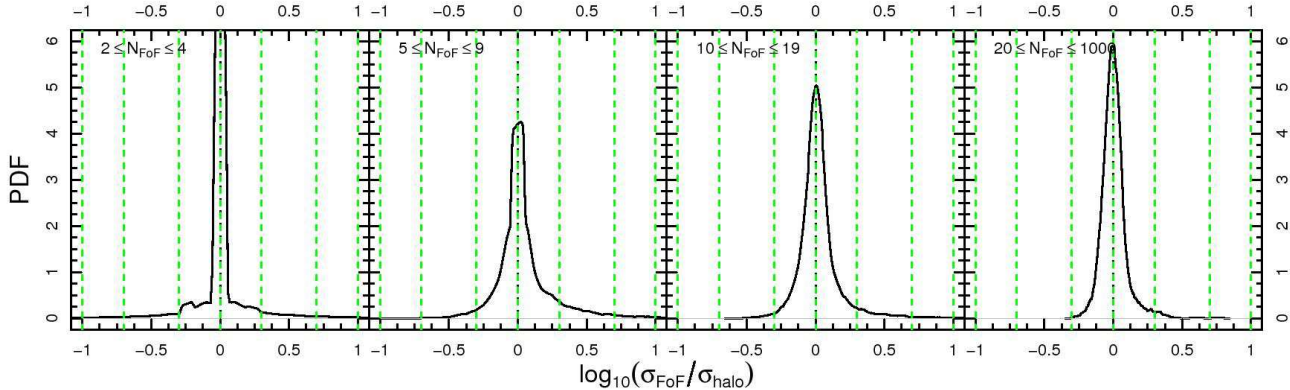}}}
\caption{\small Probability distribution function (PDF) of 
  $\log_{10} \sigma_{\rm FoF}/\sigma_{\rm halo}$, i.e. the log-ratio
  of the measured/recovered velocity dispersion ($\sigma_{\rm FoF}$)
  to the intrinsic galaxy velocity dispersion ($\sigma_{\rm halo}$), 
  for best matching FoF/ halo mock groups. Each panel shows groups of
  different multiplicities, as labelled. The vertical dashed lines
  indicate where $\sigma_{\rm FoF}$ is a factor 2/5/10 off the
  intrinsic $\sigma_{\rm halo}$. The more peaked and centred on 0 the
  PDF is, the more accurately the underlying $\sigma_{\rm halo}$ is
  recovered. 
}
\label{fig:velcomp}
\end{figure*}

Fig.~\ref{fig:velcomp} shows the distribution of the log-ratio of
the measured/recovered velocity dispersion ($\sigma_{\rm FoF}$) to the
intrinsic galaxy velocity dispersion ($\sigma_{\rm halo}$), for
best matching FoF/ halo mock groups. Explicitly $\sigma_{\rm halo}$ is
estimated using Eq.~\ref{eq:gapper} with mock GAMA galaxies belonging
to the same underlying halo, i.e. $\sigma_{\rm halo}$ 
does not correspond to the underlying dark matter halo velocity
dispersion. Furthermore $\sigma_{\rm halo}$ is estimated using only
the line-of-sight velocity information. Hence a
perfect grouping would result in $\delta_{\rm Dirac}$ distributions in
Fig.~\ref{fig:velcomp}.
The fact that these distributions are so tight is a reflection of the
quality of the FoF grouping.  
For $\sim80.4$\% ($\sim50$\%) of all mock groups, the recovered
$\sigma_{\rm FoF}$ is within $\sim50$\% ($\sim14$\%) of the intrinsic
value. 
The distributions are median unbiased for most multiplicities with the
mode close to zero as well. The symmetry of Fig.~\ref{fig:velcomp} is
a good indication that the FoF groups are as likely to
underestimate as overestimate the velocity dispersion.

\subsection{Group centre and projected radius: definitions and estimators}
\label{sec:radius}

More contentious quantities to define and estimate are the 
{\it centre} and the {\it projected radius} of a group. Firstly there is no
unique way to define the group centre (e.g. centre of mass (CoM),
geometric centre (GC), brightest group/cluster galaxy (BCG),...) from
which the projected radius is defined. Secondly the projected radius
definition will depend on what fraction of galaxies should be enclosed
within it and on what assumption is made for the distance to the
group.

To determine the most robust and appropriate definitions for the
centre and projected radius of a group a number of schemes were
investigated. Hereafter we implicitly assume projected radius when
referring to the group radius.

\subsubsection{Projected group centre}
\label{sec:gc}

For the group centre three approaches were considered. Firstly, the
group centre was defined as the centre of light (CoL) derived from the
$r_{\rm AB}$-band luminosity of all the galaxies associated with the
group, which is an easily observable proxy for the CoM.
Secondly, an iterative procedure was used where at each step the
$r_{\rm AB}$-band CoL was found and the most distant galaxy rejected. When
only two galaxies remain, the brighter $r_{\rm AB}$-band galaxy is 
used as the group centre. We refer to it as Iter. Thirdly, the
brightest group/cluster member (BCG) was assumed to be the group
centre. For mock groups with $N_{\rm FoF} \ge 5$, 95\% of the time the
iterative procedure produces the same group centre as the BCG definition.

\begin{figure*}
\centerline{\mbox{\includegraphics[width=7in]{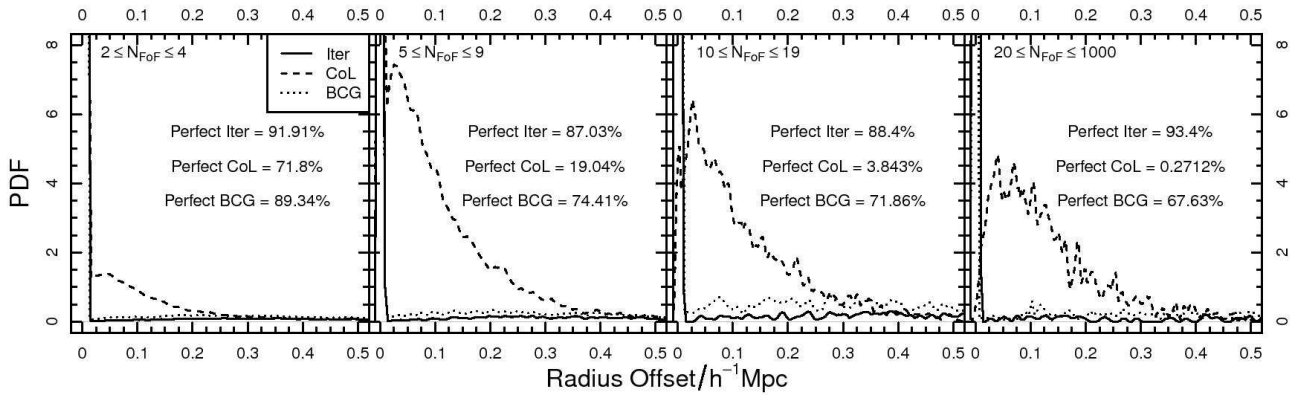}}}
\caption{\small Distribution of position offsets between different group centre
  definitions and the underlying/true group centre for bijectively
  matched mock groups. Each panel shows groups of different multiplicities, as
  labelled. Solid/dashed/dotted lines indicate the
  Iter/CoL/BCG centre definitions (see text). The nearly vertical
  lines at small radii correspond to groups which have a perfectly
  recovered centre position (i.e. zero offset). Their fraction is
  indicated in the panel as ``Perfect''.
}
\label{fig:raddistcomp}
\end{figure*}

Fig.~\ref{fig:raddistcomp} presents a comparison between three 
group centre definitions (Iter, CoL, BCG) and the true/underlying group
centre for the best matching (highest $Q_{\rm tot}$) mock groups. { In this
context ``true'' refers to the centre we obtain when running the same
algorithm on the exact mock group.}
The plot shows the distribution of the
positional offsets for the different definitions of group centre when compared to the ``truth'' for
different group multiplicities, with the fraction that agrees
perfectly stated in each panel for each group centre definition.

The iterative method always produces the best agreement for the
exact group centre and seems to be slightly more robust to the effects
of group outliers.
As should be expected, the flux weighted CoL definition is the least
good at recovering the underlying/true halo centre position.
With the CoL definition, the group needs to be recovered exactly to
get a perfect match and any small perturbations in membership
influences the accuracy with which the centre is recovered. This is
very different to the BCG or Iter centre definitions, which are only
very mildly influenced by perturbations in membership.

The iterative centre is therefore preferable over merely using the
BCG: it has a larger precise matching fraction and a smaller fraction
of groups with spuriously large centre offsets. It is very stable as a
function of multiplicity, with a fraction of precise group centre
matches of $\sim 90$\%, as indicated in the panels of
Fig.~\ref{fig:raddistcomp}. 
Hereafter we refer to the Iter centre definition as the group centre.

\subsubsection{Radial group centre}

\begin{figure*}
\centerline{\mbox{\includegraphics[width=7in]{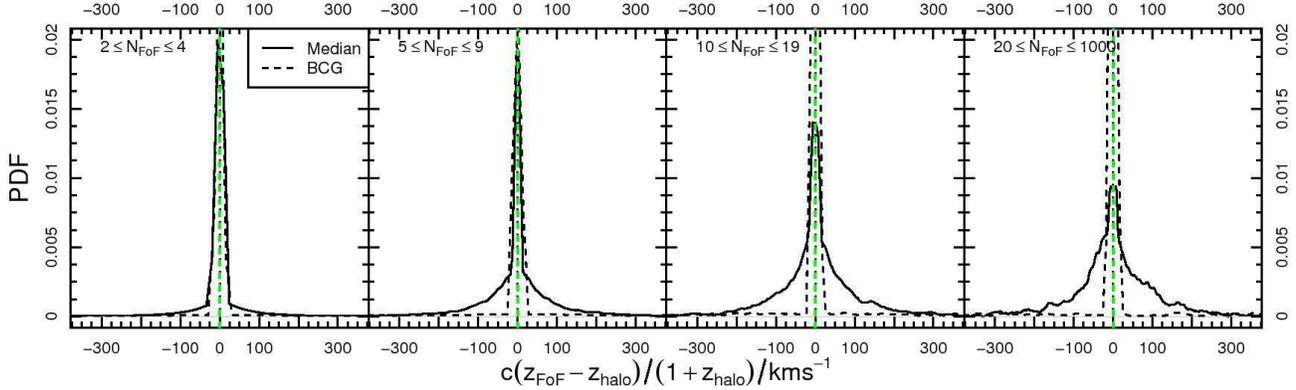}}}
\caption{\small Probability distribution function (PDF) of 
  $z_{\rm FoF}-z_{\rm halo}$ for best matching FoF/ halo mock groups,
  where $z$ is the median redshift of the group. Each panel shows groups of
  different multiplicities, as labelled. The fraction of exact matches
  is indicated in each panel, as ``Perfect''. 
}
\label{fig:medianzcomp}
\end{figure*}

The group centre definitions as considered in~\S\ref{sec:gc} do
not necessarily define what the actual group redshift should be. One
possible solution is to identify it with the redshift of the central
galaxy, as found with the Iter centre definition. An alternative
solution would be to select the group redshift as the median redshift
of the group members.
Fig.~\ref{fig:medianzcomp} presents the distribution of the difference
between the recovered median redshift and the intrinsic median
redshift for best matching FoF/ halo mock groups. The fraction of group
redshifts that agree precisely is stable as a function of  
multiplicity at $\sim55$\%, and the offset is usually less than $100\kms$.
80\% of the time the redshift differences are within the
GAMA velocity error of $\sim50\kms$ \citep[see][for details]{driv11}.
It is essential to notice that this radial centre is defined in redshift space (i.e.\ including peculiar velocities) as opposed to real space (i.e.\ based on Hubble flow redshift), as only information for the former is available from a redshift survey. A comparison between the real and the redshift space centre shows directly the importance and the impact of bulk flow motions, i.e. the galaxy groups themselves are not at rest.

\subsubsection{Projected group radius}

The radius definition must be a compromise between containing a large
enough number of galaxies to be stable statistically and 
small enough to not be overly biased by or sensitive to outliers and
interlopers (which tend to lie at larger projected distances).
Three radius definition were considered: Rad$_{50}$, Rad$_{1-\sigma}$
and Rad$_{100}$ containing 50\%, 68\% and 100\% of the galaxies in the
group respectively. The latter, Rad$_{100}$, is mainly used for illustrative
purposes, as it is extremely sensitive to outliers.
Rad$_{\rm X}$ is defined using the default quantile definition in R, i.e. the group members are sorted in ascending radius value, assigned a specific percentile (the most central 0\% and the furthest away 100\%) and finally a linear interpolation between the radii of the two relevant percentiles is performed. This implies that only the radial distance of the two galaxies bracketing the percentile definition used are considered in the estimate of Rad$_{\rm X}$, explaining why Rad$_{100}$ is expected to be the most sensitive to outliers.

\begin{figure*}
\centerline{\mbox{\includegraphics[width=7in]{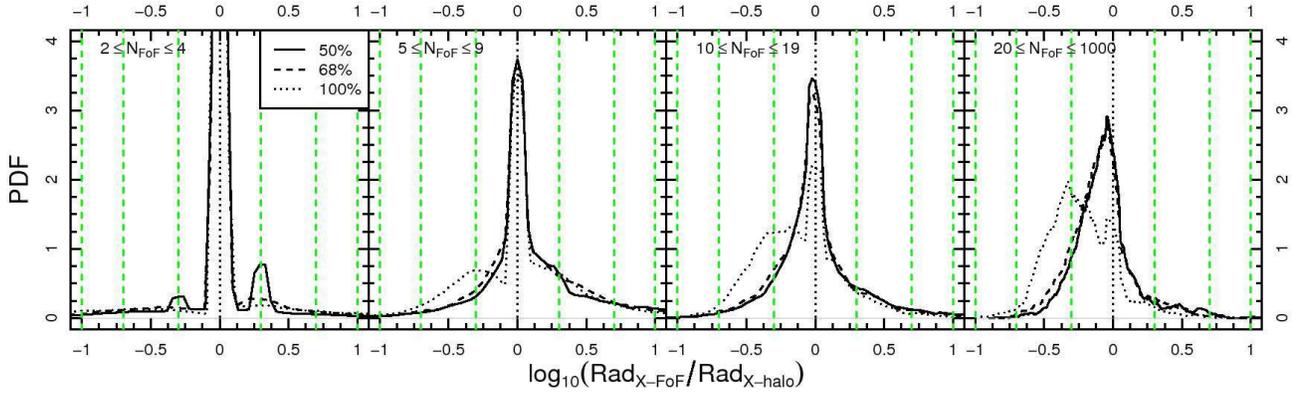}}}
\caption{\small Probability distribution function (PDF) of 
  $\log_{10} {\rm Rad_{\rm X-FoF}/Rad_{\rm X-halo}}$, i.e. the log-ratio
  of the measured/recovered radius (Rad$_{\rm X-FoF}$)
  to the intrinsic galaxy radius (Rad$_{\rm X-halo}$), 
  for best matching FoF/ halo mock groups. Each panel shows groups of
  different multiplicities, as labelled.
    Solid/dashed/dotted lines indicate the
  Rad$_{50}$, Rad$_{1-\sigma}$ and Rad$_{100}$ radii definitions
  respectively encompassing 50\%, 68\% and 100\% of the galaxies in
  the group. The solid line, Rad$_{50}$, produces the tightest
  distribution of the three considered. 
  The vertical dashed lines indicate where Rad$_{\rm X-FoF}$ is a
  factor 2/5/10 off the intrinsic Rad$_{\rm X-halo}$. 
    }
\label{fig:radcomp}
\end{figure*}

Fig.~\ref{fig:radcomp} shows a comparison between three radii
definitions as measured from the iterative centre for recovered mock
groups (Rad$_{\rm X-FoF}$) and for true mock haloes 
(Rad$_{\rm X-halo}$) for best matching FoF/ halo mock groups. Rad$_{50}$ is
marginally more centrally concentrated than Rad$_{1\sigma}$ for all
multiplicity subsets and is hence the least affected by interlopers
and outliers. 

The subsets plotted in Fig.~\ref{fig:radcomp} up to 
$10 \le N_{\rm FoF} \le 19$ are all median unbiased, although there
is a slight high-moment excess of large radius groups for 
$2 \le N_{\rm FoF} \le 9$ and a high moment excess of erroneously low radius 
groups for $10 \le N_{\rm FoF} \le 19$. This does not affect the median
of the distribution, but requires the mean to be offset from the median
in these cases.

The highest multiplicity subset (right most panel of
Fig.~\ref{fig:radcomp}) has an identifiable excess of low
radius groups, leading to a biased median that is $\sim15$\% lower
than the original aim. Hence the estimated Rad$_{\rm 50-FoF}$ for half
of the highest multiplicity groups is underestimated by at least
$\sim15$\% compared to the corresponding underlying 
Rad$_{\rm 50-halo}$. 
We note however that this definition still behaves better than any of the other two considered.

Whilst the accuracy of the measured velocity dispersion noticeably
improves as a function of multiplicity (see Fig.~\ref{fig:velcomp}),
the accuracy of the observed radius does not. 
This observation should be expected since groups
have their centres iterated towards the optimal solution. During this process
they, in effect, become lower multiplicity as the outliers are removed, and thus
will suffer from similar numerical artefacts.

Based on the improvement in radius agreement for $N_{\rm FoF} \ge 5$, 
Rad$_{50}$ was selected as the preferred definition of radius for
use in the GAMA galaxy group catalogue. For the remainder of this paper,
and in any future discussion of GAMA galaxy groups, any mention of
group radius implicitly refers to Rad$_{50}$.
However it is to be noted that Rad$_{1-\sigma}$ is better behaved for low multiplicity groups ($N_{\rm FoF} \le 4$), as the ``bumps'' at $\pm 0.3$ in the left most panel of Fig.~\ref{fig:radcomp} have vanished nearly completely in that case. The origin of these two spikes becomes clear in the discussion of Fig.~\ref{fig:splatgrid}.

\subsection{Dynamical group mass estimator and calibration}
\label{sec:mass}

Once an unbiased and robust group velocity dispersion and a nearly
unbiased group radius can be estimated, the final step 
is to combine this information into a dynamical mass estimator. To 
first order for a virialized system we expect its dynamical mass to
scale as $M \propto \sigma^2 R$, where $\sigma$ and $R$ are calculated
as described in~\S\ref{sec:veldisp} and~\S\ref{sec:radius}.

\begin{figure*}
\centerline{\mbox{\includegraphics[width=7in]{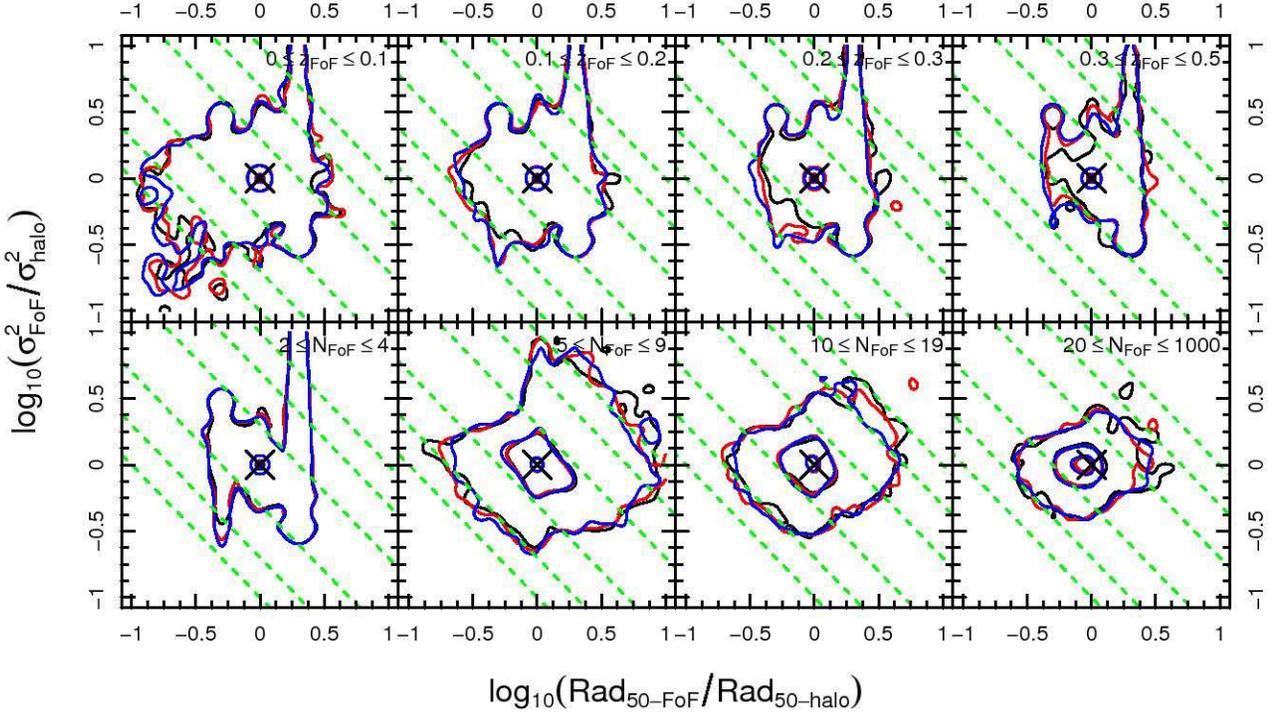}}}
\caption{\small 
2-D density distribution of the best matching FoF/ halo mock groups in the 
$\log_{10} {\rm Rad_{\rm X-FoF}/Rad_{\rm X-halo}}$
--
$\log_{10} (\sigma_{\rm FoF}/\sigma_{\rm halo})^2$ plane, split as a function of
  redshift and multiplicity (top and bottom panel respectively).
The x and y-axes show the relative accuracy of the recovered radius and
velocity dispersion (squared) respectively. The contours represent the
regions containing 10/50/90\% of the data for three magnitude limits,
i.e. $r_{\rm AB}\le19.0$ (black), $r_{\rm AB}\le19.4$ (red) and 
$r_{\rm AB}\le19.8$ (blue).
The green dashed lines delineate regions where 
$\sigma_{\rm FoF}^2 Rad_{\rm 50-FoF}$ is 2/5/10 times off the
expectation given by $\sigma_{\rm halo}^2 Rad_{\rm 50-halo}$, 
reflecting to some extent the implied uncertainty on any dynamical
mass estimate (see text for details). 
}
\label{fig:splatgrid}
\end{figure*}

To understand any correlated biases in the estimates of these two
fundamental group properties, we plot in Fig.~\ref{fig:splatgrid}
the group density distribution as a function of the relative accuracy of
the recovered group radius (x-axis) and the square of the group
velocity dispersion (y-axis). 
More precisely Fig.~\ref{fig:splatgrid} shows the group density
distribution in the 
$\log_{10} {\rm Rad_{\rm X-FoF}/Rad_{\rm X-halo}}$
--
$\log_{10} (\sigma_{\rm FoF}/\sigma_{\rm halo})^2$ plane, split as
function of redshift and multiplicity, with ranges specified
in each panel. 
The green dashed lines delineate regions where 
$\sigma_{\rm FoF}^2 {\rm Rad_{\rm 50-FoF}}$ is 2/5/10 times off the
expectation given by $\sigma_{\rm halo}^2 {\rm Rad_{\rm 50-halo}}$, 
reflecting to some extent the implied uncertainty on any dynamical
mass estimate. 
As a matter of fact, if the dynamical mass is proportional to 
$\sigma^2 R$ as expected for a virialized system and can be directly
estimated from $\sigma_{\rm halo}^2 {\rm Rad_{\rm 50-halo}}$, then the green
dashed lines indicate by what amount the halo mass as inferred from 
$\sigma_{\rm FoF}^2 {\rm Rad_{\rm 50-FoF}}$ deviates from the true
one (assuming the same proportionality factor).
Additionally any asymmetry in the density distribution
w.r.t.\ those guide lines is a sign of a bias in the inferred mass: 
a density excess in the top-right/bottom-left of any panel indicates
a bias towards incorrectly high/low dynamical masses.
Note that a density excess orthogonal to these lines is not
problematic for the mass estimates since the individual biases 
cancel out in this parametrisation.   

As a function of redshift the density distributions in Fig.~\ref{fig:splatgrid} are well behaved.
As a function of multiplicity the main effect is a tightening of the distribution, which is expected since the velocity dispersion and, to a lesser degree, the radius can be better estimated with more galaxies. The $5 \le N_{\rm FoF} \le 9$ multiplicity range shows some small bias towards high dynamical masses (the 90\% contour wing) whilst the highest multiplicity subset ($20 \le N_{\rm FoF} \le 1000$) appears to be biased to slightly low dynamical masses (offset for 10\% and 50\% contour wings). Overall the biases are small for $N_{\rm FoF} \ge 5$ multiplicity groups, and in the tails of the distributions rather than in the median or the mode. 
However, for low multiplicity groups ($N_{\rm FoF} \le 4$) the situation is rather different. First of all, there is an extensive scatter in the recovered velocity dispersion at $\log_{10} {\rm Rad_{\rm X-FoF}/Rad_{\rm X-halo}} \simeq \pm0.3$. This is entirely related to the ``bumps'' seen in Fig.~\ref{fig:radcomp} and are due to mismatches in the grouping, explaining why the velocity dispersions are so poorly recovered for some of those systems. The reason for an overdensity of groups at $\pm0.3$ (i.e. half/double the underlying radius) is related to the way Rad$_{50}$ works. { When a $N_{\rm FoF}=2$ group misses one member and when a $N_{\rm FoF}=3$ group contains one interloper this results most often in a FoF group where the calculated group centre is the same\footnote{Because the group centre is so accurately recovered, see Fig.~\ref{fig:raddistcomp}} but radius that is half and double the halo radius respectively.} Additionally any asymmetry seen in the top panels of Fig.~\ref{fig:splatgrid} can be attributed to low multiplicity groups.
Generally Fig.~\ref{fig:splatgrid} gives us confidence that measurement errors in $\sigma^2$ and $R$ are not highly correlated.

The dynamical mass of a system is estimated using
\begin{eqnarray}
\frac{M_{\rm FoF}}{\msolh} & = & \frac{A}{\frac{G}{\msol^{-1} {\rm km^{2} \, s^{-2}} {\rm Mpc}}} \, \left(\frac{\sigma_{\rm FoF}}{\kms}\right)^2 \, \frac{{\rm Rad}_{\rm FoF}}{\mpch}
\label{eq:mass_est}
\end{eqnarray}
where $G$ is the gravitational constant in suitable units, i.e.\
$G = 4.301\times10^{-9} \msol^{-1} {\rm km^{2} \, s^{-2}} {\rm Mpc}$.
$A$ is the scaling factor required to create a median unbiased mass estimate
of $M_{\rm DM}/M_{\rm FoF}$.
For a `typical' cluster with a $1\mpch$ radius and a
velocity dispersion of $1000\kms$, the mass given by
Eq.~\ref{eq:mass_est} is $\sim 2 \, A \times 10^{14}\msolh$. 
$A$ is likely to be larger than unity, since the estimated velocity
dispersion using Eq.~\ref{eq:gapper} traces the velocity dispersion
along the line-of-sight only\footnote{For isotropic systems 
$\sigma_{\rm 3D} \sim \sqrt{3} \sigma_{\rm 1D}$} and the average 
projected radius is smaller than the average intrinsic 
radius\footnote{For isotropic systems the relation depends on the exact radius definition.
Conceptually the 3D and 2D radius will agree for Rad$_100$ but increasingly disagree
as the radius measured becomes smaller due to the relative concentration of objects
towards the centre when observing a projected 2D, as opposed to intrinsic 3D,
distribution.     }.
Finally, Eq.~\ref{eq:mass_est} can only be truly valid for a system in
virial equilibrium, which many of our system will not necessarily
be. Hence the best approach is to determine $A$ in a semi-empirical
manner by requiring it to produce a median unbiased halo mass estimate
when comparing best matching FoF/ halo mock groups.

\begin{figure*}
\centerline{\mbox{\includegraphics[width=7in]{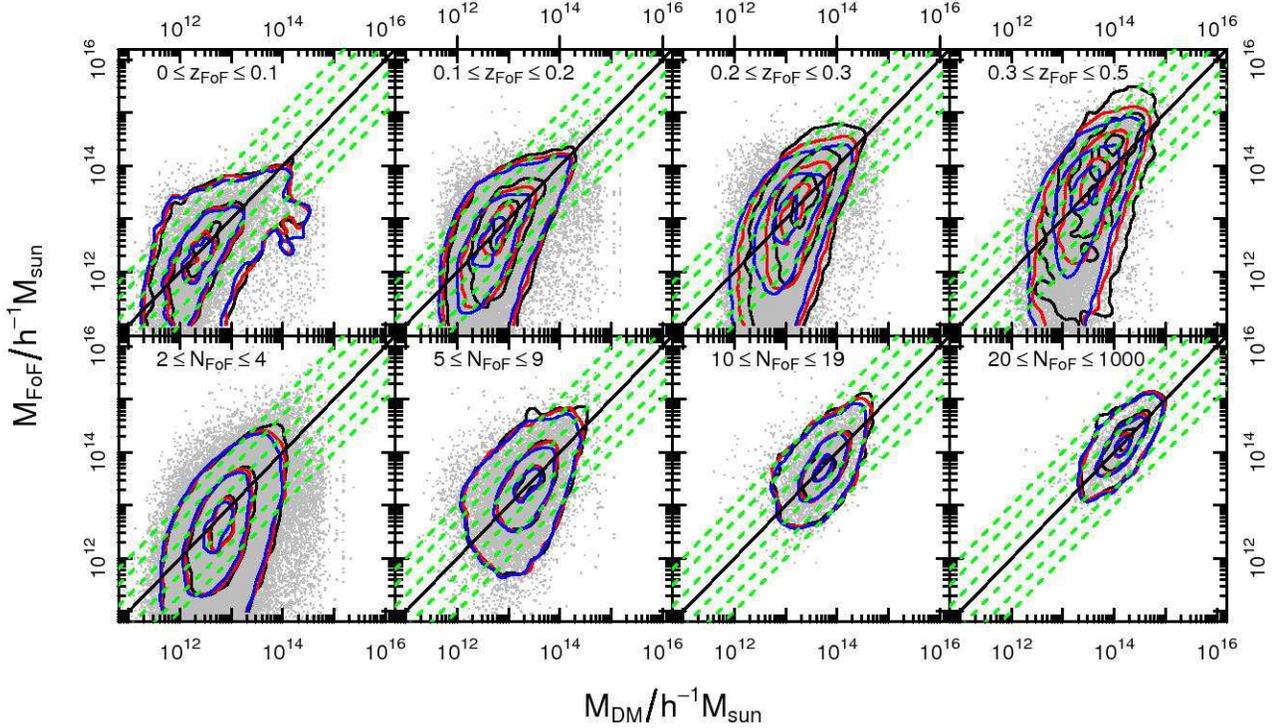}}}
\caption{\small 2-D density distribution of best matching FoF/ halo mock groups
  in the M$_{\rm FoF}$--M$_{\rm DM}$ plane, split as a function of
  redshift and multiplicity (top and bottom panel respectively). These
  panels objectively compare the recovered group masses to the
  underlying DM halo masses. The contours represent the
  regions containing 10/50/90\% of the data for three magnitude limits,
  i.e. $r_{\rm AB}\le19.0$ (black), $r_{\rm AB}\le19.4$ (red) and 
  $r_{\rm AB}\le19.8$ (blue). The dots indicate the exact 
  $M_{\rm FoF}$--$M_{\rm DM}$ pairs. For $M_{\rm FoF}$ we use
  Eq.~\ref{eq:mass_est} and $A=10.0$.
    The green dashed lines delineate regions where M$_{\rm FoF}$ is
  2/5/10 times off the underlying M$_{\rm DM}$. 
}
\label{fig:massgridcont}
\end{figure*}

Performing a single global optimisation using all bijectively matched
groups with $N_{\rm FoF} \ge 5$ results in $A=10.0$. This is somewhat different to this $A=5$ factor found
in \citet{eke04}. This should not be surprising since there are differences in the style of grouping optimisation, and
we have used a more compact definition of the group radius and a different approach to recovering the group centre.
It is interesting to note that this scaling of $A=10.0$ is identical to the dynamical mass scaling found in
\citet{chil08} for calculating the virial mass of dwarf galaxies.

Fig.~\ref{fig:massgridcont} compares the median globally calibrated dynamical
masses to the underlying DM halo mass for best matching FoF/ halo mock groups (using $A=10.0$).
Whilst the distribution is globally unbiased for $N_{\rm FoF} \ge 5$ (by definition), small deviations as a function of redshift and/or multiplicity are evident.
Offsets from the median line are evident at all multiplicities, but strongest for low multiplicity systems (i.e.\ $2 \le N_{\rm FoF} \le 4$ groups in Fig.~\ref{fig:massgridcont}).
The small biases becomes more apparent at higher redshifts, driven by the average observed group multiplicity dropping as a function of redshifts and the average mass increasing. To gauge how sensitive the scaling factor $A$ is to the specific subset of data considered combined cuts in redshift and multiplicity were made. Table~\ref{tab:Afactors} contains the various $A$ factors required for the different subsets as a function of the possible limiting magnitudes for the GAMA group catalogue.

\begin{table*}
\begin{small}
\begin{center}
\begin{tabular}{lrrrcrrrcrrrcrrr}
 &  \multicolumn{3}{l}{$2 \le N_{\rm FoF} \le 4$} & \hspace{1 mm} & \multicolumn{3}{l}{$5 \le N_{\rm FoF} \le 9$} & \hspace{1 mm} &  \multicolumn{3}{l}{$10 \le N_{\rm FoF} \le 19$} & \hspace{1 mm} &  \multicolumn{3}{l}{$20 \le N_{\rm FoF} \le 1000$} \\
 & 19.0 & 19.4 & 19.8 & & 19.0 & 19.4 & 19.8 & & 19.0 & 19.4 & 19.8 & & 19.0 & 19.4 & 19.8 \\ 
  \hline
$0 \le z_{\rm FoF} \le 0.1$ & 20.0 & 19.0 & 18.0 & & 11.8 & 10.8 & 10.9 & & 11.4 & 12.0 & 11.5 & & 12.1 & 12.6 & 12.7 \\ 
  $0.1 \le z_{\rm FoF} \le 0.2$ & 20.2 & 19.5 & 19.2 & & 10.3 & 10.5 & 10.7 & & 11.0 & 11.1 & 10.9 & & 9.2 & 10.4 & 10.9 \\ 
  $0.2 \le z_{\rm FoF} \le 0.3$ & 21.2 & 21.5 & 19.8 & & 9.0 & 10.3 & 11.2 & & 8.0 & 8.6 & 9.9 & & 6.7 & 8.3 & 9.6 \\
  $0.3 \le z_{\rm FoF} \le 0.5$ & 13.6 & 17.4 & 17.8 & & 4.4 & 6.1 & 7.9 & & 3.5 & 5.4 & 6.7 & & 4.8 & 5.6 & 6.9 \\

\end{tabular}
\caption{Values of $A$, the dynamical mass scaling factor of
  Eq.~\ref{eq:mass_est}, required to create an unbiased median mass
  estimate for different disjoint subsets of bijectively matched
  groups.
  }  
\label{tab:Afactors}
\end{center}
\end{small}
\end{table*}

Using the data in Table~\ref{tab:Afactors} the best fitting plane that accounts for the variation of A as a function of $\sqrt{N_{\rm FoF}}$ and $\sqrt{z_{\rm FoF}}$ is calculated. To prevent strong biases to low $N_{\rm FoF}$ systems purely by virtue of their overwhelming numbers, the plane was not weighted by frequency and should produce the appropriate corrections throughout the parameter space investigated. The plane function for $A$ is given by

\begin{equation}
A(N_{\rm FoF},z_{\rm FoF}) = A_{\rm c} + \frac{A_{\rm N}}{\sqrt{N_{\rm FoF}}} + \frac{A_{\rm z}}{\sqrt{z_{\rm FoF}}},
\label{eq:Ascale}
\end{equation}

\noindent where $A_{c}$, $A_{N}$ and $A_{z}$ are constants to be fitted. Table~\ref{tab:AfactorsFn} contains the parameters that produce the best fitting planes for the three different GAMA magnitude limits. { The motivation for the functional form is mainly driven to ensure positivity of $A(N_{\rm FoF},z_{\rm FoF})$ over the range of GAMA multiplicities and redshifts, and a good fit to the data within these limits.} The errors shown in Table~\ref{tab:AfactorsFn} are estimated from finding the best fitting plane for the 9 mock GAMA volumes separately and measuring the standard deviation of the individual best fitting planes, much like the approach used for Table~\ref{paramerror}.

\begin{table}
\begin{center}
\begin{tabular}{lccc}
 & $A_{\rm c}$ & $A_{\rm N}$ & $A_{\rm z}$ \\
 \hline
$r_{\rm AB}\le19.0$ & $-$4.3 $\pm$ 3.1 & 22.5 $\pm$ 1.7 & 3.1 $\pm$ 1.1 \\
$r_{\rm AB}\le19.4$ & $-$1.2 $\pm$ 1.7 & 20.7 $\pm$ 1.4 & 2.3 $\pm$ 0.6 \\
$r_{\rm AB}\le19.8$ & $+$2.0 $\pm$ 1.4 & 17.9 $\pm$ 1.1 & 1.5 $\pm$ 0.4 \\
\end{tabular}
\caption{Table of parameters that create the best fitting plane to the data in Table~\ref{tab:Afactors}. The plane is a function of group redshift and multiplicity, as given in Eq.~\ref{eq:Ascale}. Errors are estimated from running plane fits to the 9 mock GAMA volumes separately and measuring the standard deviation of the individual best fitting planes.}
\label{tab:AfactorsFn}
\end{center}
\end{table}
 
\subsubsection{Mass estimate scatter}

It is important to highlight that even though the observed dynamical mass estimates and halo masses are well correlated (in particular the scatter is approximately mirrored across the 1--1 line in Fig.~\ref{fig:massgridcont}), it is impossible to select an unbiased subset of mass unless the selection is across the mode of the distribution. This is due to Eddington bias rather than any intrinsic issue with the mass estimates--- since most haloes in GAMA will have moderate masses ($\sim10^{13}\msolh$) if simple Gaussian scatter in mass estimates is assumed, then a high mass subset must contain a larger fraction of lower mass haloes scattered {\it up} in mass, and a low mass subset must contain a larger fraction of higher mass haloes scattered {\it down} in mass, hence the medians are biased. This effect is different to a Malmquist-bias, which explains the observational bias in distribution of halo masses as a function of distance.

\begin{figure*}
\centerline{\mbox{\includegraphics[width=7in]{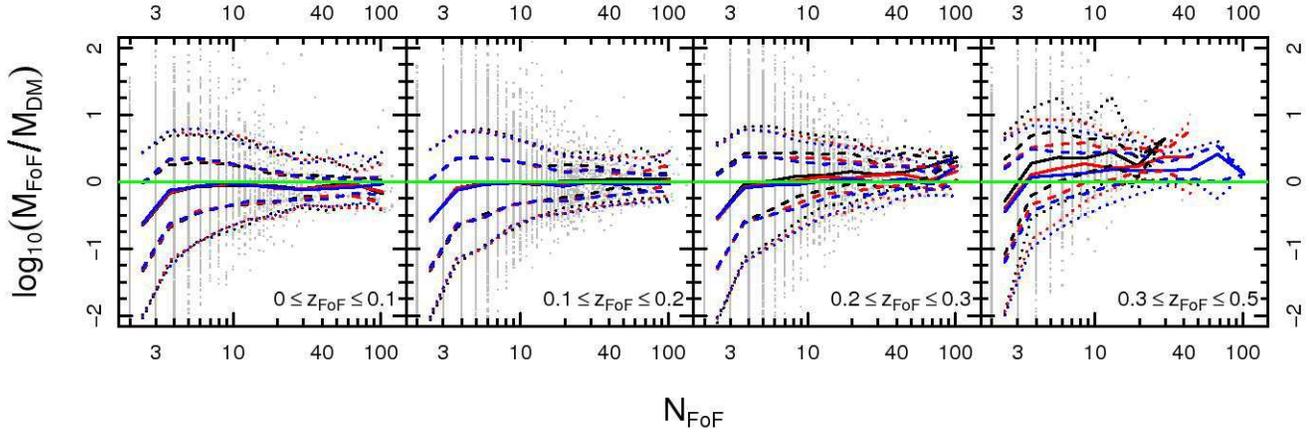}}}
\caption{\small Relative difference between measured and underlying
  group masses as a function of multiplicity for different redshift
  subsets. The improvement in the measurement of the velocity
  dispersion and the radius tightens the distribution until 
  $N_{\rm FoF} \sim 50$. The lines represent the 3 survey depths of
  interest: $r_{\rm AB}\le19.0$ (black) $r_{\rm AB}\le19.4$ (red) and 
  $r_{\rm AB}\le19.8$ (blue).  For $M_{\rm FoF}$ we use
  Eq.~\ref{eq:mass_est} and $A=10.0$.}
\label{fig:multerrorVz}
\end{figure*}

This effect can be modelled quite accurately by assuming we have median unbiased log-normal relative error in the mass estimate, where the standard deviation of the distribution ($M_{\rm err}$) is a function of system multiplicity. The effect multiplicity has on the accuracy of the mass can be seen clearly in Fig.~\ref{fig:multerrorVz}, where although median unbiased for $N_{\rm FoF} \ge 4$, the standard deviation of the distribution decreases strongly as a function of multiplicity. The approximate function for this effect is given by

\begin{equation}
\log_{10}(\frac{M_{\rm err}}{\msolh}) = 1.0 - 0.43 \log_{10}(N_{\rm FoF})  \, ,
\label{eq:mass_err}
\end{equation}

\begin{figure*}
\centerline{\mbox{\includegraphics[width=7in]{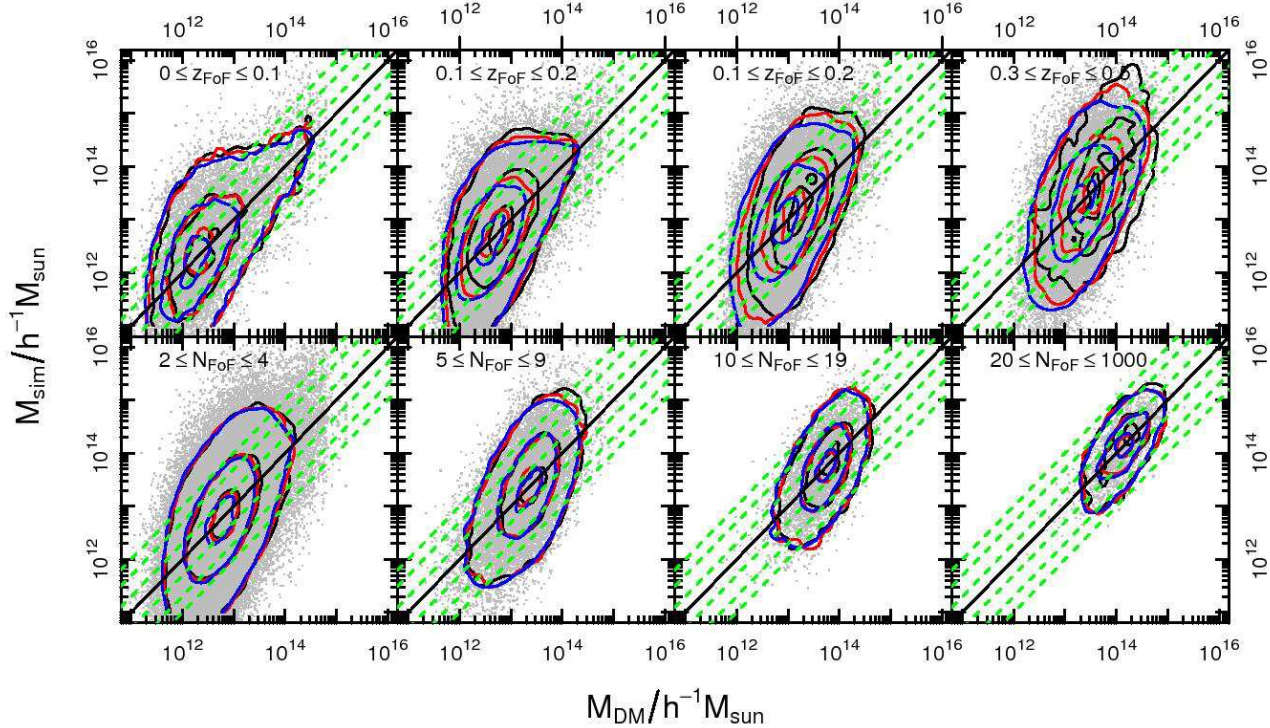}}}
\caption{\small As Fig.~\ref{fig:massgridcont}, but for the simulated relation between $M_{\rm DM}$ and $M_{\rm sim}$ ($M_{\rm DM}$ with the expected random errors applies using Eq.~\ref{eq:mass_sim}), by modelling the expected error just as a function of group multiplicity. The contours represent the regions containing 10/50/90\% of the data for three different magnitude limits, $r_{\rm AB}\le19.0$ (black), $r_{\rm AB}\le19.4$ (red) and $r_{\rm AB}\le19.8$ (blue).
 $M_{\rm sim}$ is estimated using Eq.~\ref{eq:mass_err}.}
\label{fig:multsim}
\end{figure*}

\noindent where the appropriate range of use is $2 \le N_{\rm FoF} \le 50$, beyond which the standard deviation is $\sim 0.27$. We recast this error function back onto the intrinsic mock halo masses to give a new mass with simulated dynamical mass errors:

\begin{equation}
\frac{M_{\rm sim}}{\msolh}=\frac{M_{\rm DM}}{\msolh} 10^{G(0,\log_{10}(\frac{M_{\rm err}}{\msolh}))}  \, ,
\label{eq:mass_sim}
\end{equation}

\noindent where $G(\bar{x},\mu)$ is a random sample from the normal distribution with a mean $\bar{x}$ and standard deviation $\mu$. Fig.~\ref{fig:multsim} shows how the intrinsic halo mass compares for the same halo masses but with our fiducial error function applied. This shows the main contour twisting features described above--- particular clear is the sampling bias you would expect when selecting groups based on the {\it observed} halo masses. For instance, the manner in which the mode of the contours appears to be more vertical than the 1--1 line in Fig.~\ref{fig:massgridcont} (the slight rotation of the contours) is well replicated in Fig~\ref{fig:multsim} and can be explained by the random scatter of the measured dynamical mass from the intrinsic halo mass.

\subsection{Total group luminosity estimator}
\label{sec:lumtot}

The total group luminosity is an equally important global group
property. It should not be just the total luminosity of the observed
group members but the total luminosity as inferred from an arbitrarily
faint absolute magnitude limit cut in order to address residual
selection effects. To do this we calculate the effective 
absolute magnitude limit of each group, measure the $r_{\rm AB}$-band
luminosity contained within this limit and then integrate the global GAMA galaxy LF (see
\S\ref{sec:data}) to a nominal faint limit used to correct for the missing flux.
Explicitly, for each group we calculate the following:

\begin{equation}
L_{\rm FoF} = B \, L_{\rm ob} \, \frac{\int_{-30}^{-14}10^{-0.4 M_r} \phi_{\rm GAMA}(M_r) dM_r}{\int_{-30}^{M_{r-{\rm lim}}}10^{-0.4 M_r} \phi_{\rm GAMA}(M_r) dM_r} \, , \label{eq:lum_est}
\end{equation}

\noindent where $L_{\rm ob}$ is the total observed $r_{\rm AB}$-band luminosity of the group,
$B$ is the scaling factor required to produce a perfectly median unbiased
luminosity estimate and $M_{r-{\rm lim}}$ is the effective $r_{\rm AB}$-band absolute magnitude limit
for the group. This limit depends on the redshift of observation and apparent magnitude
limit used. Corrections are only a few percent at low redshift when using $r_{\rm AB} \le 19.8$ and
can become factors of a few at $z_{\rm FoF} \sim 0.5$.
To convert magnitudes into solar luminosities we take the
$r_{\rm AB}$-band absolute magnitude of the Sun to be
$M_{r \sun}=4.67$\footnote{\tt http://mips.as.arizona.edu/{\char'176}cnaw/sun.html}.
{ The limits of $-30 \le M_r \le -14$ used in the numerator of Eq.~\ref{eq:lum_est}
are effectively limits of $-\infty \le M_r \le \infty$ since the luminosity density of
a typical LF is nearly all recovered within a couple of magnitudes of $M^*$. Assuming the
Schechter function parameters of \citet{blan03} we would expect to retrieve 99.5\% of the
intrinsic flux using these limits, assuming the LF continues down to infinitely faint galaxies. More practically,
the bright limit ($M_r \ge -30$) is much brighter than any known galaxy, and the faint limit ($M_r \le -14$)
is the limit of the GAMA SWML LF used for this work, and thus is also the effective limit of the mock
catalogues used since the galaxy luminosities were adjusted to return the GAMA LF.}

Since the median redshift of GAMA is $z\sim0.2$ and the apparent
magnitude limit is at least $r_{\rm AB}=19.4$, most groups will contain
members faintwards of $M^*_h$ 
\citep[with $M^*_h = M^* - 5 \log_{10} h = -20.44$, ][]{blan03}. Because
the luminosity density is dominated by galaxies around $M^*_h$,
the extrapolation required to get a total group luminosity will be quite conservative since most groups
are sampled well beyond $M^*_h$.

This process assumes that a global LF is appropriate for all groups
over a range of masses and environments, which is known not to be the
case \citep[e.g.][]{eke04b,crot05,robo06}. However, since the median
luminosity scaling is less than a factor 1.6,
the difference that adjusting to halo specific LFs would have to the
integrated light will usually be smaller than the statistical scatter
observed (which is many 10s of percent).

Performing a single global optimisation using all bijectively matched
groups with $N_{\rm FoF} \ge 5$ results in $B=1.04$. 
{ 
This number accounts for a number of competing effects: the shape of
the faint end slope ($\alpha$) and the characteristic magnitude
($M^*$) varying between grouped environments and the global average,
and the effects of interloper flux biasing the extrapolated
group luminosities. Overall the effects are rather small, and globally we
see a value close to 1, which implies neither a large amount of under-grouping
nor over-grouping.
}

\begin{figure*}
\centerline{\mbox{\includegraphics[width=7in]{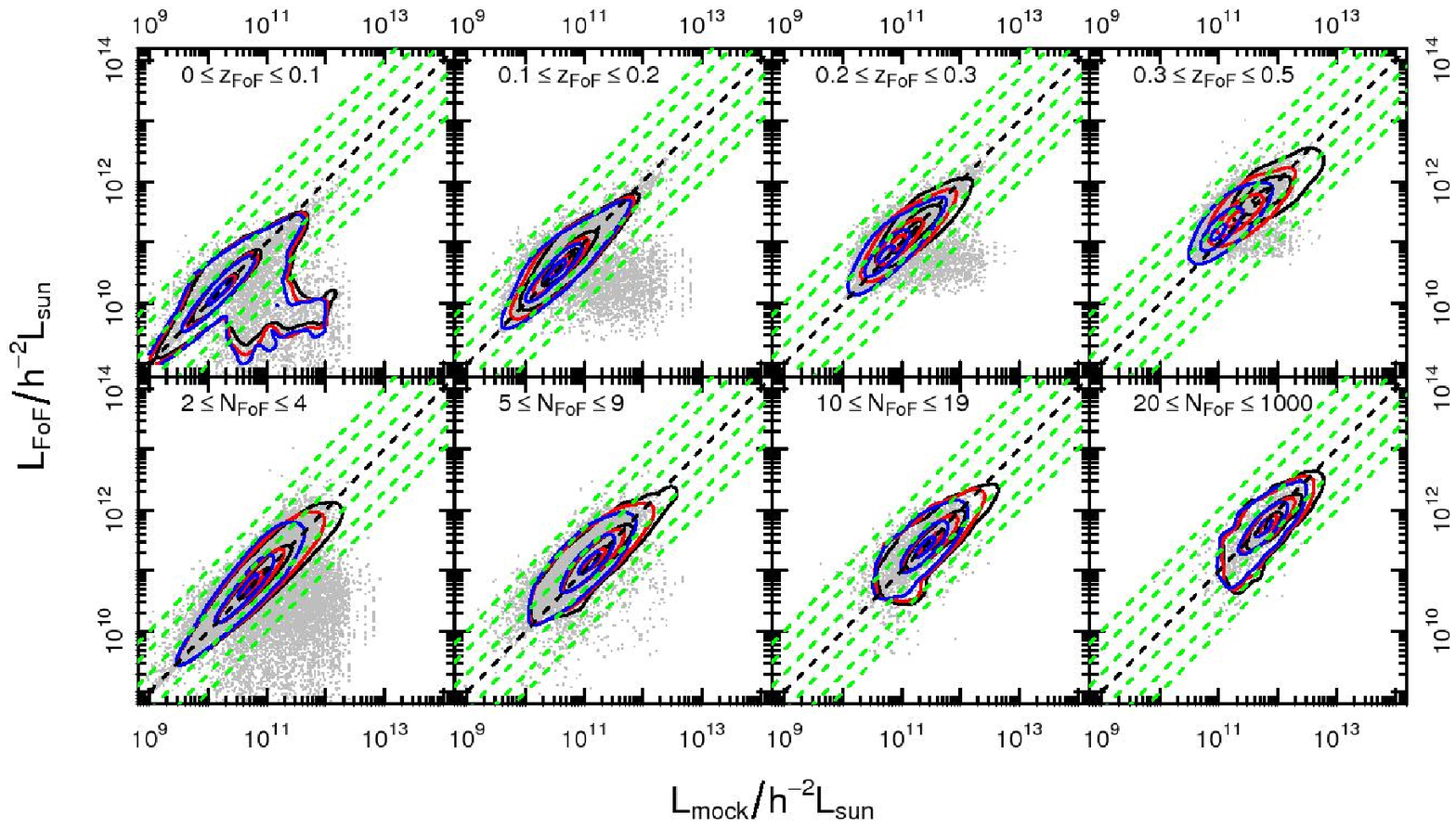}}}
\caption{\small 2-D density distribution of best matching FoF/halo mock groups
  in the L$_{\rm FoF}$--L$_{\rm mock}$ plane, split as a function of
  redshift and multiplicity (top and bottom panel respectively). These
  panels objectively compare the recovered group luminosities to the
  underlying total luminosity in the mocks. The contours represent the
  regions containing 10/50/90\% of the data for three magnitude limits,
  i.e. $r_{\rm AB}\le19.0$ (black), $r_{\rm AB}\le19.4$ (red) and 
  $r_{\rm AB}\le19.8$ (blue). The dots indicate the exact 
  L$_{\rm FoF}$--L$_{\rm mock}$ pairs.
    The green dashed lines delineate regions where L$_{\rm FoF}$ is
  2/5/10 times off the underlying L$_{\rm mock}$.
  For L$_{\rm FoF}$ we use Eq.~\ref{eq:lum_est} and $B=1.04$.
}
\label{fig:lumgrid}
\end{figure*}

Fig.~\ref{fig:lumgrid} compares the inferred total group luminosity
(L$_{\rm FoF}$) to 
the underlying mock luminosity (L$_{\rm mock}$) for best matching FoF/mock galaxy groups. 
The typical scatter as a function of mock group luminosity is quite constant regardless of group multiplicity, with only an excessive amount of scatter for the lowest multiplicity groups, as evidenced in the bottom left panel of Fig.~\ref{fig:lumgrid}. The relations are mostly unbiased, except for the two higher redshift samples (top right panels of Fig.~\ref{fig:lumgrid}).

The scatter in extrapolated group luminosity is much smaller than
seen for dynamical masses in Fig.~\ref{fig:massgridcont}. This is expected
since fewer observables are required in its estimate and the effect
of interlopers is much smaller. By their nature, interlopers are
more likely to systematically affect ``geometrical'' quantities, like
biasing the observed velocity dispersion and radius, while having a
lesser impact on e.g. total luminosities.
{ 
This is because the nature of the optimal grouping used for this work
means that on average we should miss as many true group galaxies as
add interlopers, so the net loss and gain of galaxy luminosities 
tend to balance out.
}

As with the dynamical mass estimates, scaling factors, listed in Table~\ref{tab:Bfactors}, are calculated for
various redshift and multiplicity subsets in order to properly quantify outstanding biases
that remain after scaling the observed luminosities to account for galaxies below the survey flux limit.
They
are distributed around unity, which is what we would expect if the extrapolated flux fully
accounts for all of the missing flux. The variation in the median seen in the table
is larger than seen for the dynamical mass scaling factors. This is because we have applied
a global LF correction to the data and the LF is known to vary strongly as a function
of group environment \citep[e.g.][]{robo06}. Since we are naturally more sensitive to higher
mass groups at higher redshifts, this explains the strong redshift gradient scaling factor
required, and in comparison the multiplicity variation is very small. For the dynamical $A$ factors
the dominant variable was the group multiplicity. When using the groups this is an important
consideration: the group dynamical masses are more intrinsically stable (require smaller corrections)
as a function of redshift, whilst group luminosities are more stable as a function of multiplicity.

\begin{table*}
\begin{small}
\begin{center}
\begin{tabular}{lrrrcrrrcrrrcrrr}
 &  \multicolumn{3}{l}{$2 \le N_{\rm FoF} \le 4$} & \hspace{1 mm} & \multicolumn{3}{l}{$5 \le N_{\rm FoF} \le 9$} & \hspace{1 mm} &  \multicolumn{3}{l}{$10 \le N_{\rm FoF} \le 19$} & \hspace{1 mm} &  \multicolumn{3}{l}{$20 \le N_{\rm FoF} \le 1000$} \\
 & 19.0 & 19.4 & 19.8 & & 19.0 & 19.4 & 19.8 & & 19.0 & 19.4 & 19.8 & & 19.0 & 19.4 & 19.8 \\ 
  \hline
  $0 \le z_{\rm FoF} \le 0.1$    & 1.1 & 1.1 & 1.1 & & 1.1 & 1.1 & 1.1 & & 1.4 & 1.3 & 1.2 & & 1.8 & 1.7 & 1.6 \\ 
  $0.1 \le z_{\rm FoF} \le 0.2$ & 1.0 & 1.0 & 1.0 & & 1.1 & 1.0 & 1.0 & & 1.2 & 1.1 & 1.1 & & 1.3 & 1.2 & 1.2 \\ 
  $0.2 \le z_{\rm FoF} \le 0.3$ & 1.0 & 0.9 & 0.9 & & 1.1 & 1.0 & 0.9 & & 1.2 & 1.0 & 1.0 & & 1.2 & 1.1 & 1.0 \\ 
  $0.3 \le z_{\rm FoF} \le 0.5$ & 1.1 & 0.8 & 0.7 & & 1.2 & 0.9 & 0.7 & & 1.5 & 1.0 & 0.8 & & 1.1 & 1.2 & 0.9 \\ 

\end{tabular}
\caption{Values of $B$, the luminosity scaling factor of 
  Eq.~\ref{eq:lum_est}, required to create an unbiased median halo
  luminosity estimate for different disjoint subsets of bijectively
  matched groups. 
  }
\label{tab:Bfactors}
\end{center}
\end{small}
\end{table*}

As with the dynamical masses, the total group luminosity correction factors ($B$) can be
well described by a plane that fits Table~\ref{tab:Bfactors} viz

\begin{equation}
B(N_{\rm FoF},z_{\rm FoF})= B_{\rm c} + \frac{B_{\rm N}}{\sqrt{N_{\rm FoF}}} + \frac{B_{\rm z}}{\sqrt{z_{\rm FoF}}}\; ,
\label{eq:Bscale}
\end{equation}

\noindent where $B_{\rm c}$, $B_{\rm N}$ and $B_{\rm z}$ are constants to be fitted.
Table~\ref{tab:BfactorsFn} contains the best parameters that produce the best fitting planes for the three different GAMA magnitude limits. The errors shown in Table~\ref{tab:BfactorsFn} are estimated from finding the best fitting plane for the 9 mock GAMA volumes separately and measuring the standard deviation of the individual best fitting planes, much like the approach used for Table~\ref{paramerror}.

\begin{table}
\begin{center}
\begin{tabular}{lccc}
 & $B_{\rm c}$ & $B_{\rm N}$ & $B_{\rm z}$ \\
 \hline
$r_{\rm AB}\le19.0$ & $+$1.27 $\pm$ 0.38 & -0.67 $\pm$ 0.25 & 0.08 $\pm$ 0.10 \\
$r_{\rm AB}\le19.4$ & $+$0.94 $\pm$ 0.12 & -0.67 $\pm$ 0.11 & 0.16 $\pm$ 0.04 \\
$r_{\rm AB}\le19.8$ & $+$0.65 $\pm$ 0.06 & -0.50 $\pm$ 0.06 & 0.22 $\pm$ 0.02 \\
\end{tabular}
\caption{Table of parameters that create the best fitting plane to the data in Table~\ref{tab:Bfactors}. The plane is a function of group redshift and multiplicity, as given in Eq.~\ref{eq:lum_est}. Errors are estimated from running plane fits to the 9 mock GAMA volumes separately and measuring the standard deviation of the individual best fitting planes.}
\label{tab:BfactorsFn}
\end{center}
\end{table}

\subsection{Group mass and light}

The $M/L$ observed in groups is a fundamental property of interest
in the analysis of galaxy groups. It is obviously important that any intrinsic
scatter in the estimates of both mass and luminosity of groups is not strongly
correlated.

\begin{figure*}
\centerline{\mbox{\includegraphics[width=7in]{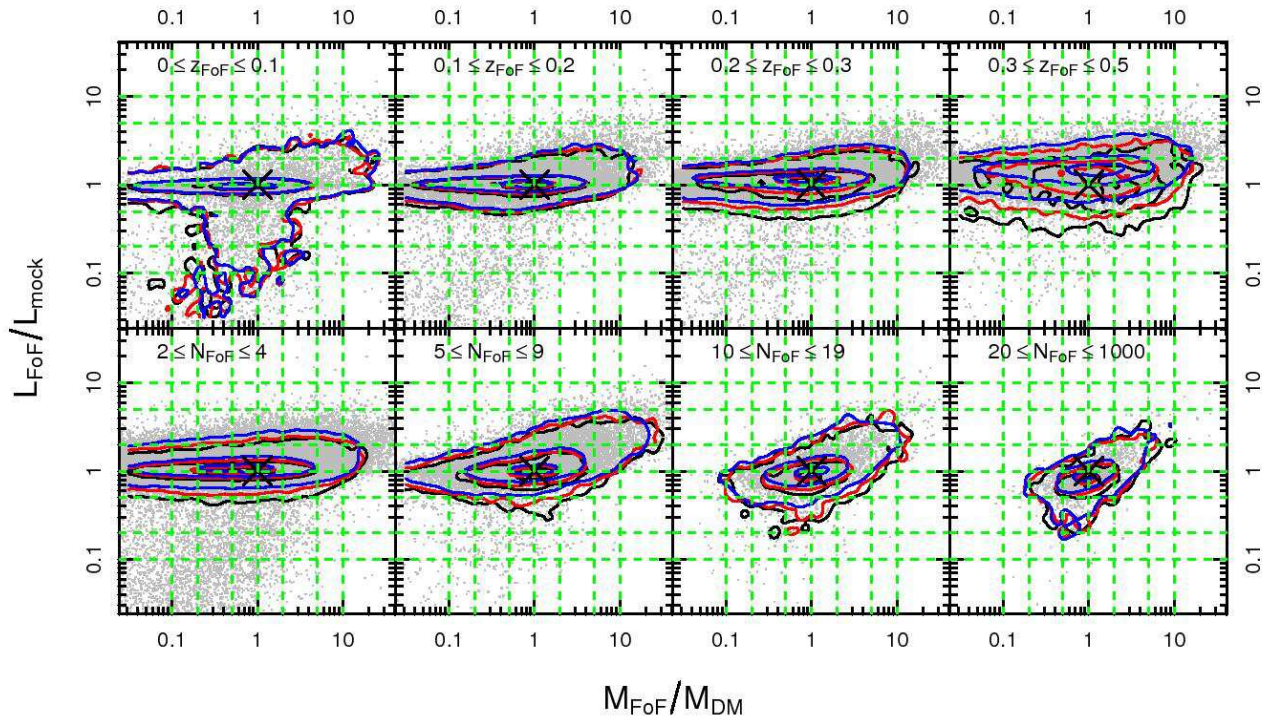}}}
\caption{\small Comparison of the fidelity of the recovered group mass (x-axis) against the group luminosity (y-axis), split as a function of
  redshift and multiplicity (top and bottom panel respectively). For both axes only a global median correction optimized for $N_{\rm FoF} \ge 5$ groups is applied, i.e. we use Eq.~\ref{eq:mass_est} and Eq.~\ref{eq:lum_est} with $A=10.0$ and $B=1.04$ for the mass and luminosity estimates respectively. The vertical (horizontal) green dashed lines present accuracy factors of 2/5/10 for mass (luminosity) estimates. The contours represent the regions containing 10/50/90\% of the data for three different magnitude limits, $r_{\rm AB}\le19.0$ (black), $r_{\rm AB}\le19.4$ (red) and $r_{\rm AB}\le19.8$ (blue).}
\label{fig:contM2L}
\end{figure*}

Fig.~\ref{fig:contM2L} shows the observed fidelity of the group dynamical
masses compared to the total group luminosities for a variety of data subsets.
Encouragingly the dynamical mass and luminosity estimates do not correlate
strongly in any direction--- the most significant concern would be strong scatter
along the $-45^\circ$ direction since this would mean that the dynamical mass
estimates tend to be erroneously small when the luminosity estimates tend to
be erroneously large (creating a very small $M/L$ ratio) and vice-versa.
Instead the two group measurements show no strong correlations in the accuracy of their recovery.

\begin{figure}
\centerline{\mbox{\includegraphics[width=3.5in]{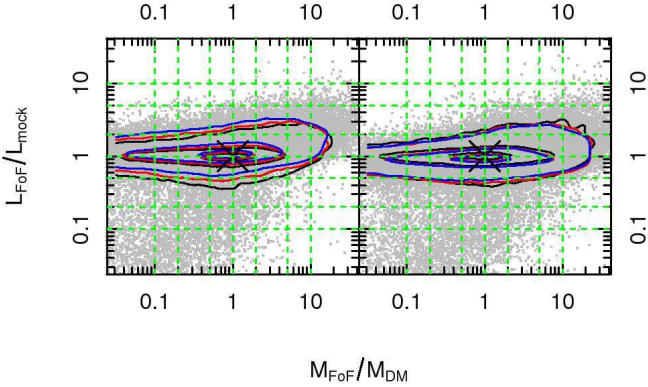}}}
\caption{\small Comparison of the fidelity of the recovered group mass (x-axis) against the group luminosity (y-axis). The left panel uses only a global median correction for mass and luminosity, optimized for $N_{\rm FoF} \ge 5$ groups (i.e. Eq.~\ref{eq:mass_est} and~\ref{eq:lum_est} with $A=10.0$ and $B=1.04$). The right panel uses the redshift and multiplicity dependent scaling functions of Eq.~\ref{eq:Ascale} and Eq.~\ref{eq:Bscale} with parameters listed in Tables~\ref{tab:AfactorsFn} and~\ref{tab:BfactorsFn} respectively. The green dashed lines show measurement accuracy factors of 2/5/10 for the mass and luminosity separately. The contours represent the regions containing 10/50/90\% of the data for three different magnitude limits, $r_{\rm AB}\le19.0$ (black), $r_{\rm AB}\le19.4$ (red) and $r_{\rm AB}\le19.8$ (blue).}
\label{fig:compscale}
\end{figure}

To demonstrate the improvement witnessed when using the multiplicity and redshift scaling relations, Fig.~\ref{fig:compscale} compares side by side the scatter expected for a simple median correction for
$N_{\rm FoF} \ge 5$ (left panel) and for a redshift and multiplicity dependent correction (right panel). The dynamical mass and luminosity scaling corrections use Eq.~\ref{eq:Ascale} and Eq.~\ref{eq:Bscale} with parameters listed in Tables~\ref{tab:AfactorsFn} and~\ref{tab:BfactorsFn} respectively. The scatter in the recovered luminosity is significantly reduced in the right panel.

It is clear that using the full multi-parameter scaling relations
offers an improved distribution of mass and luminosity scatter, as
well as creating extremely unbiased medians for the distributions. 
{ 
The three apparent magnitude limits used are brought into
closer alignment after applying the correction, and the amount of
scatter is reduced. The most significant change is for the 
90\% contour for high $L_{\rm FoF}/ L_{\rm mock}$, where we see the
contours tighten into very close agreement once the correction is made.
This means that groups extracted from regions of different
depths (e.g.\ G09 and G15 versus G12) can be compared more directly.
It is also clear that the mode and median are brought into better agreement,
moving up towards $L_{\rm FoF}/ L_{\rm mock}=1$.
}

Depending on the precise science goal the full
scaling equations should be used. Particular cases would be in any comparison
of extremely dissimilar groups over a large redshift baseline. However, in small
volume limited samples a simple median correction factor might be desirable. This
is particularly true at small redshift where the asymptotic nature of the plane function used
could produce spurious results.

\subsection{Quality of grouping}
\label{sec:qual_group}

The accuracy with which the galaxy composition of a group is recovered
is a distinct issue, but nevertheless equally important as the precise
recovery of intrinsic group properties, as considered in
\S\ref{sec:veldisp}--\S\ref{sec:lumtot}. 
For instance, even a group that has been perfectly recovered might
produce an incorrect mass estimate, the latter depending on the exact
observed configuration of galaxies on the sky and not solely on the
group membership. 
Using $Q_{\rm tot}$, as defined by Eq.~\ref{eq:quality} in
\S\ref{sec:cost_fct}, as our definition of grouping quality, we can
investigate how different aspects of grouping affect the purity of the
observed systems.

\begin{figure*}
\centerline{\mbox{\includegraphics[width=7in]{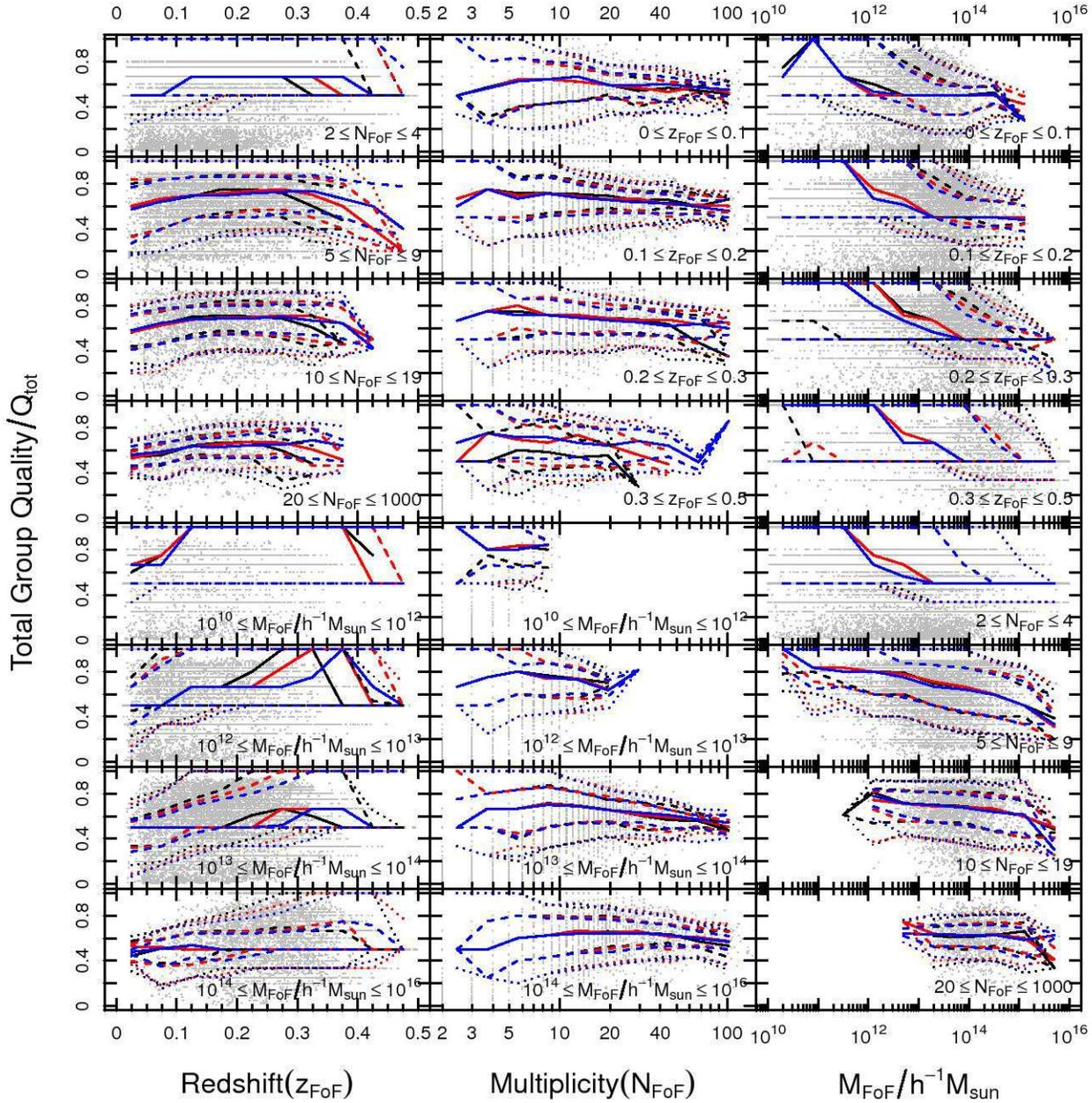}}}
\caption{\small Total group quality ($Q_{\rm tot}$) as a function of
  group redshift ($z_{\rm FoF}$), group multiplicity ($N_{\rm FoF}$)
  and group mass ($M_{\rm FoF}$).
  Each panel present a specific subsample of
  groups, as indicated by the key.
    Solid lines represent the moving median for $r_{\rm AB}\le19.0$
  (black),  $r_{\rm AB}\le19.4$ (red) and $r_{\rm AB}\le19.8$ (blue)
  survey limits. 
    Dashed (dotted) lines are for 25 and 75 (10 and 90)
  percentiles. Grey points show the $r_{\rm AB}\le 19.4$ data.
  }
\label{fig:qualitygrid}
\end{figure*}

\begin{figure*}
\centerline{\mbox{\includegraphics[width=7in]{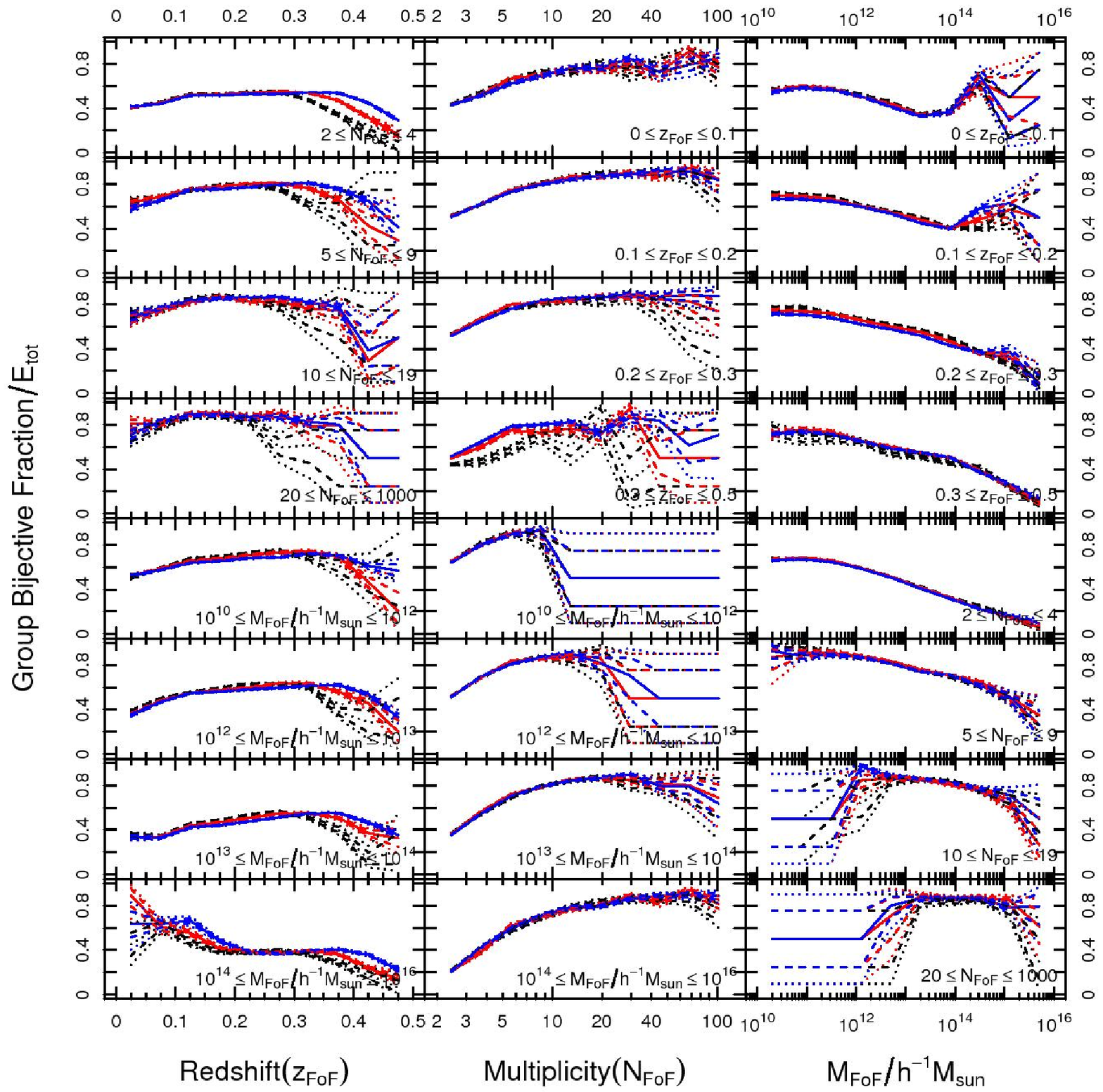}}}
\caption{\small Bijective group fraction ($E_{\rm tot}$) as a function of
  group redshift ($z_{\rm FoF}$), group multiplicity ($N_{\rm FoF}$)
  and group mass ($M_{\rm FoF}$).
  Each panel present a specific subsample of
  groups, as indicated by the key.
    Solid lines represent the moving median for $r_{\rm AB}\le19.0$
  (black),  $r_{\rm AB}\le19.4$ (red) and $r_{\rm AB}\le19.8$ (blue)
  survey limits. 
    Dashed (dotted) lines are for 25 and 75 (10 and 90)
  percentiles. Grey points show the $r_{\rm AB}\le 19.4$ data.
    }
\label{fig:bijgrid}
\end{figure*}

Fig.~\ref{fig:qualitygrid} and Fig.~\ref{fig:bijgrid} show how $Q_{\rm tot}$ and $E_{\rm tot}$ vary within
different group subsets for best matching FoF/halo mock groups.
The grouping optimisation
was not done with the whole sample, rather only groups with $N_{FoF} \ge 5$
contributed to the cost function. Hence panels that contain groups of lower multiplicity (i.e. $2 \le N_{FoF} \le 4$)
did not drive the optimisation, but demonstrate the
consequence of it.

The parameter that best constrains the group quality is the multiplicity, where the spread in observed grouping quality reduces for higher multiplicity systems. The most accurate groups tend to be at redshifts $z \sim 0.2$ and have low multiplicities. This is to be expected since the global optimisation considered will naturally be drawn to the regime where most groups are. That said, the bijective fraction of recovered groups is best for high multiplicity systems and remains very steady with redshift. The overall effect is that groups are more likely to be unambiguously discovered (i.e.\ bijective) when $N_{\rm FoF}$ is high (middle panels of Fig.~\ref{fig:bijgrid}), while the quality of the groups is, on average, quite constant with $N_{\rm FoF}$ (middle panel of Fig.~\ref{fig:qualitygrid}. Bijection and quality are obviously related, and these results should be interpreted as low multiplicity groups possessing a large amount of scatter in the quality of grouping, meaning that they can be scattered below the quality limits required for a successful bijection even though the median quality is quite high. Higher multiplicity systems possess less intrinsic scatter in the quality of grouping, meaning they are very rarely scattered below the bijection limits, and consequently the average bijection fraction remains higher.
 
The exception to this is that the lowest mass groups appear to be the most accurately recovered, even though most observed have masses $M \sim 10^{13}\msolh$. This can be understood when careful attention is paid to how the FoF algorithm constructs the groups.
It creates upper limits for the allowed difference in either the radial (velocity) or tangential (physical) separation between galaxies. It must be the case that groups that are constructed from galaxies that are at the limit of the allowed separations will be larger in terms of projected radius and observed velocity dispersion than groups with galaxy separations well within these limits. This means they will have larger dynamical masses, and assuming interlopers are spread uniformly in space they will have a lower $Q_{\rm tot}$ since they will cover a larger volume in redshift space, so be more likely to include interlopers. This is an interesting effect of the grouping, because although the masses measured are likely to be too small the actual groups are extremely secure.

\begin{figure}
\centerline{\mbox{\includegraphics[width=3.5in]{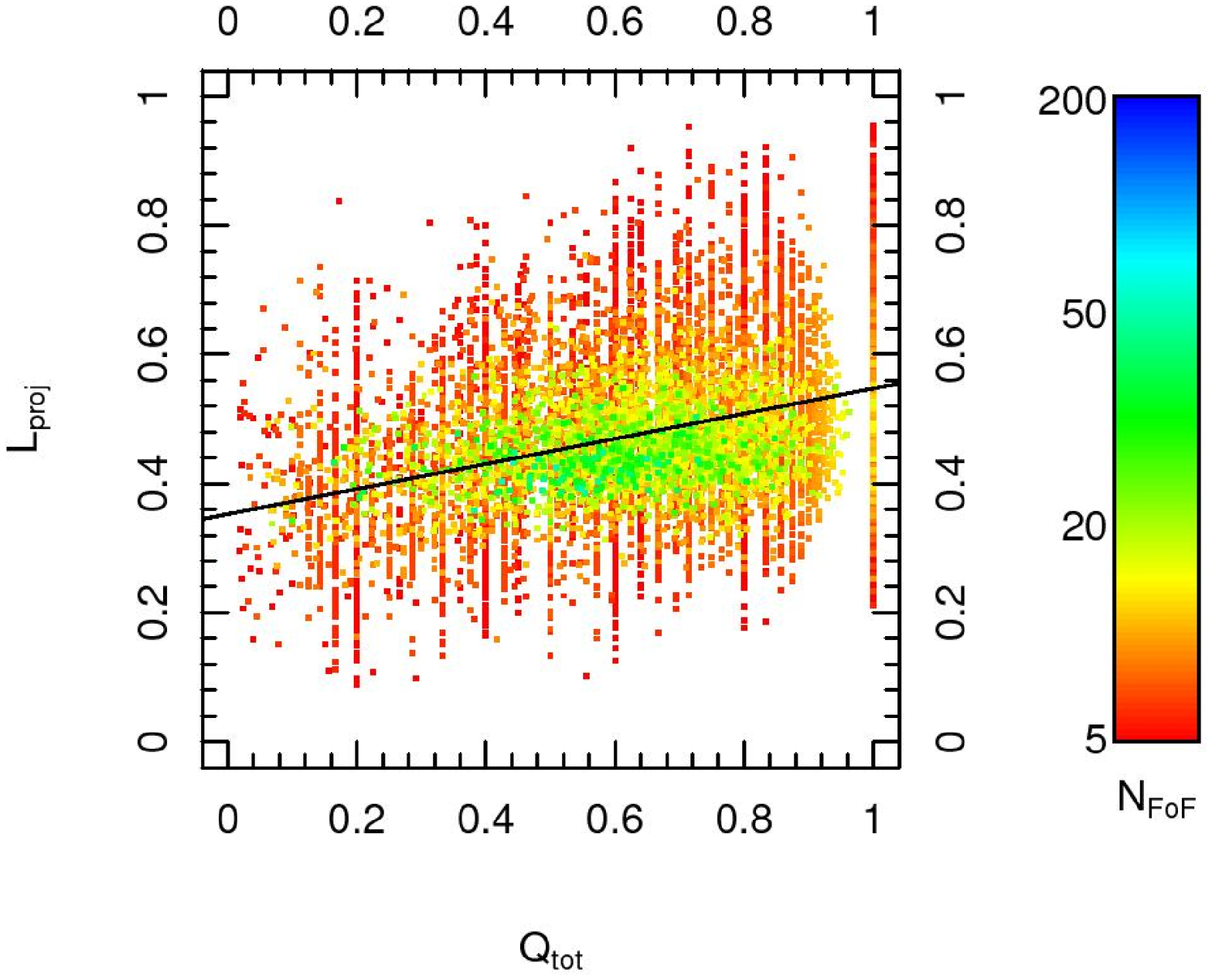}}}
\caption{\small Comparison of the observed linking strength $L_{\rm proj}$ with the intrinsic group quality $Q_{\rm tot}$. The colour of each data point represents the group multiplicity, going from $N_{FoF}=5$ (red) to $N_{FoF}=200$ (blue). The correlation is strongest for low multiplicity systems, which is important since it is these that can be pathologically bad. The black line is the linear regression fit to the entire data, so it predominantly describes the lower multiplicity systems.}
\label{fig:avstren}
\end{figure}

With this understandable effect in mind, different methods for estimating the intrinsic $Q_{\rm tot}$ using observed linking characteristics were investigated. The most successful proved to be calculating the following for each group:

\begin{equation}
L_{\rm proj} = \frac{\sum_{i=1}^{N_{\rm FoF}} \sum_{j=1}^{N_{\rm FoF}} \left[ 1- \frac{ \tan{\theta_{\rm i,j}} \, (D_{\rm com,i} + D_{\rm com,j}) } { b_{\rm i,j} \,(D_{\rm lim,i}+D_{\rm lim,j}) } \, \delta_{i,j}^c \right]}{N_{\rm links}},
\end{equation}

\noindent where $\delta_{i,j}$ is unity if $i$ and $j$ are directly linked (and zero otherwise), while all other terms are as described in Eq.~\ref{eq:D_perp_12}. 
Hence the sum is done over allowed links within the group ($N_{\rm links}$) which has a limit of $N_{\rm FoF}(N_{\rm FoF}-1)$. 
This statistic estimates how much closer than the allowed maximum separation all of the galaxies are on average, and when this number is large it indicates the group must be very compact in projection relative to the allowed size. Fig.~\ref{fig:avstren} demonstrates how $L_{\rm proj}$ correlates loosely with $Q_{\rm tot}$. Interestingly, the equivalent statistic measuring the radial linking shows very little correlation with $Q_{\rm tot}$. This means that outliers tend to fit quite comfortably in velocity space, but look anomalous in projection. To aid the selection of high-fidelity groups $L_{\rm proj}$ will be released along with the group catalogue.

\subsection{Sensitivity of grouping to mock catalogues}

So far in this work we have made the implicit assumptions that the
mocks are to a large extent a good representation of the real
Universe and that optimising the grouping algorithm to recover mock
groups as accurately as possible will have the desired effect of also
returning the best groups from the GAMA data.
Clearly we should be wary of the effects of over-tuning our algorithm
to the mocks, especially given the limitations listed
in~\S\ref{sec:mocks}. To better understand how sensitive our final
group catalogue might be to certain intrinsic mock properties,
three small variations affecting the redshift-space positions of
the mock galaxy were implemented which lead to slight changes in
the ``observed'' mocks. These perturbations
where applied to the $r \le 19.4$ mock catalogues since that should
be indicative of the impact we might expect. The modifications consist of:

\begin{itemize}

\item[1)] increasing all galaxy peculiar velocities along the line of
  sight by 10\%,
  creating groups that are less compact in velocity space than the
  default mocks: mock$_{+}$. 

\item[2)] reducing all galaxy peculiar velocities along the line of
  sight by 10\%,
  creating groups that are more compact in velocity space than the
  default mocks: mock$_{-}$. 

\item[3)] convolving all galaxy peculiar velocities along the line of
  sight with a Gaussian velocity distribution of width
  $\sigma=50\kms$, mimicking the GAMA velocity errors: mock$_{\sigma}$. 

\end{itemize}

The first two sets of mock, mock$_{+}$ and mock$_{-}$, test the sensitivity
of the grouping to the fidelity in which small scale redshift space
distortions are accounted for in the mocks. From \citep{kim09} (and
Norberg et al. in prep) we know that the~\citet{bowe06} semi-analytic
galaxy formation model do not reproduce very accurately the redshift
space clustering on \Mpch\ scales and smaller. By systematically
modifying the peculiar velocities by $\pm 10$\% and by keeping the
same FoF grouping parameters we attempt to address this mismatch
between data and mocks and measure how sensitive the grouping is such
differences. From Norberg et al. (in prep) we expect that an
additional velocity bias of $\sim +10$\% to the mock galaxies should be
enough to reconcile the redshift space clustering of the mocks and the
data. 
The third set, mock$_{\sigma}$, tests the sensitivity of the grouping
to velocity errors, which were not considered in the nominal mocks
described in \S\ref{sec:mocks} but clearly present in the GAMA data.
To fully simulate how we treat the errors for the real GAMA data the
velocity errors are taken off in quadrature as described in Eq. 17.

The FoF algorithm with the nominal parameters as listed in
Table~\ref{bestparamstab} is applied to the three sets of mocks. 
The FoF grouping of the standard and modified mocks result in pretty
similar findings: 
The first impact these perturbed mocks might have on the grouping
is on the group assignments themselves, so $S_{\rm tot}$ was calculated
for all 3 varieties of new mocks where the reference mock data is now
the original mock lightcone. This means we are only analysing how similar
the new mock FoF groupings are to the original set, not to the ``true'' mock groupings.
$S_{\rm tot}$ is found to be $\sim0.97$ for all three varieties of
mock perturbation for $N_{\rm halo} \ge 2$, and only drops slightly
for $N_{\rm halo} \ge 20$ which shows the greatest discrepancy.
In this regime mock$_{+}$ has $S_{\rm tot}=0.94$, mock$_{-}$ has
$S_{\rm tot}=0.96$ and mock$_{\sigma}$ has $S_{\rm tot}=0.93$. 

For the estimated masses, it is obvious that mock$_{-}$ and mock$_{+}$
will require slightly different scaling relations to recover unbiased
halo masses. The global mass scaling factor (where $N_{\rm FoF} \ge 5$) for mock$_{-}$,
$A_{-}$, needs to be $11.6$, so 16\% larger than $A$, while $A_{+}$ needs to be
$\sim 8.7$, so 15\% smaller than $A$. This implies that we have an underlying systematic
uncertainty of at least 15\% on all masses assuming we expect the
true physics to vary the galaxy velocities at the 10\% level. Naively we might have expected
the difference to be at the $\sim 20$\% level since $1.1^2=1.21$, but the random nature
of peculiar velocities and the slight variation in grouping conspires to reduce the variation.

For mock$_{\sigma}$ we require
exactly the same global scaling relation as before, i.e.\ $A_{\sigma}=A=10.0$. This implies
that removing the velocity error in quadrature is the correct procedure, and means we certainly do not
expect the uncertainty in radial velocities to have a significant effect on the implied
masses.

The implication for the
group luminosities are, as expected, very marginal w.r.t.\ these
modifications of the mocks, which is a result of the grouping still
being rather good for all three set of mocks (as evidenced by the
marginal change in $S_{\rm tot}$) despite the algorithm not being
tuned to them.

\section{Global properties of G$^3$C\lowercase{v}1}
\label{sec:properties}

Having run extensive optimisations and calculated refinements based on the mock catalogues, the algorithm was run over the real GAMA data. In total, taking the deepest version of each GAMA survey region possible, 14,388 groups were formed containing 44,186 galaxies out of 110,192 galaxies in our volume limited selection, meaning ~40\% of all galaxies are assigned to a group. This is almost identical to the average grouping rate found in the mocks, also ~40\%. 

The headline group number statistics are listed in Table~\ref{tab:groupstat} for each of the GAMA regions, i.e. G09, G12 and G15. $r_{\rm AB}\le 19.0$ and $r_{\rm AB}\le 19.4$ catalogues were made for G09, G12 and G15, and an extra $r_{\rm AB}\le 19.8$ catalogue was created for G12 (the only region that has deep enough spectroscopy). This table also includes the expectation from the mocks with the minimum and maximum numbers of groups in the 9 GAMA lightcone mocks. Subsets that have numbers that are outside the min-max range of the mocks are flagged with an asterisk. 

From Table~\ref{tab:groupstat} we conclude that for most multiplicity ranges and survey limits the number of GAMA groups detected is very comparable to the predictions from the GAMA lightcones. Over the full GAMA lightcones G12 and G15 are very close to the mean counts recovered from the mocks whilst G09, although very much at the underdense extreme, is not outside the min-max range expected. The comparison between data and mocks seems less favourable when splitting the groups by redshift and survey depth, where 5 GAMA subsets lie outside of the min-max limits of the mocks. The difference becomes less and less significant the deeper the survey is and seems to be most significant in G09, which is underdense in all subsets investigated.

It is well established that G09 is underdense below $z<0.2$ compared to the whole of SDSS \citep{driv11}, whilst G12 and G15 are closer to the large scale average. Overall, this underdensity accounts for why we find fewer groups in G09. G09 is similar to the most underdense and group sparse GAMA area found in the mocks, suggesting it is a rare event in the mocks but at least not completely unmatched. G12 is most like the typical mock distribution, and the GAMA $r_{\rm AB}\le19.8$ group catalogue is the most similar to the mocks of all catalogues. This catalogue tends to contain fewer large multiplicity groups than predicted by the mocks. These inconsistencies are not highly significant overall, but they reflect similar findings in the 2PIGG catalogue \citep{eke04}.

\begin{table*}
\begin{center}
\tiny
\begin{tabular}{lccccccccccccccc}
					& \multicolumn{4}{l}{$r \le 19.0$}					& \hspace{1 mm}	& \multicolumn{4}{l}{$r \le 19.4$}					& \hspace{1 mm}	& \multicolumn{2}{l}{$r \le 19.8$}	\\
                       			& G09 	& G12    		& G15	&  Mocks (low, high)	&	& G09 	& G12    		& G15	&  Mocks (low, high)	&	& G12   		&  Mocks $\pm \sigma$ (low, high)	\\
\hline
$N_{\rm group}$ 2--4	& 2051	& 2409        	& 2436     	& 2334 (3154, 4100)	&	& 3334	& 3703        	& 3776     	& 3623 (3154, 4100)	&	& 5687     		& 5520 (4861, 6101) 	\\
$N_{\rm group}$ 5--9       & 190   	& 233         	& 234       	& 253 (188, 294)   		&	& 329   	& 395         	& 339       	& 390 (322, 455)   		&	& 539       		& 584 (509, 661) 		\\
$N_{\rm group}$ 10--19  	& 45       	& 55          	& 59       	& 66 (43, 82)    		&	& 75       	& 79          	& 102       	& 102 (69, 133)    		&	& 121      		& 155 (98, 189) 		\\
$N_{\rm group}$ 20+     	& 8*       	& 16          	& 16         	& 26 (15, 39)        		&	& 17*      	& 26          	& 25         	& 40 (20, 55)        		&	& 44        		& 62 (34, 88) 			\\
\hline
$z_{\rm group}$ 0--0.1	& 419	& 577		& 512	& 531 (318, 856)		&	& 514	& 705		& 597	& 634 (379, 1028)		&	& 857		& 746 (437, 1204)		\\
$z_{\rm group}$ 0.1--0.2	& 973	& 1369		& 1450*	& 1144 (803, 1381)		&	& 1338	& 1829		& 1967*	& 1552 (1076, 1841)	&	& 2331		& 2024 (1424, 2424)	\\
$z_{\rm group}$ 0.2--0.3	& 725	& 640		& 633	& 814 (606, 996)		&	& 1372	& 1217		& 1198	& 1377 (1074, 1683)	&	& 1997		& 2124 (1683, 2584)	\\
$z_{\rm group}$ 0.3--0.5	& 178	& 127		& 100*	& 189 (125, 258)		&	& 531	& 452		& 480	& 593 (421, 730)		&	& 1206		& 1426 (1044, 1708)	\\
\hline
Total					& 2294	& 2713		& 2745	& 2678 (2204, 3107)	&	& 3755	& 4203		& 4242	& 4156 (3578, 4728)	&	& 6391		& 6321 (5535, 7025)	\\
\hline
\end{tabular}
\end{center}
\caption{Number of galaxy groups as a function of multiplicity,
  redshift and survey depth. The GAMA groups are split by GAMA
  regions, i.e. G09, G12 and G15. For the mocks, the mean number of
  groups between all 9 mock GAMA lightcones in a single GAMA region of
  $\sim48$~\sqdeg\ is listed together with their low and high extreme
  across all mocks (within brackets). Samples with an asterisk are
  those which are outside the min-max range of the mocks.}
\label{tab:groupstat}
\end{table*}

\begin{figure*}
\centerline{
  \mbox{\includegraphics[width=7in]{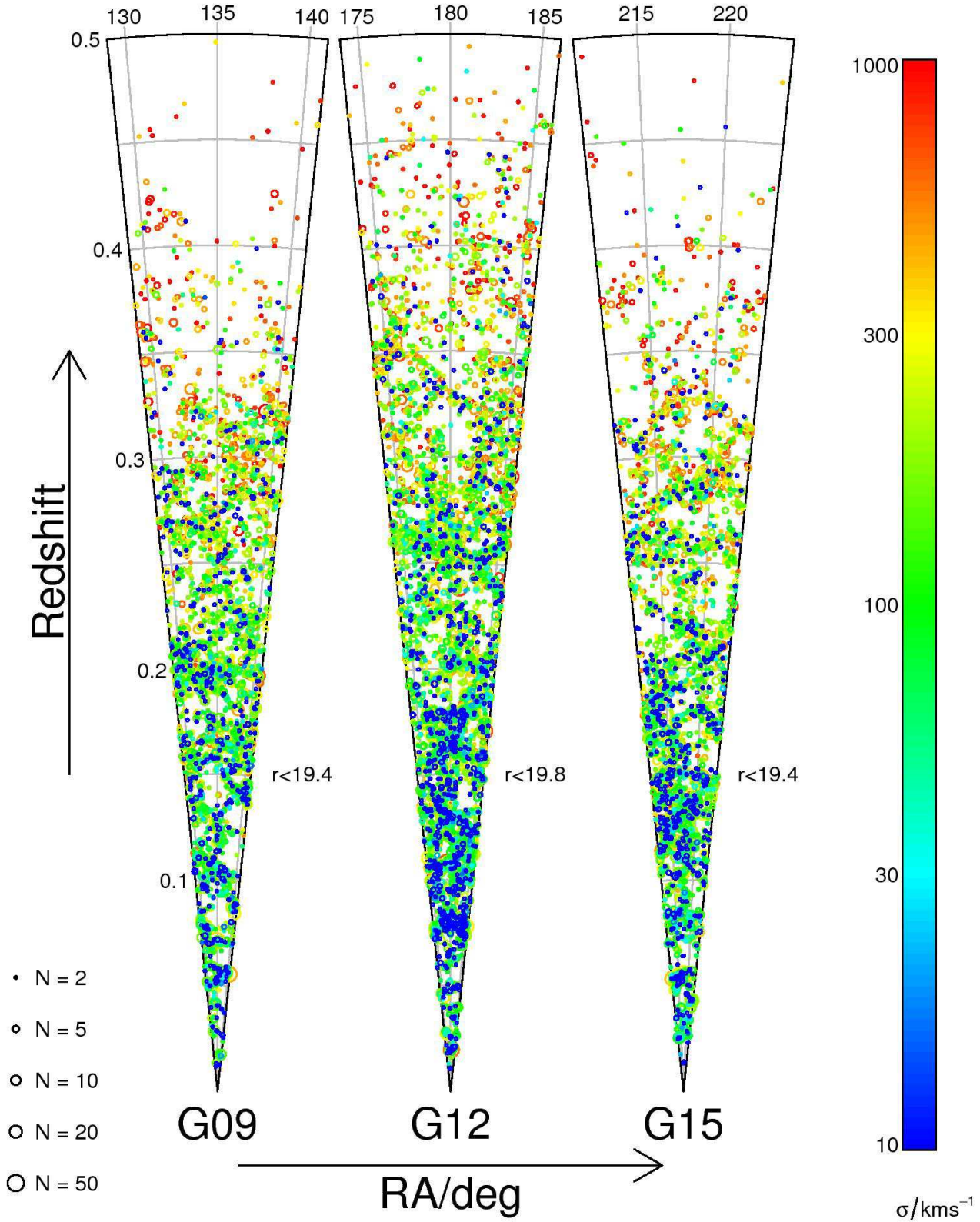}}
  }
\caption{\small Redshift space position of GAMA galaxy groups projected onto the equatorial plane, split by survey area and with symbol size reflecting the group multiplicity and symbol colour the group velocity dispersion (see figure keys for exact values). G09 and G15 are for a survey depth of $r_{\rm AB}\le 19.4$, while G12 is for $r_{\rm AB}\le 19.8$, explaining why the number of groups detected at higher redshifts is larger in G12 compared to G09 and G15. At low redshifts where the projection effects are the smallest, groups are still visually strongly associated with the filaments and nodes of the larger scale cosmic structure. Fewer groups are found beyond at higher redshift, a result of GAMA survey being magnitude limited.
}
\label{coneplot}
\end{figure*}

Fig.~\ref{coneplot} shows the position of the GAMA groups in redshift space projected onto the equatorial plane, with the symbol size reflecting the group multiplicity and colour the group velocity dispersion. The highest multiplicity groups are at lower redshifts, as should be expected in an apparent magnitude limited sample. This figure particularly highlights the sample variance seen between regions, as already mentioned in the discussion of Table~\ref{tab:groupstat}. There are vast regions of space that contain massive clusters and an assortment of groups, overlapping so tightly as to produce patches of solid colour in the plot. However, between these large filamentary regions there are voids that, whilst still possessing galaxies (in lower densities), barely contain a single significant group. At low redshifts ($z<0.1$) where the mean galaxy number density is the highest, such voids are still very evident in the GAMA data.

We still see groups of significant size ($N_{\rm FoF} \sim 20$) beyond a redshift of 0.3 in G09, and there is evidence of filamentary structure in the under-lying galaxy population beyond $z\sim0.4$ in G12 (G12 being 0.4 mags deeper than G09/G15 probes structure to slightly higher redshifts). In G12 there are a number of low multiplicity systems beyond a redshift of 0.4--- these groups appear to be associated with nodes in filamentary structure and have been visually identified as large clusters. This means that GAMA is able to measure the evolution of group properties and filamentary structure over a redshift baseline of 0--0.5, which is $\sim5$ Gyrs, or 36\% the lifetime of the Universe--- an evolutionary time span for large scale structure analysis that is unprecedented in a single coherent survey.

\begin{figure*}
\centerline{
	\mbox{\includegraphics[width=3in]{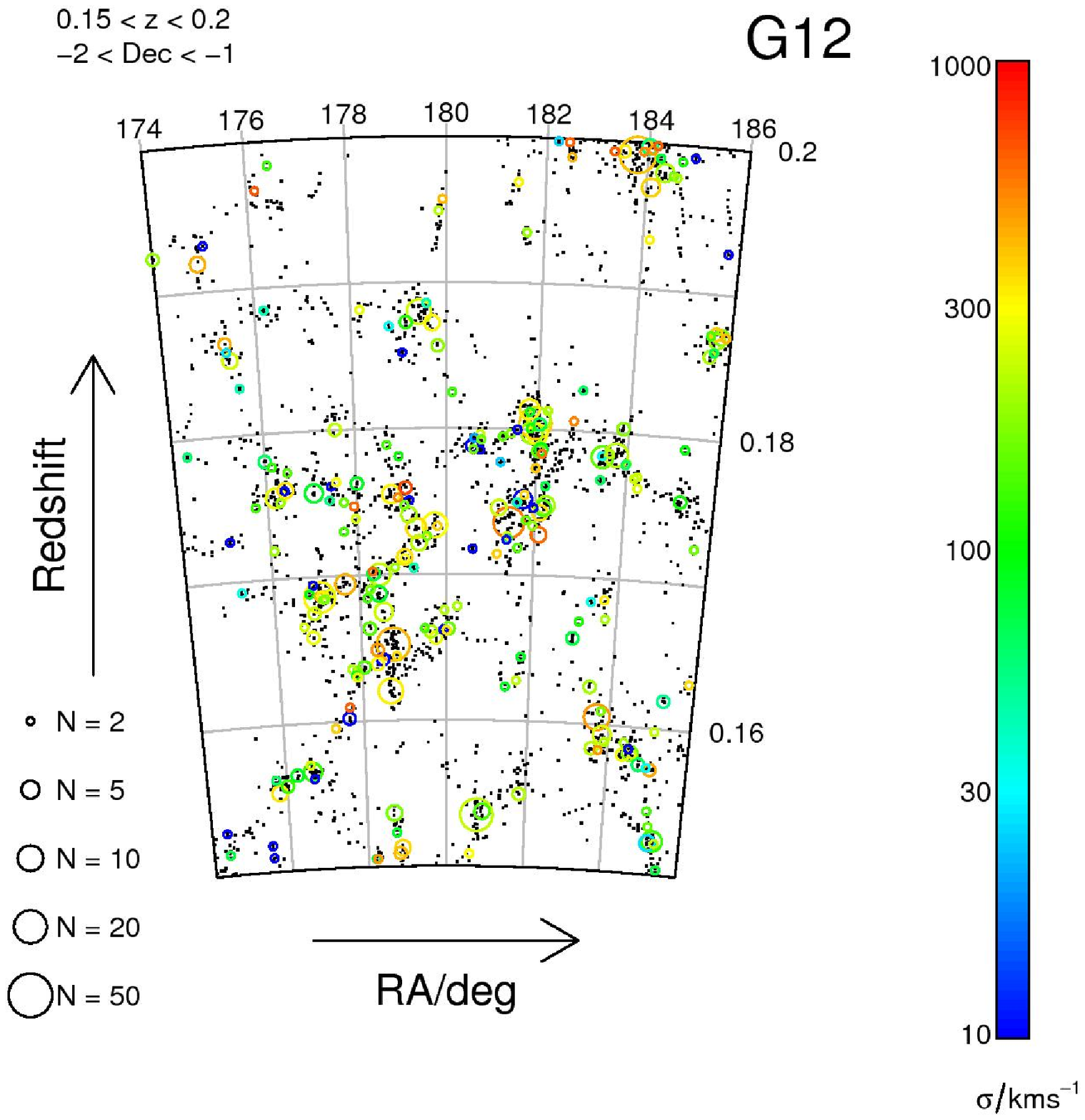}}
	\mbox{\includegraphics[width=3in]{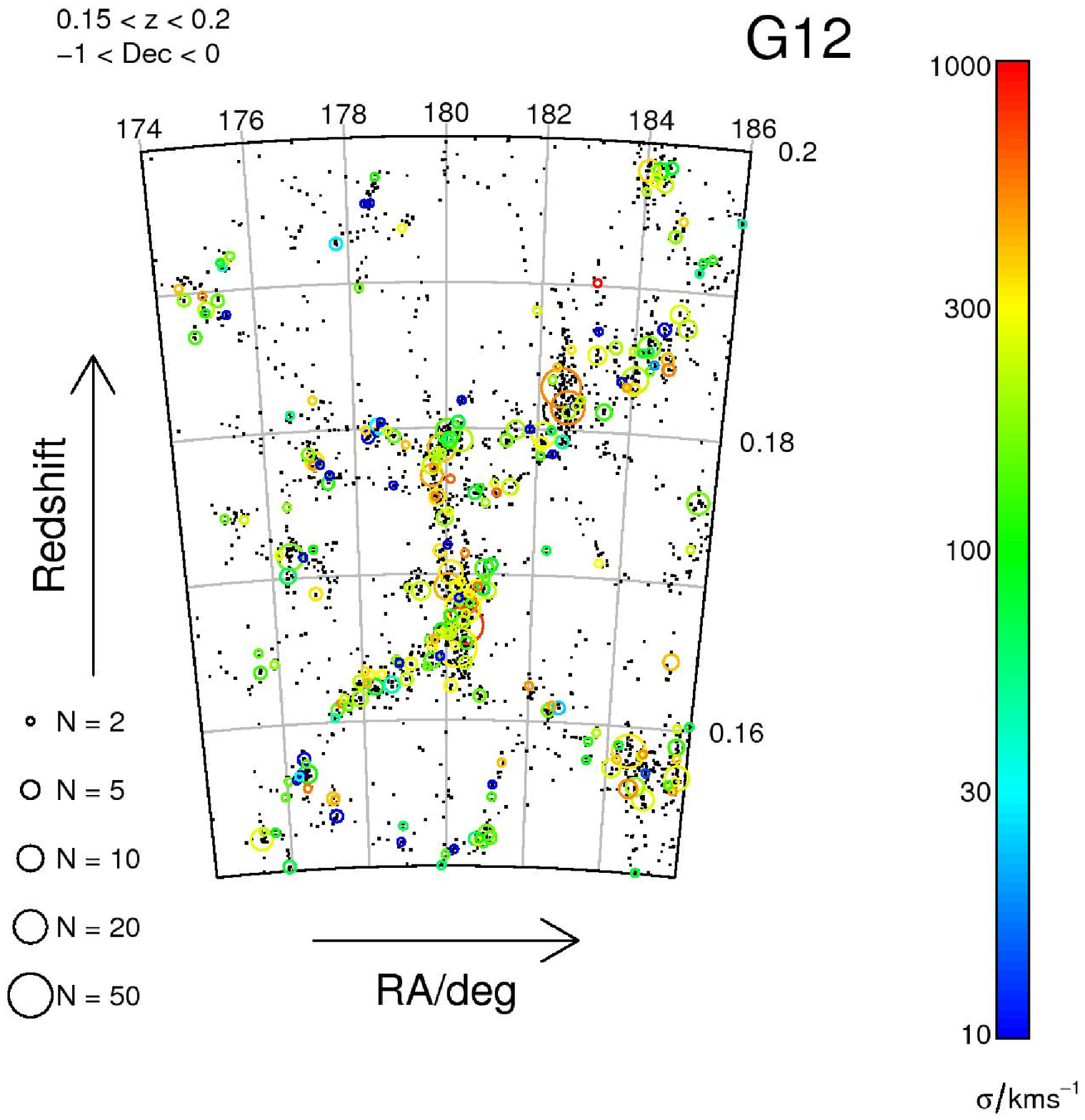}}
	}
\centerline{
	\mbox{\includegraphics[width=3in]{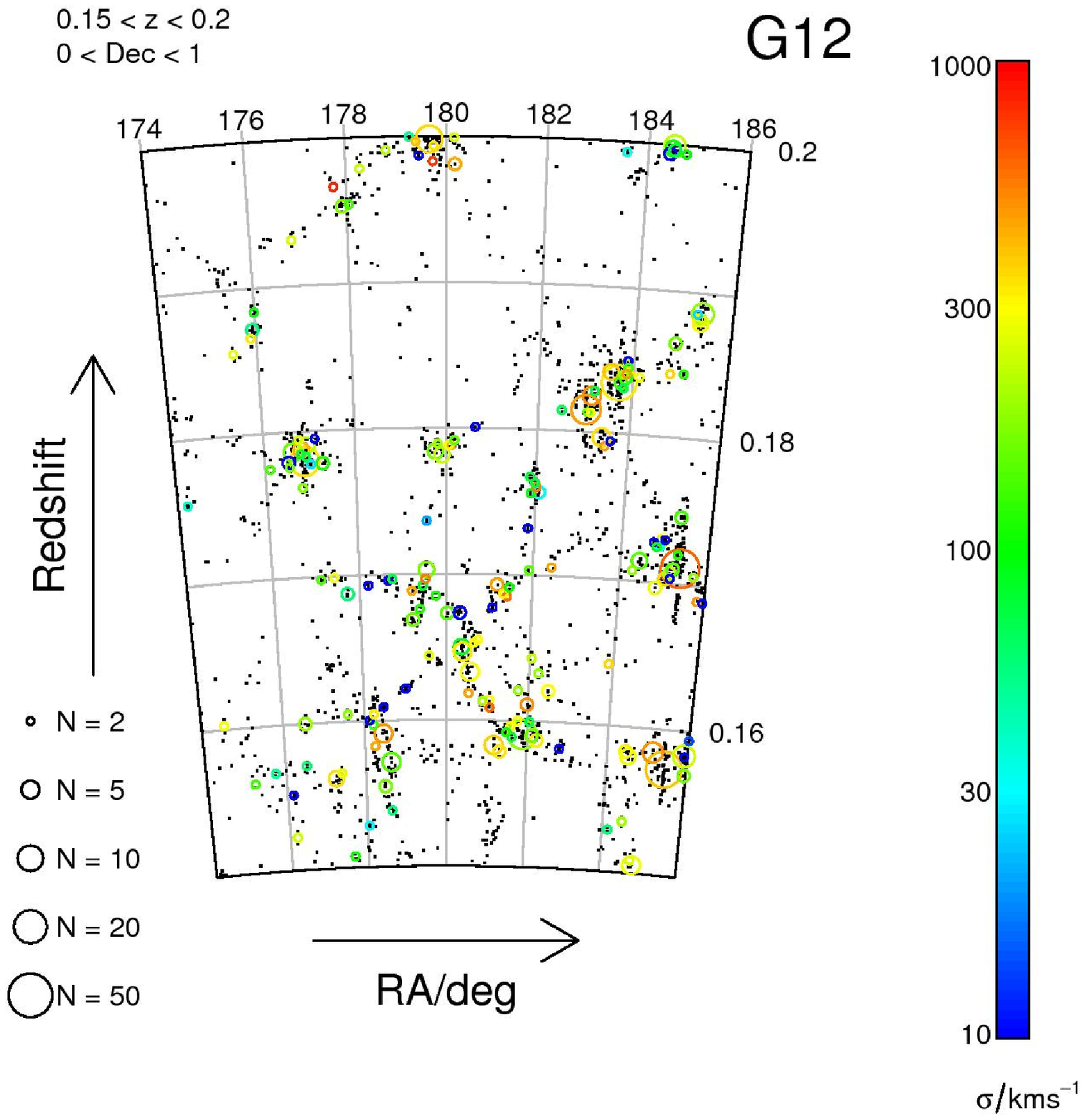}}
	\mbox{\includegraphics[width=3in]{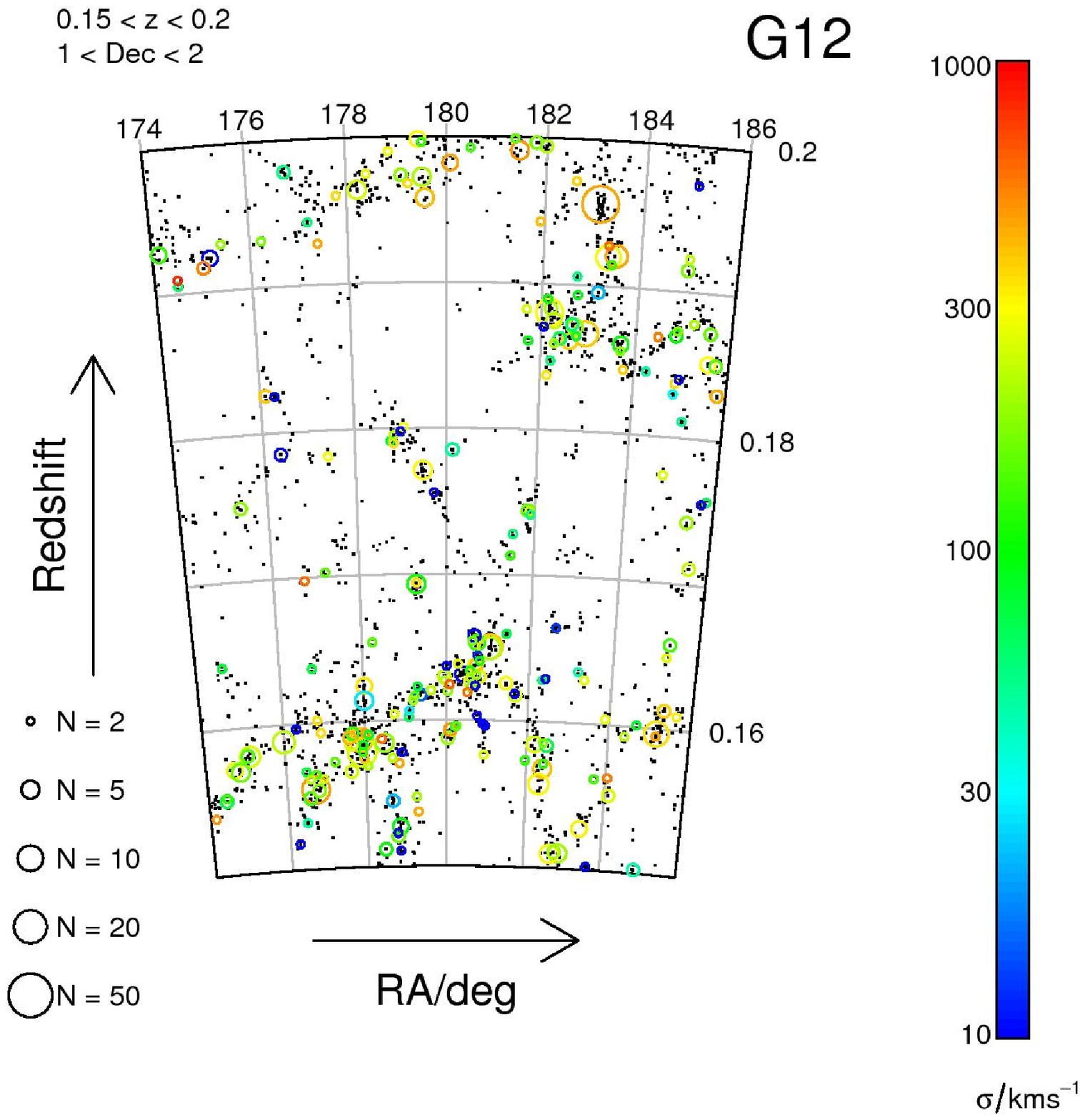}}
	}
\caption{\small Four one degree wide declination slices of the GAMA G12 region covering the $0.15<z<0.20$ redshift range. Declination increases from left to right and top to bottom, as indicated by the panel key. Galaxies are shown with black dots, and galaxy groups with the same symbols as in Fig.~\ref{coneplot}.
}
\label{fig:zoomcones}
\end{figure*}

Fig.~\ref{fig:zoomcones} shows a series of one degree wide declination slices in G12 that cover $0.15 \le z_{\rm group} \le 0.2$.
The black points show the location of individual galaxies, and as expected the groups closely trace overdensities seen in the
galaxy distribution. Intriguingly, we see evidence of extremely fine filamentary structure that is not associated with any of the
defined groups. If these structures were purely radial in direction then they could be claimed as misidentified systems, for which the filamentary structure merely betrays the velocity dispersion along the line of sight. Instead we witness gentle sweeping arcs that
move round steadily radially and in projection, implying that they are real fine filamentary structure that connects group nodes. This is probably one of the first times that one sees the galaxy distribution mimicking so closely the filamentary distribution which is so commonly seen in large Dark Matter dominated numerical simulations.

The most striking of these filaments can be found in the top-right panel of Fig.~\ref{fig:zoomcones} where fine strands
can be seen extending out from $\alpha \sim 180$deg.\ and $z \sim 0.18$, and also from $\alpha \sim 182$deg.\ and $z \sim 0.19$.
In both of these cases it is possible to identify group and cluster nodes that connect the filaments together, but there are
no groups detected within the filaments themselves.
It is important to highlight that without GAMA redshifts these regions would have previously been identified as void like, and
that the additional galaxies are not randomly distributed `field' galaxies, but appear to be in extremely well defined environments,
but non-grouped w.r.t.\ the GAMA mean galaxy number density.

\begin{figure*}
\centerline{
  \mbox{\includegraphics[width=3in]{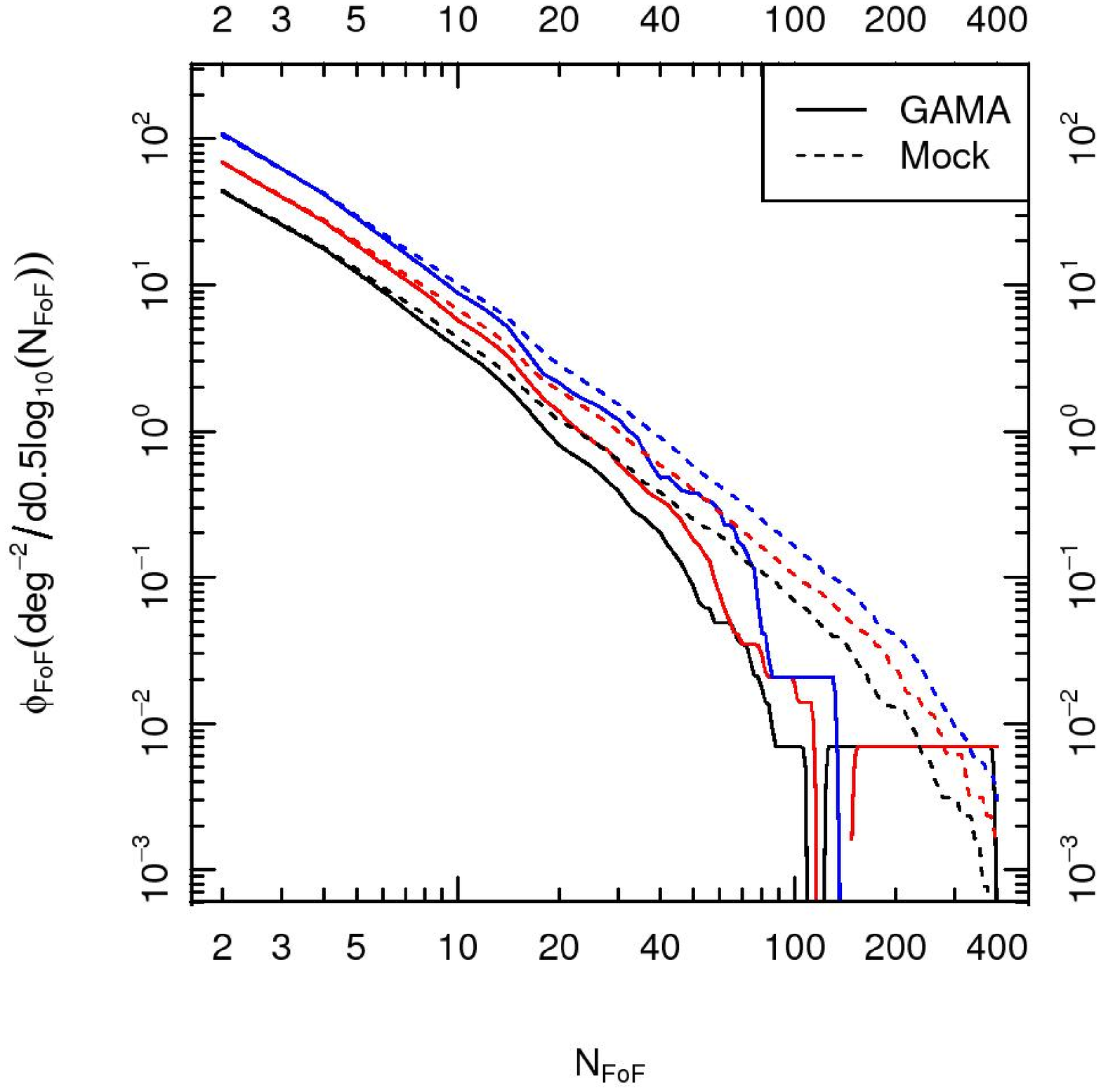}}
  \mbox{\includegraphics[width=3in]{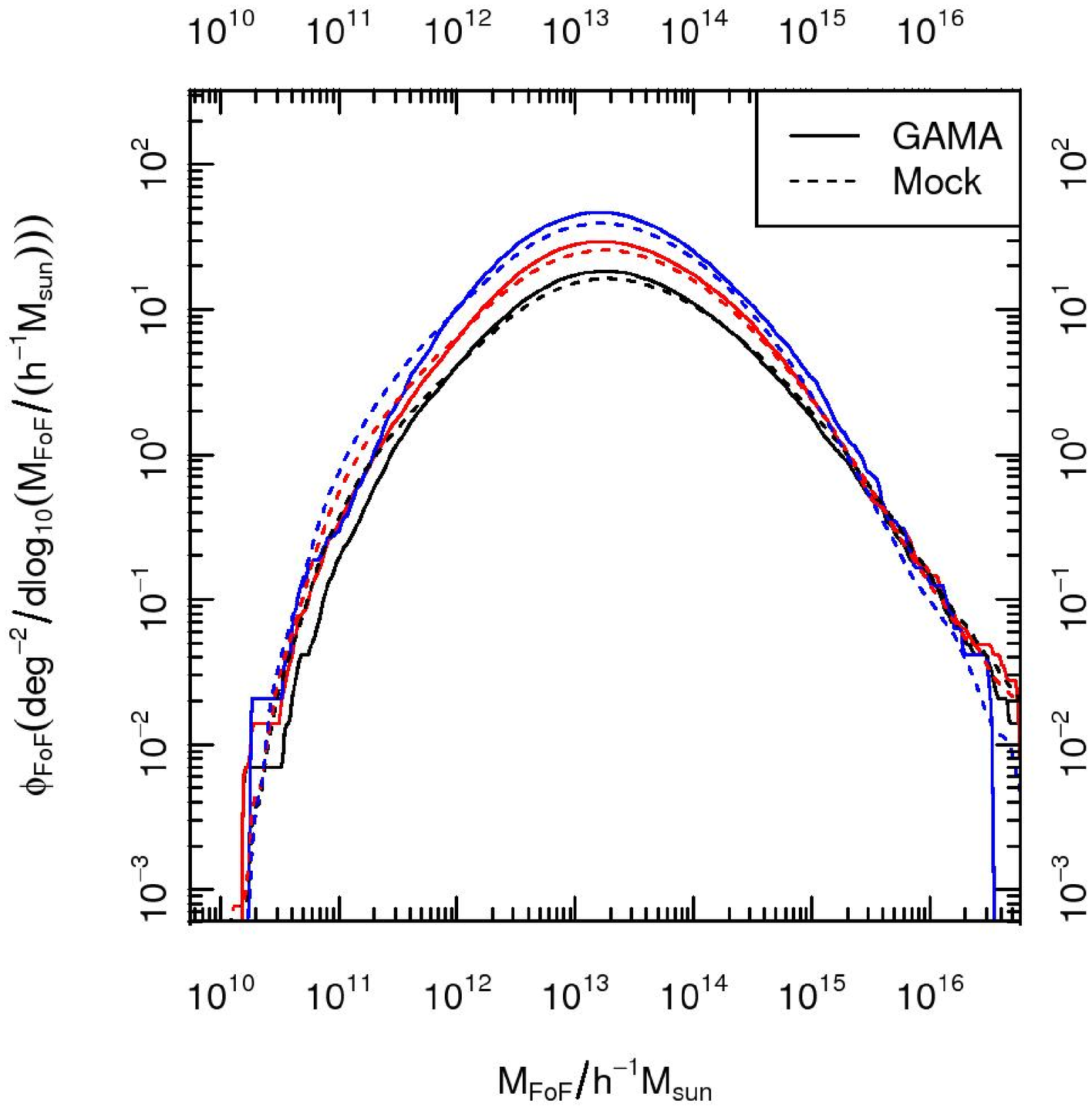}}
  }
\centerline{
  \mbox{\includegraphics[width=3in]{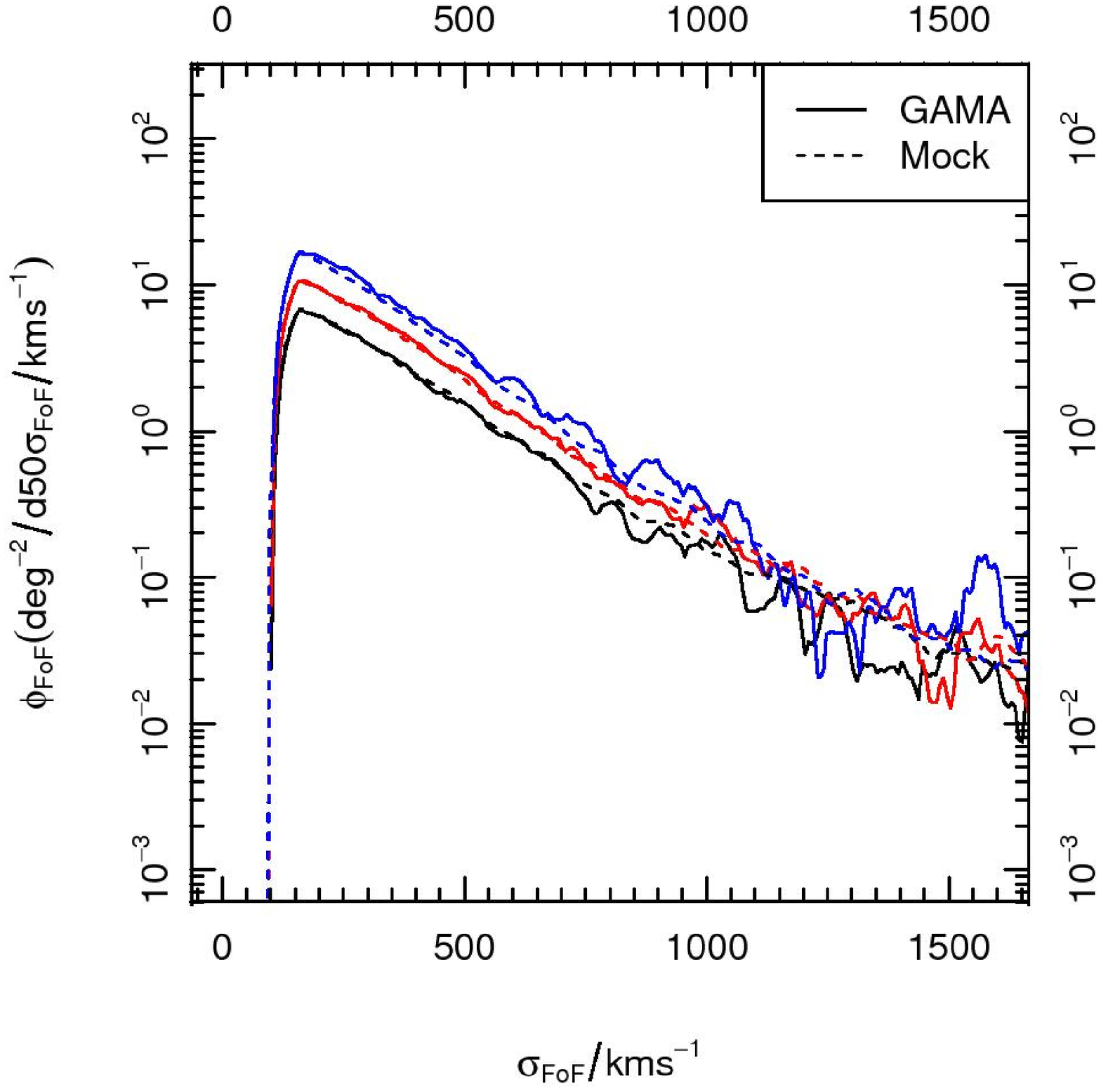}}
  \mbox{\includegraphics[width=3in]{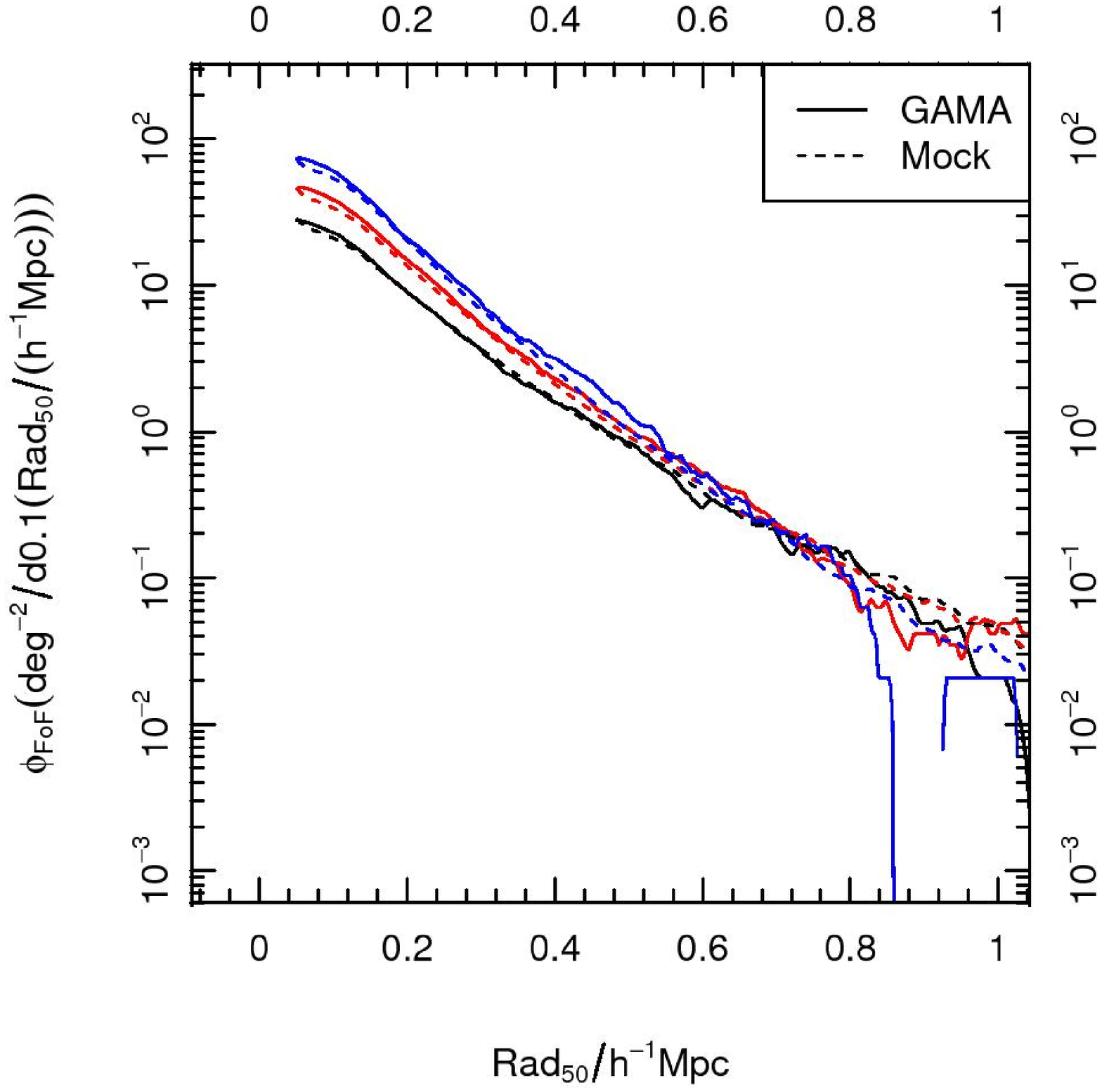}}
  }
\caption{\small Global group properties of the GAMA galaxy group catalogue (G$^3$Cv1) compared to the corresponding mock group catalogue: group multiplicity distribution (top left), dynamical group mass distribution limited to $\sigma_{\rm FoF} \ge 130\kms$ (top right), group velocity dispersion distribution limited to $\sigma_{\rm FoF} \ge 130\kms$ (bottom left) and group radius distribution (bottom right). Solid (dashed) lines for GAMA (mock) for $r_{\rm AB}\le19.0$ (black), $r_{\rm AB}\le19.4$ (red) and $r_{\rm AB}\le19.8$ survey limits. The denominator shown in the y-axis is the bin width applied, so numbers quoted are per the stated denominator. See text for discussion.
}
\label{fig:histograms}
\end{figure*}

After considering the spatial distribution of GAMA galaxy groups, Fig.~\ref{fig:histograms} shows the distributions of four basic properties of the GAMA galaxy group catalogue (G$^3$Cv1): the observed group multiplicity, mass, velocity dispersion and radius distributions. We now discuss them in turn.

The top left panel of Fig.~\ref{fig:histograms} presents the distribution of group multiplicities for three survey depths (coloured solid lines) to be compared to the equivalent average mock multiplicity distributions (dashed lines).
Unsurprisingly the raw number of groups increases with survey depth explaining why the three coloured curves are ordered as a function of survey depth, i.e. $r_{\rm AB}\le19.0$ (black), $r_{\rm AB}\le19.4$ (red) and $r_{\rm AB}\le19.8$.
More importantly, the number of high multiplicity systems is significantly different between data and mocks, a result already discussed in Table~\ref{tab:groupstat}, while their numbers are much more similar for low multiplicity systems. The difference at the high multiplicity end is important and put key constraints on the galaxy formation model used. The group multiplicity distribution is mostly sensitive to the Halo Occupation Distribution (HOD), as for a given number of haloes the group multiplicity distribution is entirely dependent on its HOD. A known feature of the GALFORM~\citet{bowe06} galaxy formation model is its tendency to populate the more massive haloes with an excess of faint satellite galaxies \citep[e.g.][]{kim09}. 

The top right panel of Fig.~\ref{fig:histograms} presents the distribution of group masses for three survey depths (coloured solid lines) to be compared to the equivalent average mass distributions from the mocks (dashed lines). 
For the comparison to be as fair as possible, the group masses used for the mocks is estimated in exactly the same way as the data. Because velocities uncertainties have not been included in the mocks it is essential to remove from this comparison all groups which velocity dispersion estimate is significantly affected by this uncertainty, as the group mass is proportional to $\sigma^2$ (see Eq.~\ref{eq:mass_est}) and would bias the distribution. To achieve this we simulated mock $\sigma$ groups with 80$\kms$ velocity error and calculated the velocity dispersion at which more than 95\% of the population should be robust to being scattered below the presumed GAMA group velocity error (which would give a corrected $\sigma$ of  0$\kms$). This velocity dispersion limit was found to be 130$\kms$. Thus the top-right panel only shows a comparison of groups where this selection has been applied.

The agreement between data and mocks beyond $\sim 10^{13}\msolh$ is remarkably good for all survey depths, with possibly only the normalisation that is slightly lower for GAMA data than for the mocks (however within the typical scatter expected from sample variance). The relative profiles are all very similar. We note that this mass distribution has been convolved with the error distribution on the group masses which have been estimated using a single correction factor ($A=10$). This explains why unrealistically large group masses are found (e.g. greater than $10^{16}\msolh$). More detailed work on estimating the group masses is underway (Alpaslan in prep).

The bottom left panel of Fig.~\ref{fig:histograms} presents the distribution of group velocity dispersions for three survey depths (coloured solid lines) to be compared to the equivalent average group velocity dispersion distributions from the mocks (dashed lines).
For the comparison to be as fair as possible, the velocity dispersion used for the mocks is estimated in exactly the same way as the data. Because velocities uncertainties have not been included in the mocks, it is essential to remove from this comparison all groups those for which the velocity dispersion estimate is significantly affected by this uncertainty. This can be straightforwardly done by ignoring groups with $\sigma \le 130\kms$ (as discussed above).
Beyond that limit in the velocity dispersion distribution, the data and mock distributions are very comparable, showing yet again how closely matched the mocks and the data are. For smaller velocity dispersion system a more careful modelling of the velocity errors (and hence velocity dispersion errors) is needed before any conclusions can be drawn on how appropriate the mocks are. Work is currently ongoing within GAMA to better understand the precise nature, and distribution, of the redshift velocity errors. A full comparison is deferred until these errors have been fully characterised.

Finally, the bottom right panel of Fig.~\ref{fig:histograms} presents the distribution of group radius for three survey depths (coloured solid lines) to be compared to the equivalent average group radius distributions from the mocks (dashed lines).
Considering the full sample of groups, the mocks and the data seem to be very comparable.

\begin{figure*}
\centerline{
  \mbox{\includegraphics[width=7in]{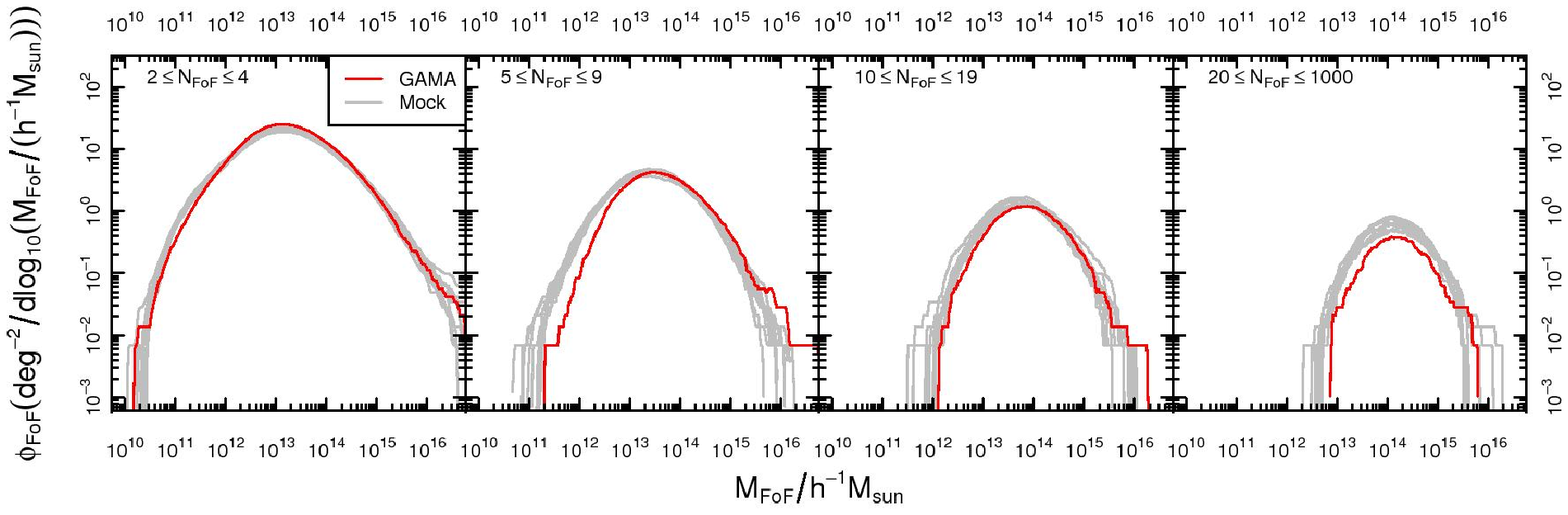}}
  }
\centerline{
  \mbox{\includegraphics[width=7in]{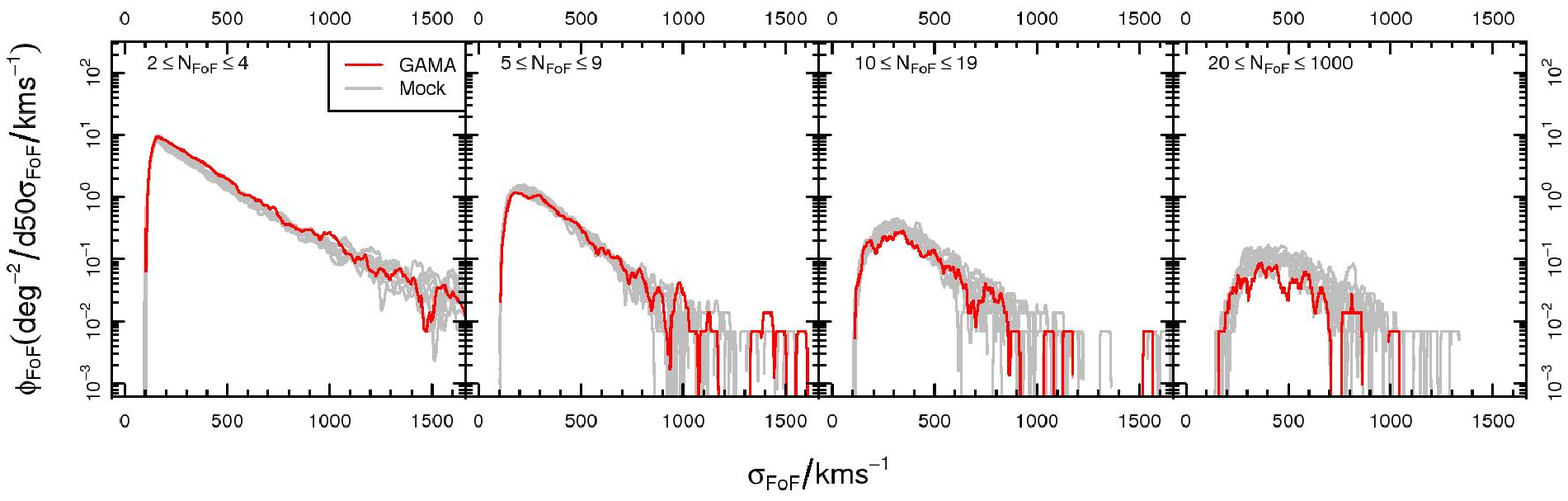}}
  }
\centerline{
  \mbox{\includegraphics[width=7in]{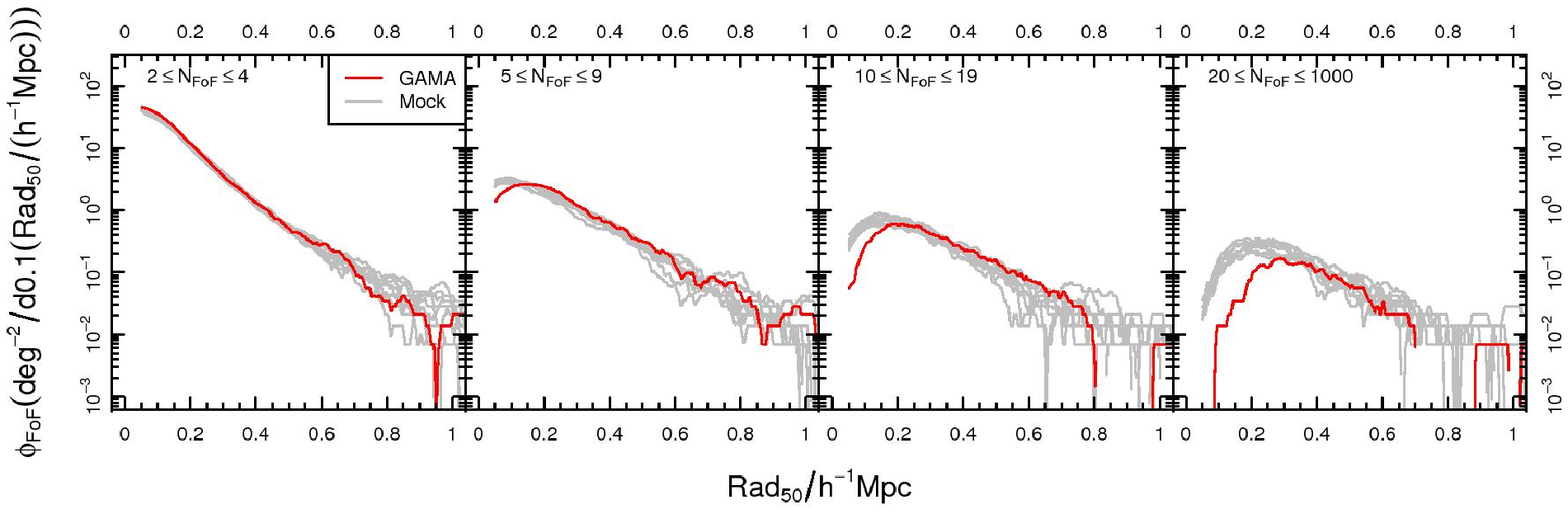}}
  }
\caption{\small Distribution of GAMA and mock galaxy group mass (top panels), velocity (middle panels) and radius (bottom panels), for a survey depth of $r_{\rm AB}\le 19.4$. GAMA is shown in red while the mocks are in grey. Multiplicity subsets are as stated in each panel. For the mass and velocity panels the mocks and GAMA data are limited to $\sigma \ge 130 \kms$, required to avoid the effects of velocity errors in the GAMA data biasing the results. For the mass and velocity plots the clearest differences are normalisation offsets, and for $N_{\rm FoF}\ge5$ there is a clear tendency for GAMA groups to have smaller $M_{\rm FoF}$ and $\sigma_{\rm FoF}$ for a given multiplicity subset. The distributions are significantly different for compact systems (Rad$_{50}\le0.2\mpch$) with $N_{\rm FoF}\ge5$, where GAMA groups are less compact in projection. This effect becomes more significant for higher multiplicity subsets.
}
\label{fig:histogramradius194}
\end{figure*}

To investigate in more detail where differences between the GAMA data and the mocks may reside we divided the mass, velocity dispersion and radius distributions into multiplicity subsets (Fig.~\ref{fig:histogramradius194}).
For clarity, Fig.~\ref{fig:histogramradius194} only uses the $r_{\rm AB}\le19.4$ survey limit, the deepest limit appropriate for all GAMA regions. Also, mock distributions for each of the 9 mock lightcones are shown with grey lines rather than the sample mean shown in Fig.~\ref{fig:histograms}. This makes allows us to see where the GAMA group distributions lie in the context of the full range of mock distributions, and therefore how significant the differences are as a function of each parameter. Plotting in this manner makes comparison much clearer than showing the error bars.
The agreement is very good for $2 \ge N_{\rm FoF} \ge 4$ for all three group properties plotted, however discrepancies are apparent for higher multiplicities both in normalisation and to a lesser extent in shape.

For the mass distributions (top panel of Fig.~\ref{fig:histogramradius194}) it is clear that GAMA possesses a lower normalisation in counts compared to the mock groups, an effect that is more noticeable for larger multiplicities. The largest deviations in the shapes of the distribution are seen for $M_{\rm FoF} \le 10^{13} M_{\sun}$, where we see excess number counts for the mock groups. This difference is most evident for  $5 \ge N_{\rm FoF} \ge 9$. The most likely explanation for this low mass excess comes from the finding that mock groups are typically more compact than GAMA groups, which will naturally cause a lower estimation of the mass. The radial discrepancies are discussed in more detail below.

The velocity dispersion (middle panel of Fig.~\ref{fig:histogramradius194}) only shows strong evidence of a normalisation offset, where the agreement is excellent for low multiplicity systems but as this increases we find the GAMA groups have a general count deficit. Since the strength of the normalisation offset varies with multiplicity the difference cannot be simply due to sample variance, where all multiplicity subsets would betray the same deficit.

The differences between GAMA and the mocks is most pronounced for the group radius (bottom panel of Fig.~\ref{fig:histogramradius194}). The most significant deviations are seen where Rad$_{50} \le 0.2\mpch$: GAMA finds many fewer systems, and the effect is much more significant for higher multiplicities where the mocks contain a significant excess of compact systems not seen at all in the data.
At the GAMA median redshift ($z \simeq 2$), $0.1\mpch$ (comoving) radius corresponds to an angular separation of 25$''$ on the sky. Whilst the simplest explanation might be the GAMA survey suffers from significant close pair incompleteness, Fig.~19 of \citet{driv11} suggests this not be the case: GAMA is better than 95\% complete for systems with up to 5 neighbours within 40$''$ (on the sky). These separations are much larger than the expected optical confusion limit (1--2``), so photometric bias (i.e.\ close pairs not being deblended) cannot explain the discrepancies we find. Since the main variance witnessed for velocity dispersions between the mocks and GAMA data is the normalisation, the more compact mock groups appear to be the origin of the low mass population we find in the top panels of Fig.~\ref{fig:histogramradius194}.

The differences seen in Fig.~\ref{fig:histogramradius194} could well be due to limitations in the physics implemented in the GALFORM~\citet{bowe06} semi-analytic galaxy formation model, where the exact distribution of galaxies within a halo depends on their dynamical friction timescale and which dark matter particle the galaxy was originally associated with. Despite the high numerical resolution of the Millennium simulation, the vast majority of the satellite galaxies in the galaxy formation model are not resolved in subhaloes, implying that their merging timescales are governed by an analytic calculation and their position is given by the most bound dark matter particle of their parent halo. A consequence of a too long merging timescale is an overabundance of galaxies at small distances away from the centre of the halo. This, together with the definition of group radius adopted for this work (i.e. Rad$_{50}$), is the most likely explanation for the apparent excess of compact groups in the mocks compared to the data. This has the consequence of also creating a deficit of low mass groups in the GAMA data in comparison to the mocks since the dynamical masses are directly proportional to the group radius measured.

In summary, the GAMA group catalogue (G$^3$Cv1) and its mock counterpart are similar in many respects, but not all. In the discussion of Fig.~\ref{fig:histograms} and Fig.~\ref{fig:histogramradius194} it has become clear that already G$^3$Cv1 is providing new constraints to the galaxy formation model used to construct the mocks and will be implemented in the next generation of mocks. 
Investigating the discrepancies between GAMA and mock group catalogues, and the impact this has on any measured HMF, 
is a complex and important task. A full analysis is deferred to a GAMA paper in preparation, which will present a more in depth analysis of a series of statistically equivalent mocks as well as galaxy formation based mocks as used here. Only with a large variety of mocks will it be possible to put realistic constraints on the underlying dark matter model. The analysis in the present paper is entirely limited to one family of mock realisations, which explains why the constraints from the GAMA groups are so far mostly limited to possible constraints on the galaxy formation model rather than on the underlying dark matter physics.

\section{Group examples}
\label{sec:examples}

\begin{figure*}
\centerline{
  \mbox{\includegraphics[width=3in]{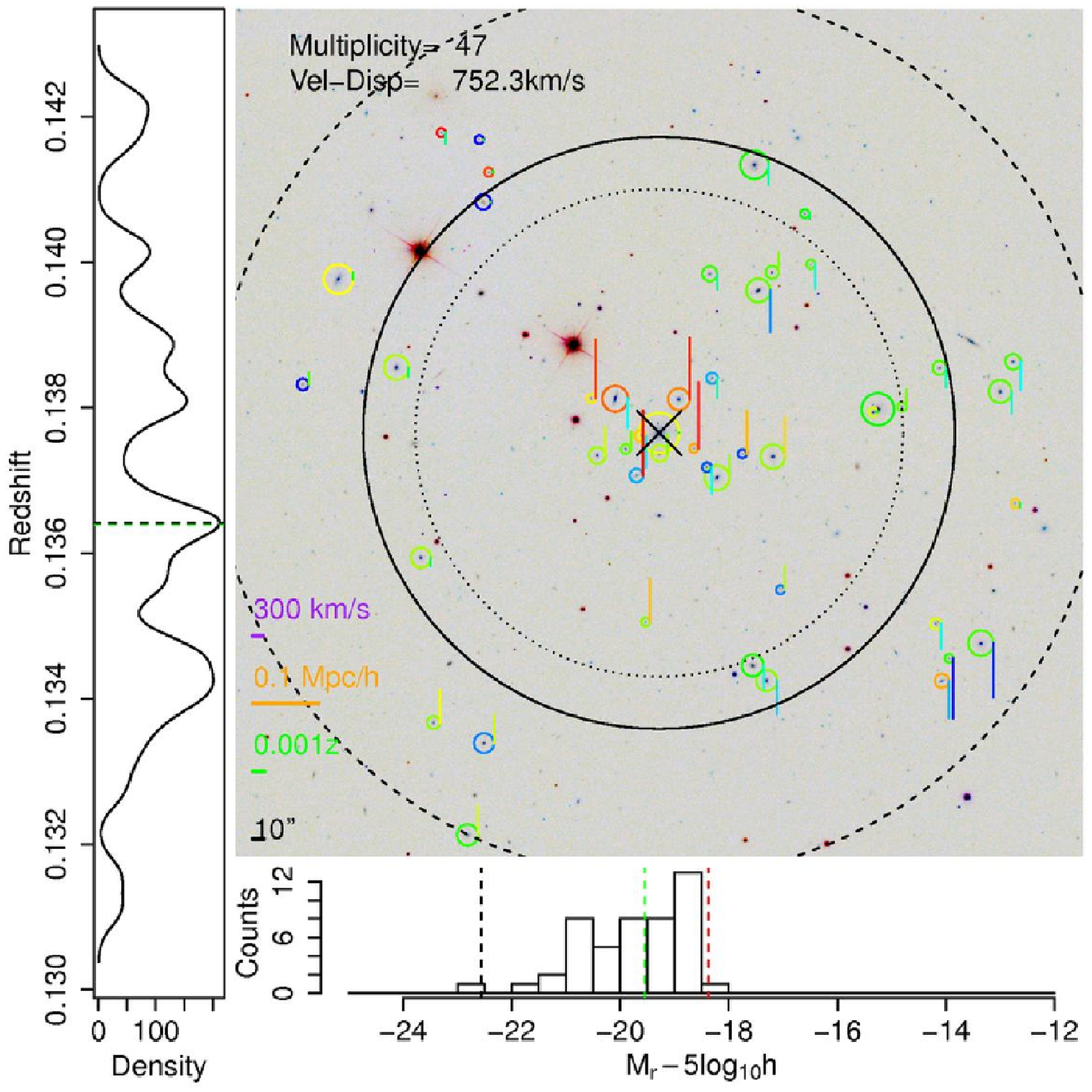}}
  \mbox{\includegraphics[width=3in]{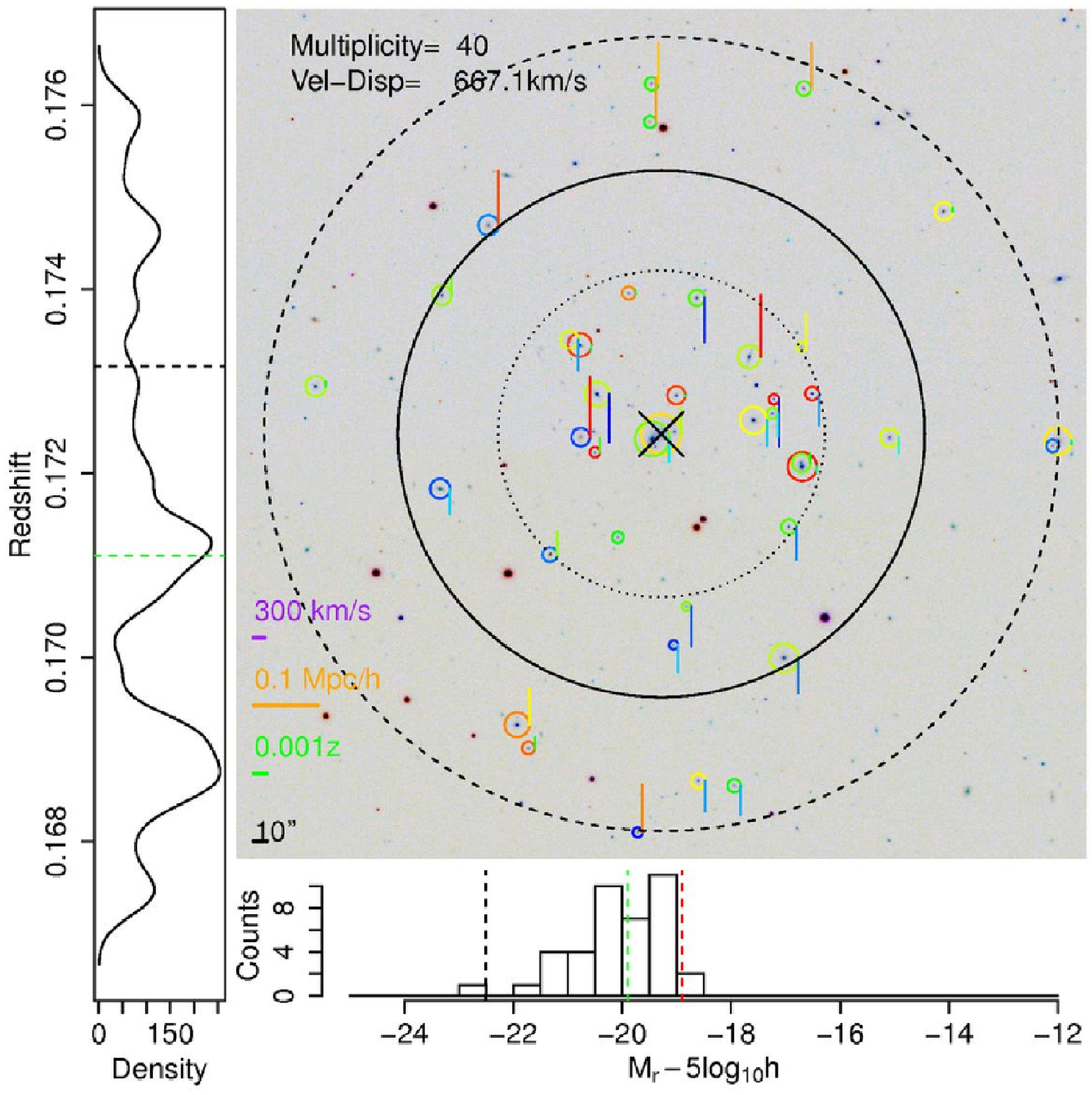}}
  }
\centerline{
  \mbox{\includegraphics[width=3in]{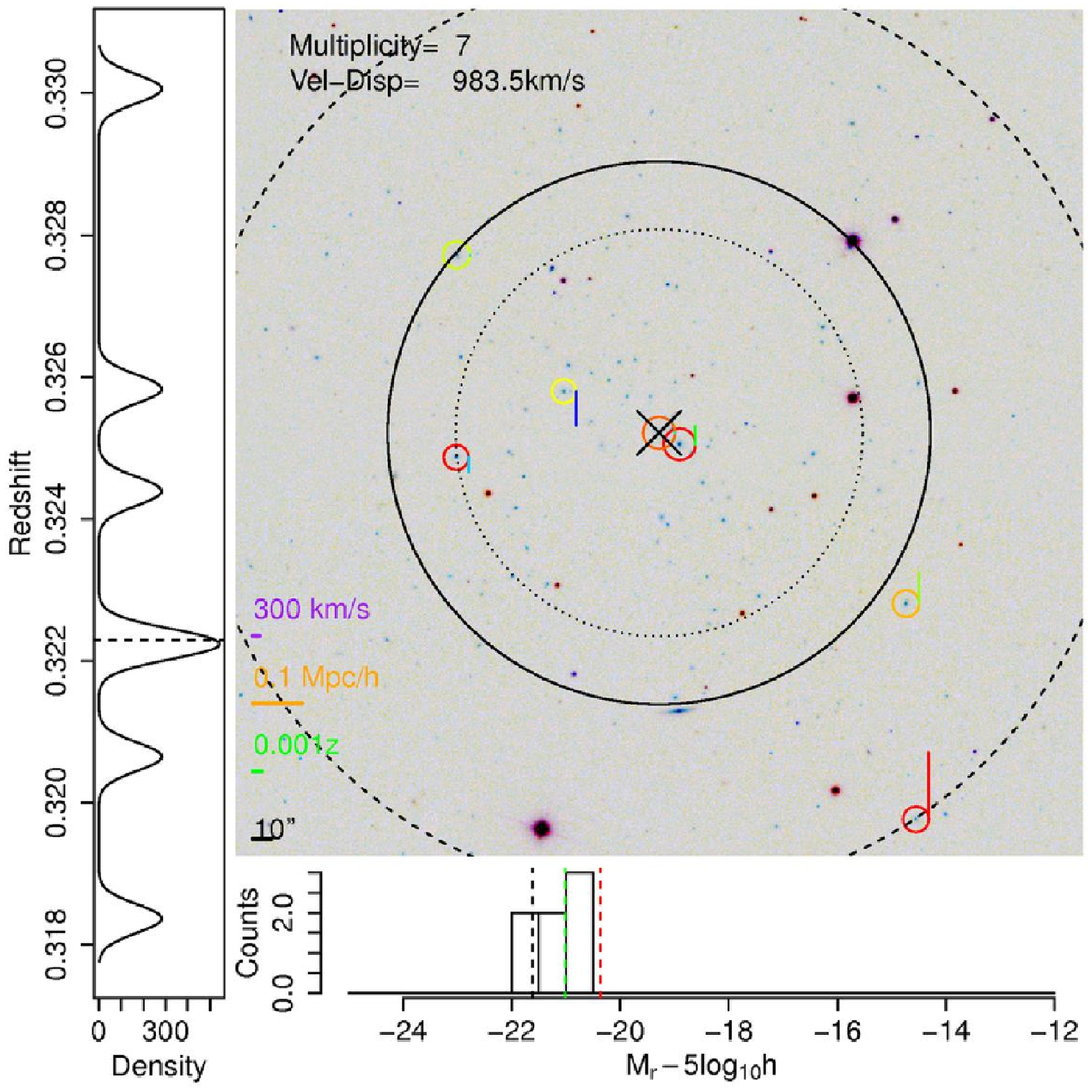}}
  \mbox{\includegraphics[width=3in]{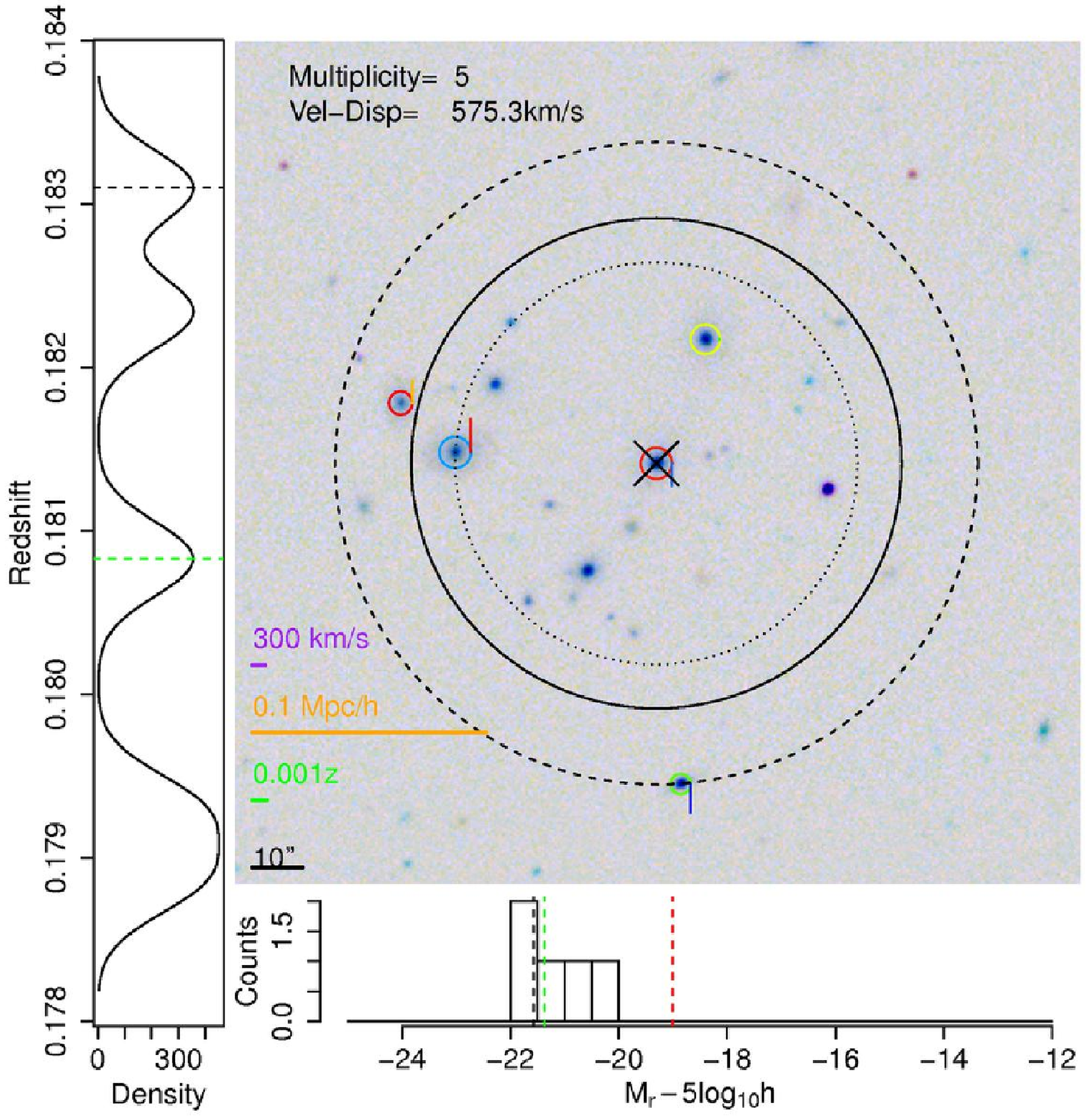}}
  }
\caption{\small Top panels show two cluster scale groups confirmed spectroscopically. Bottom panels show low multiplicity groups with significant, possibly associated, background galaxies. The rgb image is a $K_{\rm AB}$-$r_{\rm AB}$-$u_{\rm AB}$-band composite. The size of the circle marking group members scales with the $r_{\rm AB}$-band flux and its colour reflects the galaxy $u_{\rm AB}-r_{\rm AB}$ colour. A galaxy redshifted w.r.t.\ the group median redshift has a red upwards pointing line which length scales with the velocity difference, while for a blueshifted one the line is blue and points downwards. The rings represent the $50^{th}$, $68^{th}$ and $100^{th}$ percentiles of the radial galaxy distributions relative to the iterative group centre. { The velocity PDF smoothed with a Gaussian kernel of width $\sigma=50\kms$ (the typical GAMA velocity error) is shown on the left of each panel, where the group median is shown with a green dashed-line and the BCG with a black dashed line.} The bottom plot presents the raw absolute $r_{\rm AB}$ magnitude distribution of the group, with the effective GAMA survey limit shown with a red dashed-line, the group median absolute magnitude with green, and the BCG absolute magnitude with black.}
\label{clusterimages}
\end{figure*}

\begin{figure*}
\centerline{
  \mbox{\includegraphics[width=3in]{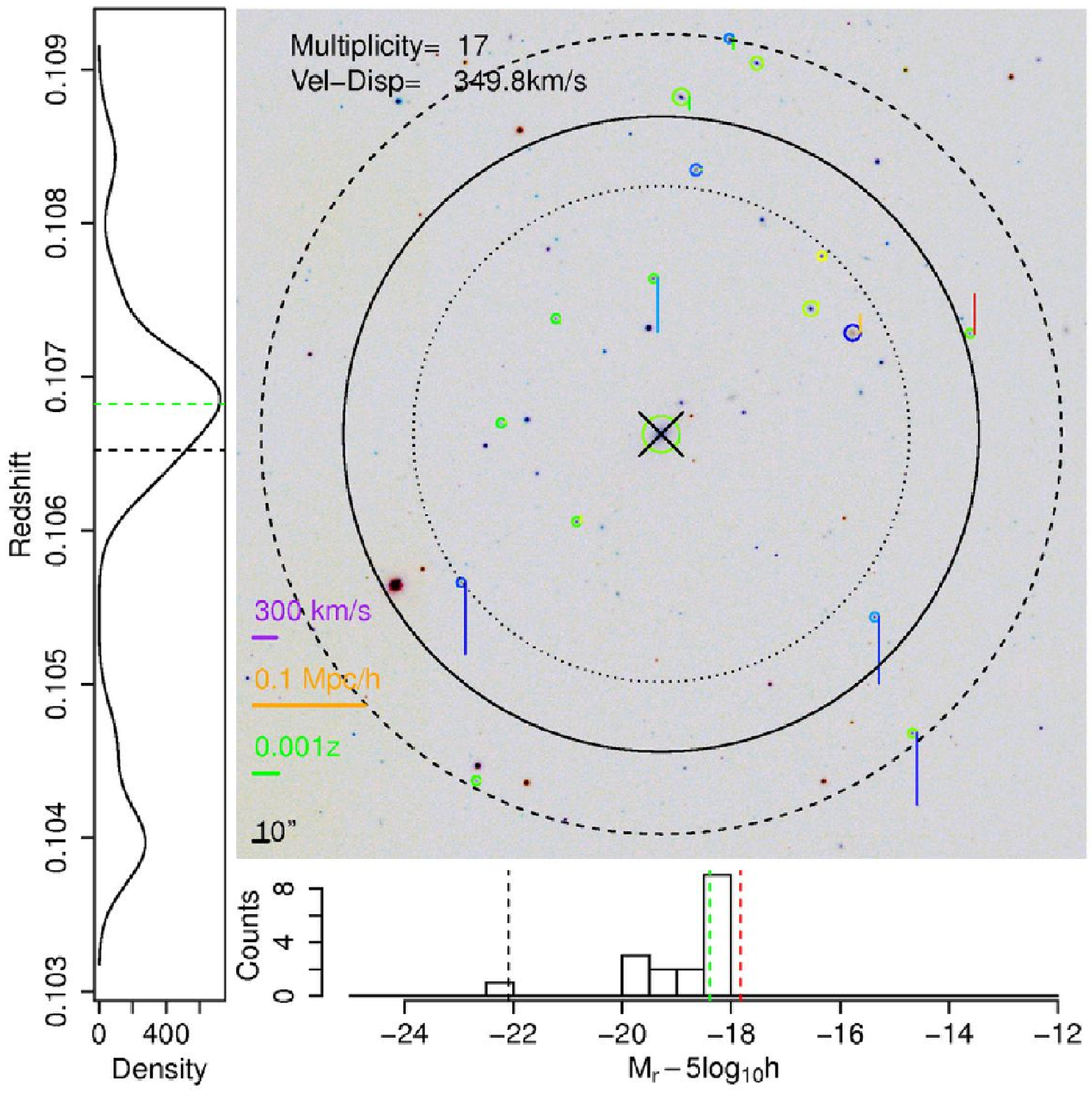}}
  \mbox{\includegraphics[width=3in]{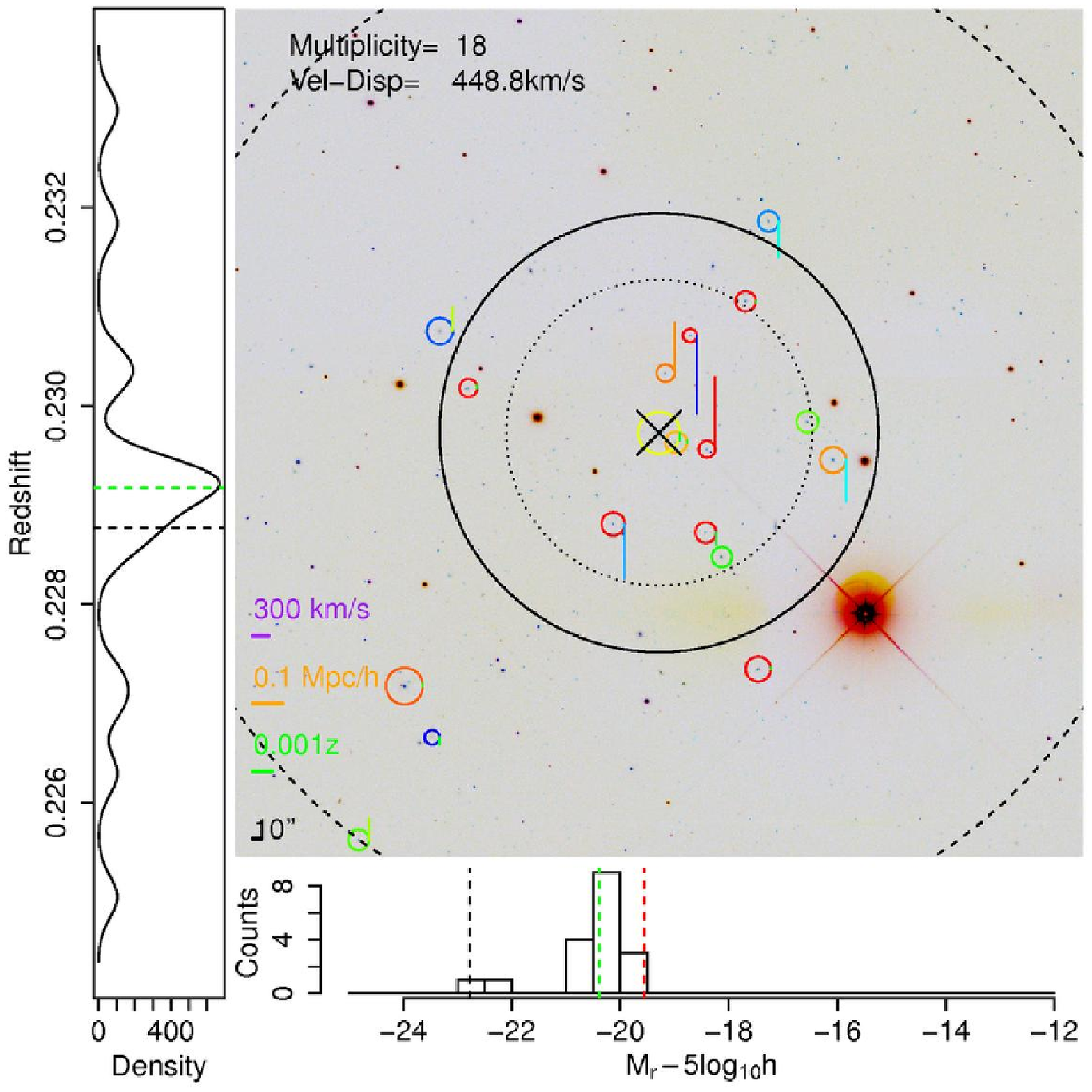}}
  }
\centerline{
  \mbox{\includegraphics[width=3in]{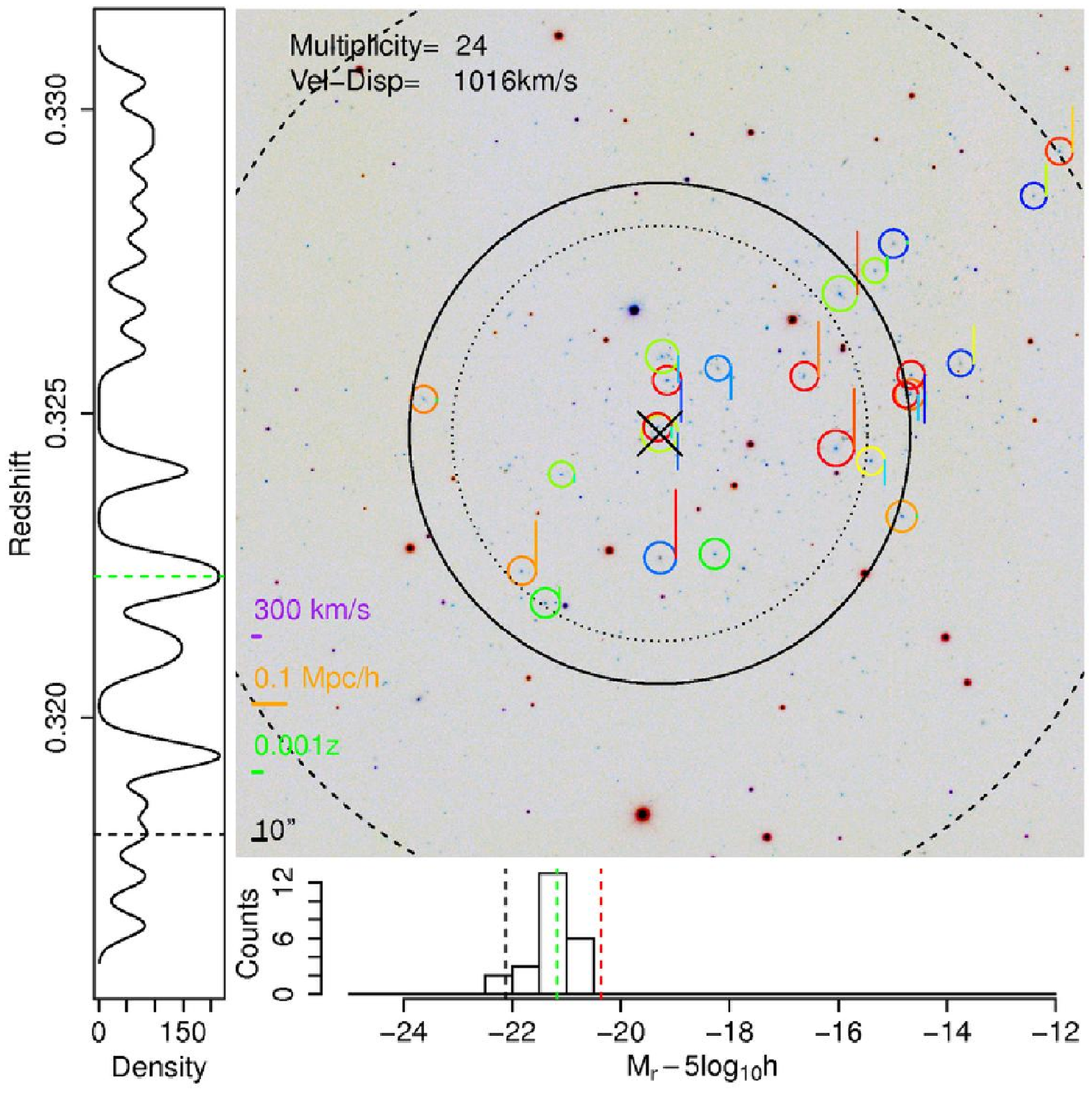}}
  \mbox{\includegraphics[width=3in]{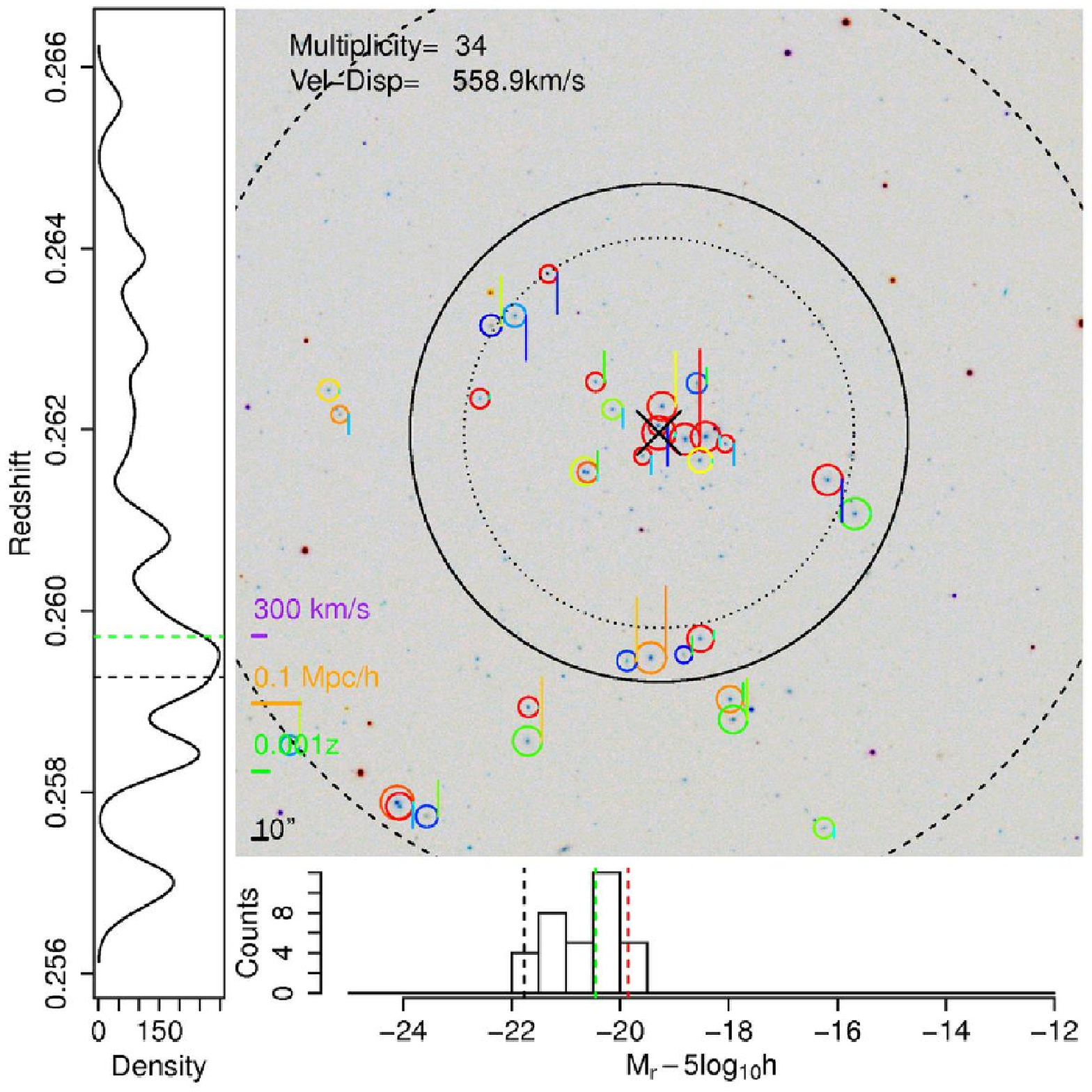}}
  }
\caption{\small Top panels are potential fossil-groups, where the BCG is at least 2 magnitudes brighter than the second ranked galaxy in the $r_{\rm AB}$-band (in the case of the top-right groups the second rank galaxy is nearer in magnitude than this, but it is separated a large distance in projection). Bottom panels show groups with complex in-fall structure. See Fig.~\ref{clusterimages} for figure description.}
\label{evolveimages}
\end{figure*}

\begin{figure*}
\centerline{
  \mbox{\includegraphics[width=3in]{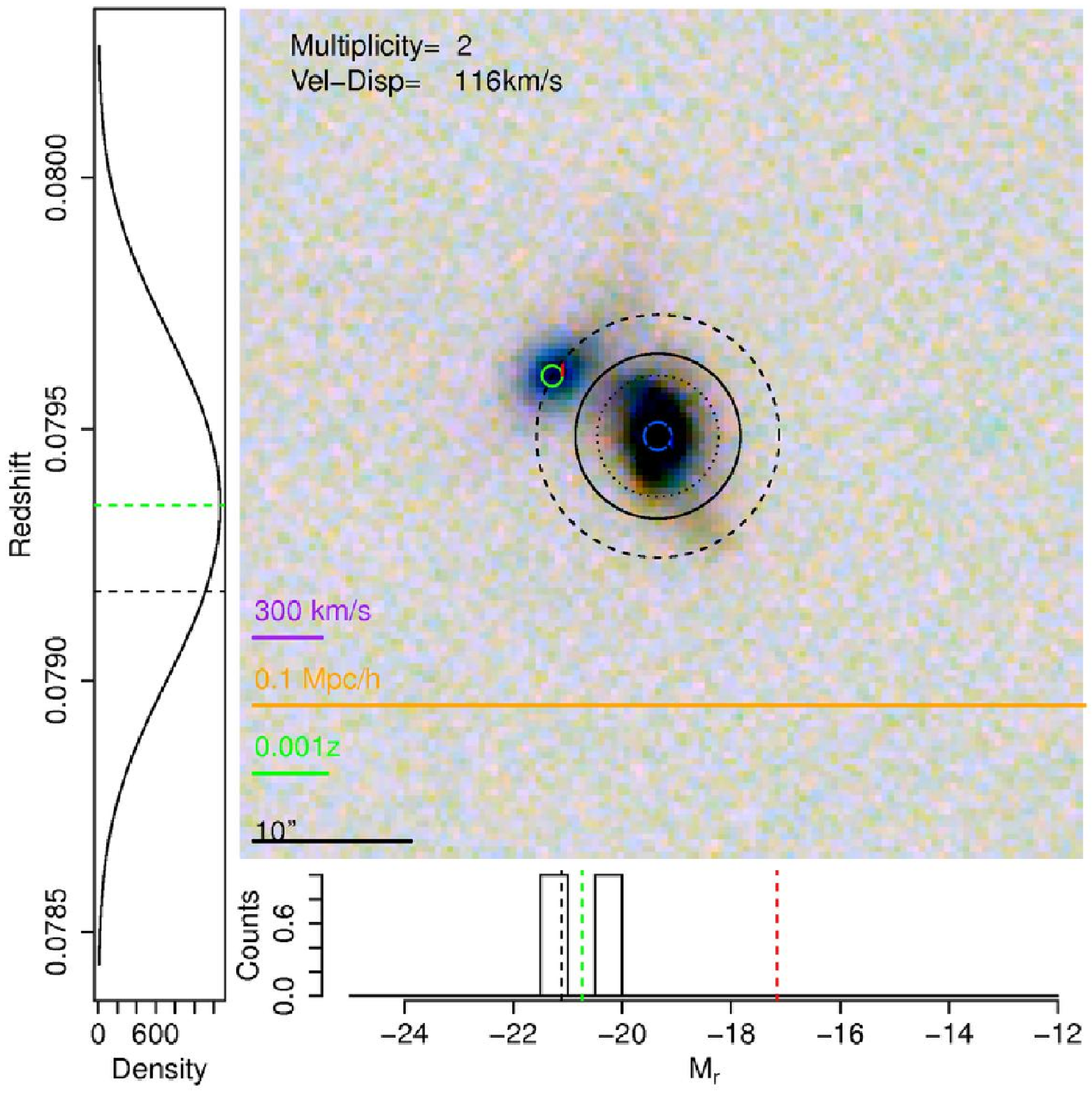}}
  \mbox{\includegraphics[width=3in]{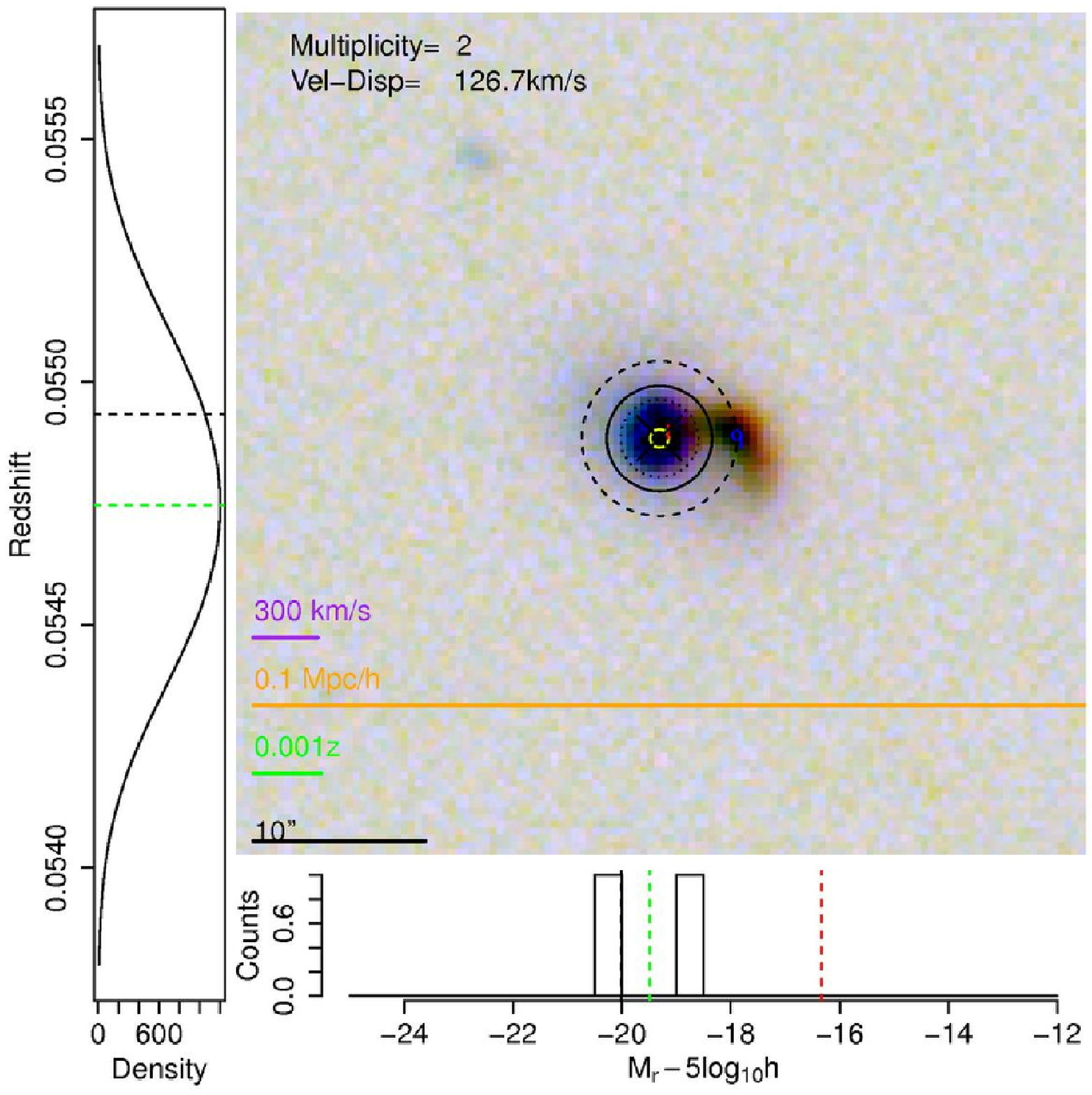}}
  }
\centerline{
  \mbox{\includegraphics[width=3in]{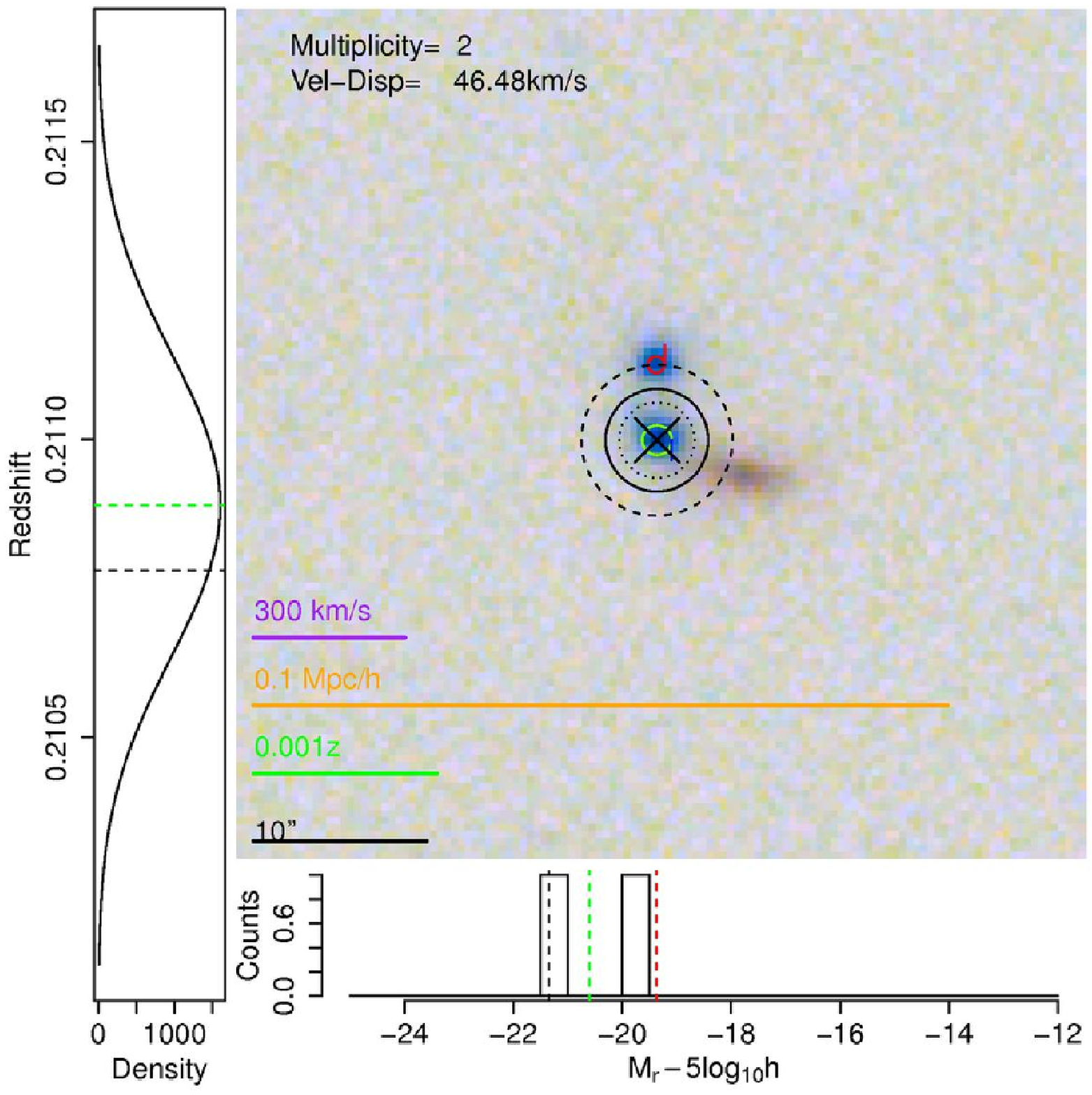}}
  \mbox{\includegraphics[width=3in]{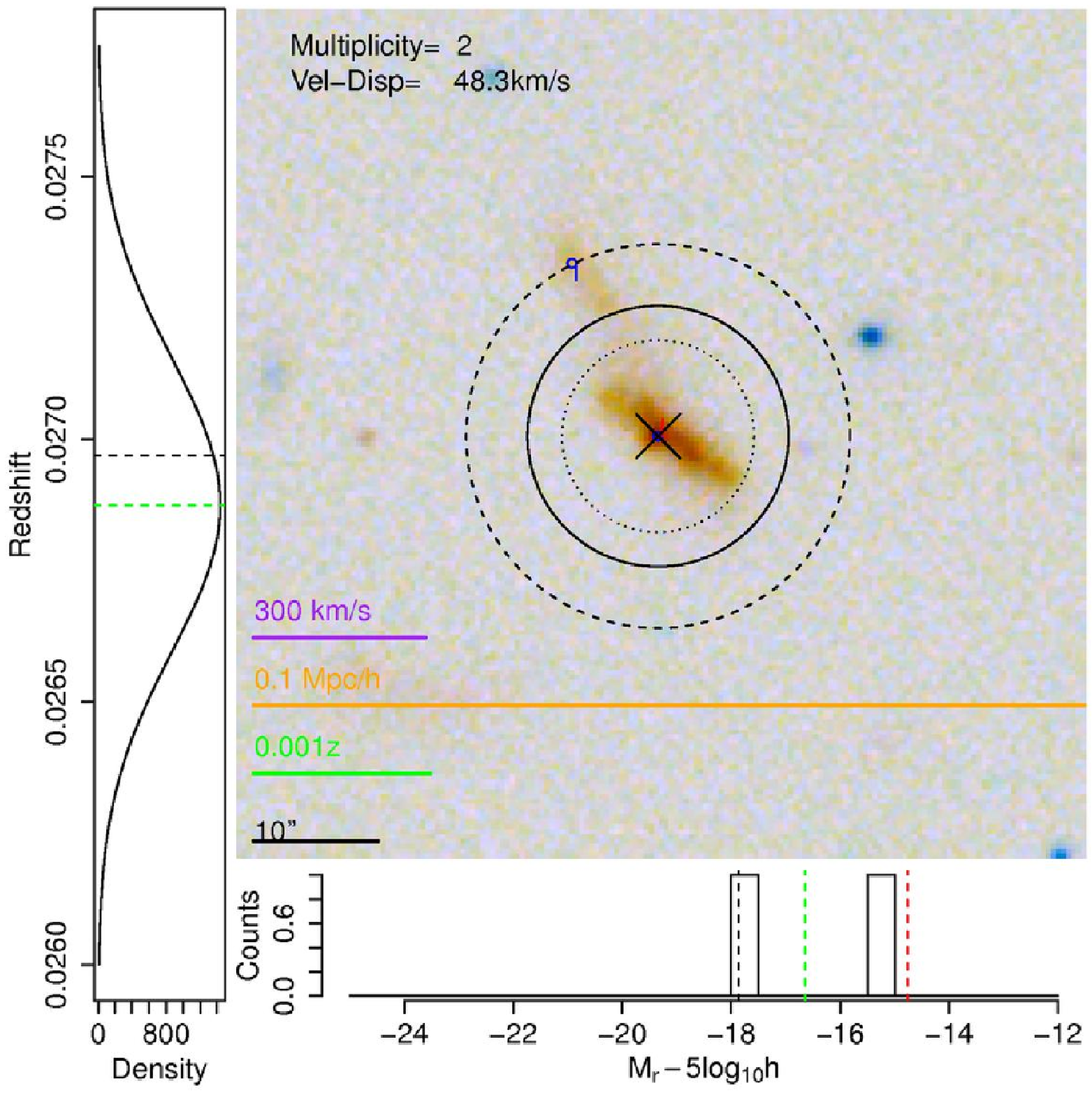}}
  }
\caption{\small Examples of ultra low-mass groups that are also excellent candidates for merging systems. The bottom plots are groups that are within the nominal SDSS $r_{\rm AB} \le 17.77$ limit, but one or both galaxies are missing from that survey due to fibre collisions. The bottom plots are groups that are both too faint and too close together to be present in a spectroscopic SDSS catalogue. See Fig.~\ref{clusterimages} for figure description.}
\label{mergerimages}
\end{figure*}

For every group we create a rgb image as a $K_{\rm AB}$-$r_{\rm AB}$-$u_{\rm AB}$-band composite, 
along with visual diagnostics that allow interesting features to be easily identified. Example images are shown in Fig.~\ref{clusterimages}, Fig.~\ref{evolveimages} and Fig.~\ref{mergerimages} and discussed hereafter.

Fig.~\ref{clusterimages} highlights 4 cluster-scale groups extracted from the GAMA data. The top panel shows relatively low redshift clusters with high multiplicities, whilst the bottom panels are examples of low multiplicity groups that show evidence for a lot of associated galaxies that are fainter than the GAMA survey limits (shown by a dashed red line on the luminosity distribution plotted in each panel). All of these groups are quite circularly symmetric and concentrated towards the centre, both of which are indicators of a well virialised population of galaxies.

Fig.~\ref{evolveimages} shows groups at radically different stages of evolution. The top panels show examples of fossil groups with one exceptionally dominant BCG. In both cases only the BCG had a known redshift before GAMA, and the large peak in the redshift distribution suggests particularly strong radial linking--- an indication that the grouping is reliable. The bottom panels show groups with very loose association in comparison. Both groups are quite massive (in the cluster regime) and have identifiable background galaxies, but neither exhibits a centrally concentrated distribution of galaxies or a dominant BCG. Both of these groups have a relatively uniform redshift distribution, showing none of the strong central peak seen for the fossil groups in the top panel. The bottom-right group in particular has a very flat luminosity distribution and an extremely non-circular distribution of galaxies. The most likely scenario is that this group has two distinct sub-structures (top and bottom) collapsing into each other, where the bottom structure is physically nearer to us in space and thus exhibits a large extra component of recessional velocity towards the CoM.

Fig.~\ref{mergerimages} shows particularly pleasing examples of galaxy-galaxy merging/interactions. A natural outcome from the GAMA group catalogue is that nearly all possible close-pairs will be grouped (modulo a very small amount of incompleteness). Often these merging systems will be found in higher multiplicity systems, but here are examples of two member groups that exhibit evidence for mergers. The top-left and top-right panels show quite similar looking systems: a red (likely passive) galaxy interacting with a blue (late-type) galaxy. The top-left panel has larger tidal tails and more of the flux is in the late-type system, suggesting it is at an early stage of the merging process. The top panels are examples where the multi-pass nature of GAMA has overcome the problems of fibre collisions to give us redshifts for both galaxies in the merging system. The bottom panels show merging systems that are both too faint and too close to be obtainable with SDSS data. The bottom-left panel system appears to be a triple merger system, where the blue galaxy to the right does not have GAMA redshift because it is too faint. The bottom-right panel shows two extremely faint and relatively $u$-band bright galaxies merging--- a tidal connection can be seen between them. In both of these bottom panels the groups in question have extremely low velocity dispersions ($\sim 45\kms$) and very low implied dynamical masses ($\sim 10^{10}\msolh$).

In such systems dynamical friction is acting in such a manner that the dynamical mass will likely not be a good indicator of the intrinsic halo mass, rather it highlights a system where the energy has been transferred from group scale kinematics (energy in galaxies) to galaxy scale kinematics (energy in the stars/ gas). Dynamical friction conspires to reduce the velocity difference and physical distance between merging galaxies, and since we use $M_{\rm FoF} \propto \sigma^2 R$ this will also reduce the implied dynamical mass that we measure.

\subsection{GAMA group catalogues}
\label{sec:catalogues}

The generation of a group catalogue produces a myriad of outputs, most of which are not of interest to the typical user. To ease interpretation for the average user, a deliberately simplified set of outputs will be made available. For each GAMA region two tables are released. The first one is a two column link list that identifies which group every galaxy belongs. The second is a table of group properties with the most important attributes of each group. This includes the group radius Rad$_{50}$, the velocity dispersion $\sigma_{\rm FoF}$, the implied dynamical mass.
Other metrics related to each group are also calculated to aid the analysis and interpretation of individual grouping quality. As well as the L$_{\rm proj}$ linking quality discussed above, the kurtosis of the radial separation of all galaxies from the group centre is calculated and the `modality' of the system is also computed using $(1+{\rm skewness}^2)/(3+{\rm kurtosis}^2)$. This will be 1/3 for a normal distribution and 0.555 for a uniform, and is a useful metric since it does not just provide information on how non-Gaussian the velocity profile of each system is--- it also provides information on the whether the velocity profile is more cusped or cored than a Gaussian distribution. Additionally, in a similar manner to how the local over-density was calculated in a comoving cylinder centred around each galaxy, the local relative density is calculated for each group. This is calculated using a comoving cylinder of radius $1.5\mpch$ and total radial depth of $36\mpch$, and gives a measure of how isolated the groups are relative to much larger scale structure.

Finally, as a separate but useful output from creating the GAMA galaxy group catalogue, a full pair catalogue will be released. This is a natural output of the galaxy--galaxy linking stage of the grouping algorithm, and includes all pairs that are within a common velocity separation of 1000$\kms$ and a physical projected separation of 50$\,{\it h}^{-1}\, {\rm kpc}$. This will be used within the team for work involving the study of galaxy pairs.

\section{Conclusions}
\label{sec:conclusions}

In this paper we have presented a new group catalogue based on the spectroscopic component of the GAMA survey. The FoF based grouping algorithm used has been extensively tested on semi-analytic derived mock catalogues, and has been designed to be extremely robust to the effects of outliers and linking errors. The velocity dispersion and radius of the groups are median unbiased, even when allowing for the possibility of catastrophic grouping errors. Globally, 77\% of the recovered FoF groups bijectively (unambiguously) match a mock group, and 89\% of all mock groups are bijectively recovered. The purity of all FoF groups is 80\%, and for mock groups the equivalent figure is 73\%. This suggests that the FoF algorithm is quite well balanced and does not have a strong preference to over-grouping or to conservatively recovering just the strongly bound core of systems.

The overall number of groups within from $0\le z \le 0.5$ is remarkably consistent between the mocks and real groups, and for the most part comfortably within the range expected given the large sample variance that can affect galaxy surveys such as GAMA. The histograms of raw group multiplicity and dynamically estimated group mass show a large amount of agreement between the GAMA data and the mock catalogues for the most part. Discrepancies appear at the high multiplicity end, where GAMA finds fewer high multiplicity systems than recovered from the mock volumes. 
A more in depth analysis of the discrepancies between GAMA and mock groups
is deferred to a later paper, still in preparation.

The showcase examples of a small number of GAMA groups highlight the parameter space that is now opened up, and demonstrate the advantages brought by having extremely high spatial completeness. Accurate group dynamics and a full sample of close pairs will be of key importance for determining the Halo Mass Function in upcoming work, and for finding new constraints on the galaxy merger rate in the local Universe, two of the main goals of the GAMA survey.

{ The G$^3$C will be made publicly available on the GAMA website ({\tt http://www.gama-survey.org}) as soon as the associated redshift data are made available. Interested parties should contact the author at {\tt asgr@st-and.ac.uk} if they wish to make use of the group catalogue data before the full public release.}

\section*{Acknowledgments}
We thank Vincent Eke for his helpful refereeing comments. These added clarity to various aspects of the paper.
ASGR acknowledges STFC and SUPA funding that was used to do this work.
PN acknowledges a Royal Society URF, an ERC StG grant (DEGAS-259586) and STFC funding.
GAMA is a joint European-Australasian project based around a spectroscopic campaign using the Anglo-Australian Telescope. The GAMA input catalogue is based on data taken from the Sloan Digital Sky Survey and the UKIRT Infrared Deep Sky Survey. Complementary imaging of the GAMA regions is being obtained by a number of independent survey programs including GALEX MIS, VST KIDS, VISTA VIKING, WISE, Herschel-ATLAS, GMRT and ASKAP providing UV to radio coverage. GAMA is funded by the STFC (UK), the ARC (Australia), the AAO, and the participating institutions. The GAMA website is {\tt http://www.gama-survey.org/}.

\label{lastpage}

\bibliographystyle{mn2e}
\setlength{\bibhang}{2.0em}
\setlength\labelwidth{0.0em}
\bibliography{/Users/aaron/Work/refs/library}

\end{document}